\documentstyle[prb,aps,tighten,epsfig]{revtex}

\begin{document}
\author{Pierre Le Doussal}
\address{CNRS-Laboratoire de Physique Th\'eorique de l'Ecole\\
Normale Sup\'erieure, 24 rue Lhomond, F-75231
Paris\cite{ledoussal_labpro}}
\author{Thierry Giamarchi}
\address{Laboratoire de Physique des Solides, Universit{\'e} Paris-Sud,\\
B{\^a}t. 510, 91405 Orsay, France\cite{giam_labasc}}
\title{Moving glass theory of driven lattices with disorder}
\date{\today}
\maketitle

\begin{abstract}
We study periodic structures, such as vortex lattices, moving
in a random pinning potential under the action of an external
driving force. As predicted in [T. Giamarchi, P. Le Doussal
Phys. Rev. Lett. {\bf 76} 3408 (1996)] the periodicity in the direction
{\bf transverse} to motion leads to to a new class of 
driven systems: the Moving Glasses. 
We analyse using several renormalization
group techniques the physical properties of such systems
both at zero and non zero temperature.
The Moving glass has the following 
generic properties (in $d \leq 3$ for uncorrelated disorder)
(i) decay of translational long range order 
(ii) particles flow along static channels (iii) the 
channel pattern is highly correlated along the direction
transverse to motion through elastic compression modes 
(iv) there are barriers to transverse motion. 
We demonstrate the existence of the ``transverse critical force''
at $T=0$ and study the transverse depinning. A ``static random force''
both in longitudinal and transverse directions is shown
to be generated by motion. Displacements are found to grow logarithmically
at large scale in $d=3$ and as a 
power law in $d=2$. The persistence of quasi long range
translational order in $d=3$ at weak disorder,
or large velocity leads to the prediction of the
topologically ordered ``Moving Bragg Glass''.
This dynamical phase which is a continuation of the static Bragg glass
studied previously, is shown to be stable to a non zero temperature.
At finite but low temperature, the channels broaden but
survive and strong non linear effects still exist in the 
transverse response, though the asymptotic behavior
is found to be linear. In $d=2$, or in $d=3$ at intermediate
disorder, another moving glass state exist,
which retains smectic order in the transverse direction: the Moving Transverse
Glass. It is described by the Moving glass equation introduced in
our previous work.
The existence of channels allows to naturally describe the transition
towards plastic flow. We propose a phase diagram in temperature, force 
and disorder for the static and moving structures.
For correlated disorder we predict a ``moving Bose glass'' state
with anisotropic transverse Meissner effect,
localization and transverse pinning. We discuss the effect of additional
linear and non linear terms generated at large scale in the equation of motion.
Generalizations of the Moving glass equation
to a larger class of non potential glassy systems described by zero temperature
and non zero temperature disordered fixed points (dissipative glasses)
are proposed. We discuss experimental consequences
for several systems, such as anomalous Hall effect in the Wigner crystal,
transverse critical current in the vortex lattice, and solid friction.

\end{abstract}
\pacs{to be added}


\widetext

\section{Introduction}

Interacting systems which tend to form spontaneously periodic
structures can exhibit a remarkable variety of complex phenomena
when they are driven by an external force over a
disordered substrate. Many of these phenomena, which arise
from the interplay between elasticity, periodicity, quenched disorder,
non linearities and driving, are still poorly understood
or even unexplored. For numerous such experimental systems,
transport experiments are usually a
convenient way to probe the physics
(and sometimes the only way when more direct methods -
e.g. imaging are not available). It is thus an important and challenging
problem to obtain a quantitative description of their
driven dynamics. Vortex lattices in type II superconductors
are a prominent example of such systems \cite{blatter_vortex_review}.
The motion of the
lattice under the action of the Lorentz force (associated to a transport
supercurrent) in the presence of pinning impurities has been
studied in many recent experiments 
\cite{charalambous_melting_rc,safar_transport_tricritical,kwok_electron_defects,%
danna_steps_ybco,yaron_nature2,higgins_second_peak,hellerqvist_ordering_long}
There are other examples of well studied driven systems where quenched disorder
is known to be important, such as the
two dimensional electron gas in a magnetic field which forms
a Wigner crystal
\cite{andrei_wigner_2d,perruchot_prl,perruchot_thesis} moving under an
applied voltage, lattices of magnetic bubbles
\cite{seshadri_bubbles_thermal,seshadri_bubbles_long}
moving under an applied magnetic field gradient, Charge Density Waves (CDW)
\cite{gruner_revue_cdw} or colloids \cite{murray_colloid_prb}
submitted to an electric field, driven josephson junction arrays
\cite{vinokur_josephson_short,balents_josephson_long} etc..
This problem may also be important in understanding tribology
and solid friction phenomena
\cite{cule_hwa_friction}, surface growth of crystals with quenched bulk
or substrate disorder \cite{toner_log_2}, domain walls in incommensurate solids
\cite{pokrovsky_talapov_prl}.
One striking property exhibited
by all these systems is pinning, i.e
the fact that at low temperature there is no macroscopic motion
unless the applied force $f$ is larger than a
threshold critical force $f_c$. Dynamic properties have thus been studied
for some time, quite extensively near the depinning threshold
\cite{fisher_depinning_meanfield,narayan_fisher_depinning,%
nattermann_stepanow_depinning},
but mostly in the context of CDW
\cite{littlewood_sliding_cdw,sneddon_cross_fisher,narayan_fisher_cdw}
or for models based on driven manifolds 
\cite{kardar_review_lines,ertas_kardar_anisotropic}
and their relation to
growth processes \cite{krim_growth_review} described by the Kardar Parisi
Zhang (KPZ) equation\cite{kardar_parisi_zhang,kardar_review_lines}
They are however, far from being fully understood. In addition,
the full problem of a periodic {\it lattice}
(with additional periodicity transverse to the direction of motion)
was not scrutinized until very recently \cite{giamarchi_book_young}.

 A crucial question in both the dynamics and
the statics is whether topological defects in the periodic structure
are generated by disorder, temperature and the driving force
or their combined effect. Another important issue is to characterize
the degree of order (e.g translational order, or temporal order)
in the structure in presence of quenched disorder.
In the absence of topological defects
it is sufficient in the statics to consider only elastic deformations.
In the dynamics this leads to {\it elastic flow}.
On the other hand, if these defects exist (e.g
unbound dislocation loops) the {\it internal} periodicity
of the structure is lost and one must consider also plastic
deformations. In the dynamics the flow will then not be elastic but turn into
{\it plastic flow} with a radically different behaviour.

The {\it statics} of lattices with impurity disorder has been much
investigated recently, especially in the context of type II
superconductors. It was generally agreed that disorder
leads to a {\it glass phase} (often called \cite{fisher_vortexglass_short} a
{\it vortex glass}) with many metastable
states, diverging barriers between these states
\cite{feigelman_collective,blatter_vortex_review}, pinning
and loss of translational order. Indeed, general arguments
\cite{fisher_vortexglass_long,villain_cosine_realrg}, unchallenged
until recently, tended to show that disorder would
always favor the presence of dislocations destroying the Abrikosov lattice
beyond some length scale. In a series of recent works
\cite{giamarchi_vortex_short,giamarchi_vortex_long,giamarchi_vortex_comment,%
giamarchi_diagphas_prb},
we have obtained a different picture of the {\it statics} of
disordered lattices (including vortex lattices) and
predicted the existence of a new thermodynamic phase,
the {\it Bragg glass}. The Bragg glass has the following properties:
(i) it is topologically ordered
(ii) relative displacements grow only logarithmically
at large scale
(iii) translational order decays at most algebraically
and there are divergent Bragg peaks in the structure
function in $d=3$ (i.e quasi long range order survives).
(iv) it is nevertheless a true static glass phase with diverging barriers.
There has been several analytical 
\cite{carpentier_bglass_layered,kierfeld_bglass_layered,fisher_bragg_proof}
and numerical studies
\cite{ryu_diagphas_numerics,gingras_dislocations_numerics} confirming this theory.
The predicted consequences for
the phase diagram of superconductors compare well with 
the most recent experiments \cite{giamarchi_diagphas_prb}

While some progress towards a consistent theoretical treatment
has been made in the statics, it is still further removed
in the dynamics. Determining the various phases of driven
system is still a widely open question.
Evidence based mostly on experiments, numerical simulations
and qualitative arguments indicates that quite generally
one expects plastic motion for either strong disorder situations,
high temperature, or near the depinning threshold in low dimensions
(for CDW see e.g.\cite{coppersmith_defects_cdw}).
Indeed there has been a large number of studies on
plastic (defective) flow \cite{shi_berlinsky,pla_nori,reichhardt_nori_periodic}.
In the
context of superconductors a $H$-$T$ phase diagram
with regions of elastic flow and regions of
plastic flow was observed\cite{bhattacharya_peak_prl,higgins_second_peak}.
Several experimental effects have been attributed to
plastic flow, such as the peak effect
\cite{bhattacharya_peak_effect1,higgins_second_peak,wordenweber_kes_peak,%
berghuis_kes_peak},
unusual broadband noise \cite{marley_broadband_noise} and
fingerprint phenomena in the I-V curve
\cite{bhattacharya_fingerprints,hellerqvist_ordering_short,%
hellerqvist_ordering_long}.
Steps in the $I-V$ curve were also observed in YBCO near melting in
\cite{danna_steps_ybco}.
Close to the threshold and in strong disorder situations
the depinning is observed to proceed through what can be called ``plastic
channels'' \cite{jensen_plasticflow_short,jensen_plasticflow_long}
between pinned regions. This type of filamentary flow has been found
in \cite{gronbech_jensen_filamentary} in simulations
of 2D (strong disorder) thin film geometry (with $c_{11} \gg c_{66}$).
Depinning then proceeds via
filamentary channels which become increasingly denser.
Filamentary flow was
proposed as an explanation for
the observed sharp dynamical transition
observed in MoGe films \cite{hellerqvist_ordering_short,%
hellerqvist_ordering_long}
characterized by abrupt steps in the differential resistance. Also,
interesting effects of synchronization of the flow in different channels
were observed in \cite{gronbech_jensen_filamentary}. Despite the
number of experimental and numerical data \cite{pla_nori,reichhardt_nori_periodic}
a detailed theoretical
understanding of plastic motion remains quite a
challenge \cite{watson_fisher_plastic}.

As in the statics, one is in a better position to describe
the elastic flow regime, which is still a difficult problem.
This is the situation on which we will focus in this paper.
Though elastic flow in some cases extends to all velocities,
a natural idea was to start from the large velocity region and
carry perturbation theory in $1/v$. At large velocity
one may think at first that since the sliding system averages
enough over disorder one recovers a simple behavior, in fact much
simpler than in the statics. Indeed it was observed experimentally, some time
ago in neutron diffraction experiments \cite{thorel_neutrons_vortex},
and in more details recently \cite{yaron_neutrons_vortex},
that at large velocity the vortex lattice is more translationally ordered
than at low velocity. This tendency to {\it dynamical reordering} has also been
seen in numerical simulations \cite{shi_berlinsky,pla_nori,koshelev_dynamics_first}.
The $1/v$ expansion has been fruitful to compute the corrections to the
velocity itself in \cite{larkin_largev,schmidt_hauger,sneddon_cross_fisher}.
Recently it was extended by Koshelev and Vinokur
in \cite{koshelev_dynamics} to compute the vortex displacements
$u$ induced by disorder and leads to a description in term
of an additional effective shaking temperature induced by motion.
This description suggests bounded displacements
in the solid and thus a perfect moving crystal at large velocity.

Recently we have investigated \cite{giamarchi_moving_prl} 
the effects of the {\it periodicity} of the moving lattice in the direction
transverse to motion, in the same spirit which led to the prediction
of the Bragg glass in the statics. It was still an open problem
how much of the glassy properties remain once the lattice is set in motion.
We found that, contrarily to the naive expectation, some modes of the disorder
are not affected by the motion even at large velocity. Thus, 
the large $v$ expansion of \cite{koshelev_dynamics} breaks down
and the effects of non linear static disorder persists at all
velocities, leading to new physics. As a result the moving lattice
is not a perfect crystal but a {\it moving glass}.

The aim of this paper is to provide a detailed
description of the moving glass state predicted in
\cite{giamarchi_moving_prl} and to present our
approach to the general problem of moving lattices.
A brief account of some of the new results contained 
here (e.g the $T=0$ renormalization
group equations RG and fixed points) has already appeared
in \cite{giamarchi_m2s97_vortex,giamarchi_pld_bookyoung}.
We will use several RG approach at zero and 
at non zero temperature. Since several Sections
of this paper are rather technical
we have chosen to expose all the results
about the physics of the moving glass
in Sections \ref{sec:structures} and \ref{sec:physics}
in a self contained manner, avoiding all technicalities.
The reader can find there the results for 
the existence of channels (\ref{sec:channels})
the transverse I-V curves at $T=0$
and the dynamical Larkin length ( \ref{sec:larkin} ),
the random force and the correlation functions ( \ref{sec:correlations} )
the various crossover lengths and the 
the finite temperature results (\ref{sec:temperature}).
Decoupling scenarios
which distinguish between the Moving Bragg glass and the
Moving transverse glass ( \ref{sec:decoupling}) 
as well as predictions for the 
dynamical phase diagrams are also given in ( \ref{sec:diagrams}) 
Finally we discuss how the moving glass theory
stands presently compared to numerical simulations ( \ref{sec:numerics}) 
and experiments (\ref{sec:experiments}) and present
some suggestions of further observables which would be interesting to measure.

The following Sections are devoted to making progress
in an an analytic description of the moving state
of interacting particles in a random potential. Since this
is a vastly difficult problem, it is potentially
dangerous (and unfruitful) to try to attack this
problem by treating all the effects at the same time
(dislocations, non linearities, thermal effects etc..).
Already within the simplifying assumption of an elastic flow
two main types of phenomena are missed in a naive large
$v$ approach. 
The first one is a direct consequence of previous works on
driven dynamics of CDW and elastic manifolds
\cite{kardar_parisi_zhang,kardar_review_lines}. It is expected
on symmetry grounds \cite{hwa_pld} that new non linear KPZ terms
$(\nabla u)^2$ will be generated by motion, an effect which
was studied in the driven liquid \cite{hwa_driven_liquids}.
Another important effect, studied so far only within
the physics of CDW, is the generation of a static
{\it random force} convincingly argued by
Krug \cite{krug_random_mobility} and explored
in \cite{balents_dynamics_vortex}. If both effects
are assumed to occur simultaneously, they may
lead to interesting interplays which have been
explored only recently and only in simple CDW models
\cite{chen_kpz_dynamics}.
However there still no explicit RG derivation of 
those terms even in CDW
models. In the context of driven lattices, there are 
not even discussed yet. Our aim in this paper is
to remedy this situation. We derive these terms
explicitly and show that other linear terms, a priori even more
relevant are generated.
Though these additional linear, non linear and random 
force terms certainly complicate seriously the problem, focusing
exclusively on these terms only obscurs the physics of the present
problem. Indeed the second and as we show here more 
important effect in the moving structure is the crucial
role of transverse periodicity to describe the dynamics.

A rigorous study of the problem of 
moving interacting particules would be to 
first study the fully elastic flow of a lattice.
Once the main elastic physics is understood
a second step is then to allow for topological excitations
(vacancies, interstitials, dislocations).
In principle the results obtained within the
elastic only approach can, as in the statics,
be used to check self consistently the
stability of the elastic flow itself. It is hard to see
how one can do that in a controlled way 
without some detailed understanding the elastic flow first !
Here we carry most of the first step and
propose an effective description of the second.

Even the purely elastic model turns out to be
difficult to treat when all sources of anisotropies,
non linear elasticity, cutoff effects are included.
There are no analogous terms in the statics and
thus in that sense the dynamics is more difficult.
Our strategy has thus been to simplify the problem in
several stages and resort to simplified models.
The simplified models of Moving glasses that we have obtained
turn out to exhibit some new physics and become interesting
in their own. They call for interesting generalizations
to other models exhibiting dissipative glassy behaviour,
as we will propose.

We will call Model I the full model of an elastic flow 
of a lattice containing all the above mentionned relevant
linear and non linear terms. Such models can also be written
for other elastic structures with related kind of order
(such as liquid crystal order). This
model will be discussed in Section \ref{beyond}. However
its complete study goes beyond the present paper.

Fortunately, a useful and further simplified model can be
constructed (Model II). It corresponds to considering the
above full elastic model in the continuum limit.
It will certainly give a very good approximation of the full model
at least up to some very large scale.
This model was discussed in \cite{giamarchi_moving_prl}
and is studied in details here. 
It has both longitudinal degrees of freedom (along the direction
of motion) and transverse ones. Though it is quite
difficult, it can be handled by perturbative renormalization
group studies, as we will show here. It has a new and 
non trivial fixed point which gives a detailed description
of the moving Bragg glass phase.

It turns out that most of the physics of Moving glass
is contained in a further simplification of 
Model II which retains only the transverse
degrees of freedom (displacements). This model,
which here we call Model III,
was introduced in Ref. \onlinecite{giamarchi_moving_prl}
and is described by the equation of motion:
\begin{eqnarray}   \label{movglasseq1}
\eta \partial_t u +\eta v\partial_x u &=& c\nabla^2 u
+ F^{\text{stat}}(r,u(r,t)) + \zeta (r,t) 
\end{eqnarray}
which we call {\it the moving glass equation}.
$F^{\text{stat}}$ is a non linear static pinning force and
we have denoted $x$ the direction of motion,
$y$ the transverse direction(s) and $r=(x,y)$.
The model retains only the transverse displaceemnt
$u\equiv u_y$.

Equation (\ref{movglasseq1}) was obtained simply by considering
the density modes of the moving structure which are
{\it uniform in the direction of motion}. Indeed,
the key point of Ref. \onlinecite{giamarchi_moving_prl}
is that the transverse physics is
to a large extent independent of the details
of the behaviour of the structure along the
direction of motion. This is because the
transverse density modes, which can be termed
smectic modes, see an almost static disorder
and thus are the most important one 
to describe the physics of moving structures with
a periodicity in the direction transverse to
motion! Let us emphasize that this is 
explicitly confirmed here by the detailed RG analysis
of the properties of Model II. Note that
to obtain Model III one sets {\it formally}
$u_x=0$ \cite{footnote_balents}. 

\begin{figure}
\centerline{\epsfig{file=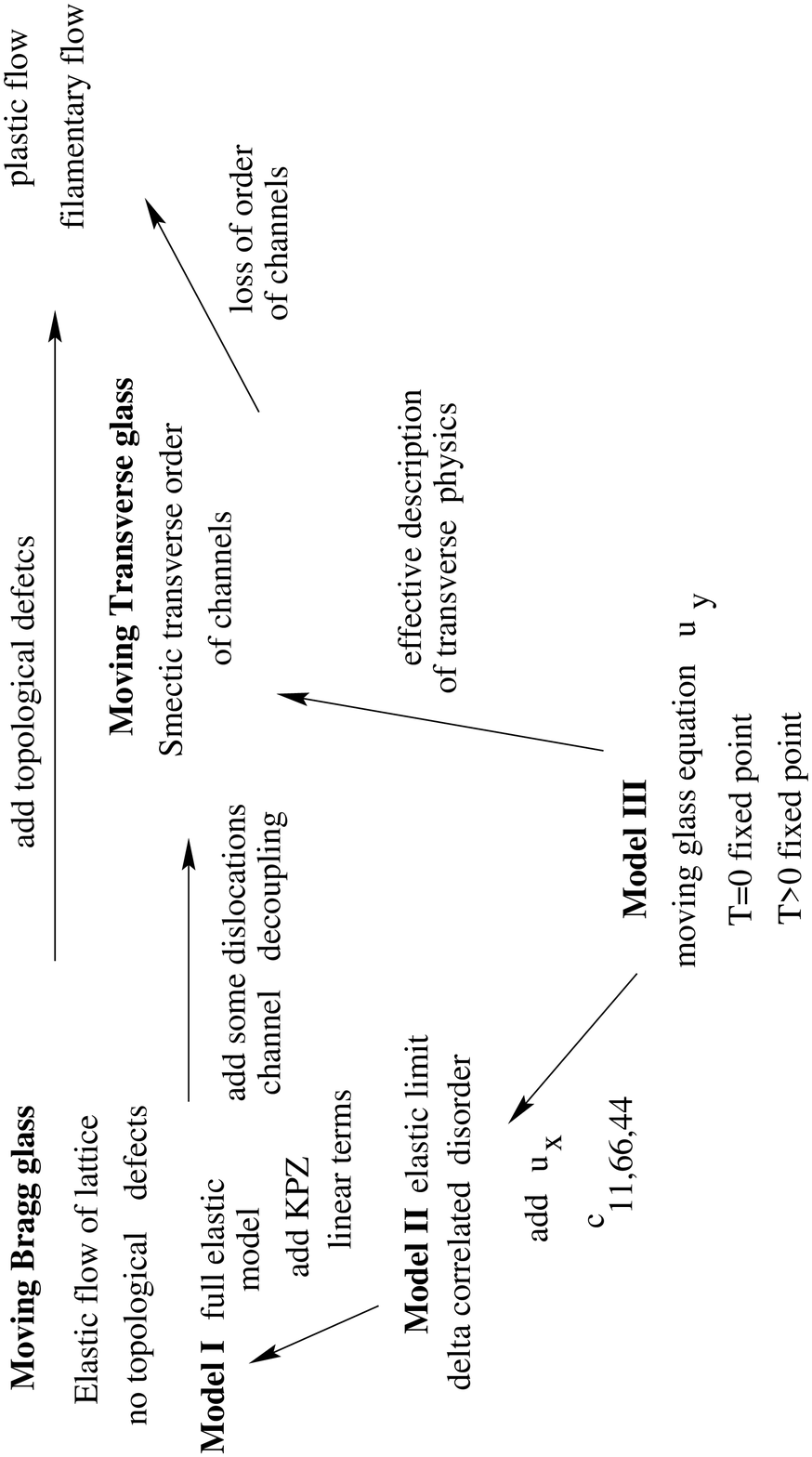,angle=-90,width=14cm}}
\caption{Various models studied here to describe
with various levels of approximation the (i) fully elastic
flow of a lattice (ii) intermediate phase with ordering transverse
to motion (iii) plastic flow \label{fig:strategy}}
\end{figure}

The hierarchy of models introduced here is
represented in Fig. \ref{fig:strategy}.

The outline of the paper is as follows. 
After the Sections \ref{sec:structures} and
\ref{sec:physics} where we give a non technical 
discussion of the physical results,
we start in Section \ref{model} by deriving
an equation of motion and, carefully examining its
symmetries, we introduce the Models I,II,III
and explain the approximations leading to them.
In Section \ref{sec:pert} we perform perturbation theory
on the full dynamical problem, focusing on Model II.
In Section \ref{renormalizationtransverse} we
use the functional RG to study Model III
and thus the transverse physics
in $d=3$ and $d=3-\epsilon$. We study $T=0$ and
$T>0$. In Section \ref{dynamicalco} we study a two dimensional
version of the moving glass equation Model III.
This allows to obtain results in $d=2$ at $T>0$ and
in $d=2+\epsilon$. Having obtained a good understanding
of the transverse physics in the previous sections,
we attack in Section \ref{frgcomplete} the RG of Model II.
Finally in Section \ref{beyond} we examine the full
Model I, show that linear terms and KPZ terms are generated
at large scales and discuss some consequences.

\section{Moving structures and moving glasses}

\label{sec:structures}

\subsection{Moving structures: general considerations}

All the structures we consider share the same basic
features. The static system in the absence of quenched 
substrate disorder consists of a network of interacting objects
at equilibrium positions
$R_i^0$, forming either a perfect lattice (periodic case)
or elastic manifolds (non periodic case). Depending on the
system the objects can be either pointlike (e.g. electrons
in a Wigner crystal) or lines (vortex lines in superconductors ).
Deformations away from equilibrium positions are described by
displacements $u_i$ or in a coarse grained description 
$u(r,t)$ where $r$ is the internal coordinate.
A complete characterization of the structure 
in motion uses three parameters (i) the internal dimension $D$
(ii) the number of components $n$ of the displacement field $u_{\alpha}$
and (iii) the embedding space dimension $d$. Two examples
are shown in Figure~\ref{lattice} and more details
are given in the Appendix. Since we are mostly interested 
here in periodic structures (though not exclusively) we
can set $D=d$. We will consider motion along one direction
called $x$, and we parametrize throughout all this paper
the space variable $r$ as $r = (x , y , z)$ where $x$
is one dimension, $y$ has a priori $n-1$ dimensions
and $z$ has $d_z=d-n$ dimensions, and the displacements
along motion as $u_x$ and transverse to motion as
$u_y$. Three dimensional triangular flux lines
lattices driven along a lattice direction
thus have $d=3,n=2$, $r=(x,y,z)$, $u=(u_x,u_y)$
where $z$ denotes the direction of the magnetic fields.
Two dimensional triangular lattices of point vortices
have $d=2,n=2$, $r=(x,y)$.

\begin{figure}
\centerline{\epsfig{file=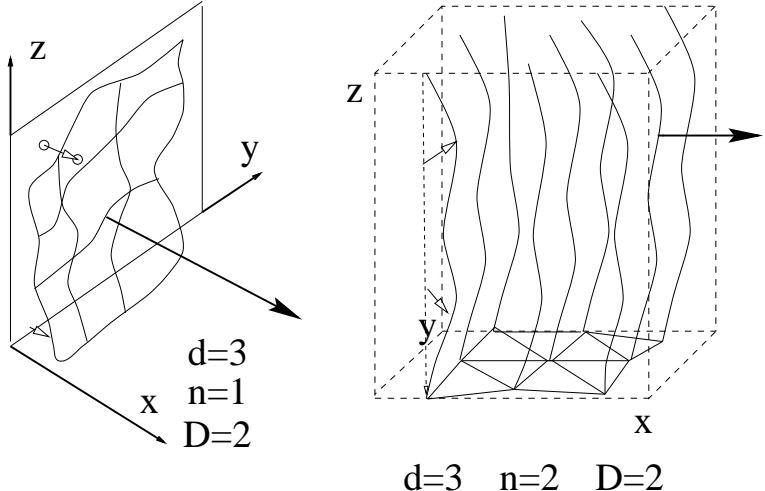,angle=-90,width=10cm}}

\caption{Two cases of a driven structure. An interface
$D=2$, $n=1$, $d=3$, driven orthogonal to its internal space.
A triangular line lattice $D=3$, $n=2$, $d=3$ driven within
its internal space
\label{lattice}}
\end{figure}

At finite temperatures or in the presence of quenched
substrate disorder the structure is deformed. 
An important issue is then to characterize the degree
of order. This can be expressed in terms of displacements correlation
functions. The simplest one measures the relative displacements of two points
(e.g two vortices) separated by a distance $r$.
\begin{equation} \label{relative}
\tilde{B}(r)= \frac1n \overline{\langle [u(r)-u(0)]^2 \rangle}
\end{equation}
where $\langle \rangle$ denotes an average over thermal fluctuations and
$\overline{\qquad}$ is an average over disorder. The growth of $\tilde{B}(r)$ with
distance is a measure of how
fast the lattice is distorted. For thermal fluctuations alone in $d>2$,
$\tilde{B}(r)$ saturates at finite values, indicating that the lattice
is preserved. Intuitively it is obvious that in the presence of disorder
$\tilde{B}(r)$, will grow faster and can become unbounded.
$\tilde{B}(r)$ can directly be extracted from direct imaging of the
lattice, such as performed in decoration experiments
of flux lattices.

\begin{figure}
\centerline{\epsfig{file=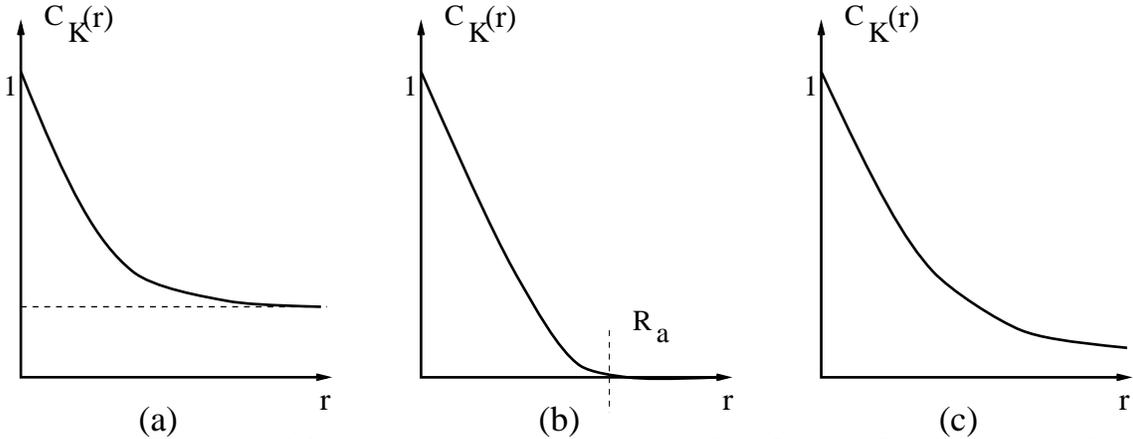,angle=-90,width=15cm}}
\caption{Various possible decays of $C_K(r)$. (a) For thermal
fluctuations alone $C_K(r)\to {\rm Cste}$, one keeps perfect $\delta$
function Bragg peaks, albeit with a reduced weight (the Debye-Waller
factor). (b) $C_K(r)$ decays exponentially fast. The structure factor
has no divergent peak any more, so translational order is destroyed
beyond length $R_a$, although some degree of order persists at short
distance. (c) $C_K(r)$ decays as a power law. The structure factor still
has divergent peaks but not sharp $\delta$ function ones. One retains
quasi-long range translational order. This is for example the case in
$d=2$ at small temperature (Kosterlitz-Thouless) or
in the Bragg glass.
\label{fig1}}
\end{figure}

\begin{figure}
\centerline{\epsfig{file=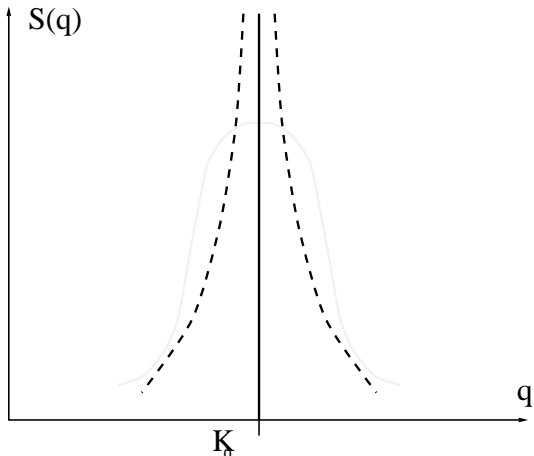,angle=-90,width=7cm}}
\caption{ Depending on the translational order remaining in the lattice
the structure factor has different shapes. The thick line is the
$\delta$ function Bragg peak of a perfect lattice (including thermal
fluctuations).
The dashed line is the power-law Bragg peak of the Bragg glass (which retains
quasi long range order and has no topological defects), the dotted
line is the lorentzian-like shape of a system loosing its translational
order exponentially fast.
\label{structure}}  
\end{figure}

Related to $\tilde{B}(r)$ is the structure factor of
the lattice, obtained by computing the Fourier transform of the density of
objects $\rho(r) = \sum_i \delta^d(r - R^0_i -u_i)$.
The square of the modulus $|\rho_k|^2$ of the Fourier transform of
the density is measured directly in diffraction (Neutrons, X-rays)
experiments. For a perfect lattice the diffraction pattern consists of
$\delta$-function Bragg peaks at the reciprocal vectors. 
The shape and width of any single peak around $K$ can 
be Fourier transformed to obtain the translational order correlation function
given by
\begin{equation}
C_K(r) = \overline{\langle e^{i K.u(r)} e^{-i K.u(0)} \rangle}
\end{equation}
$C_K(x)$ is therefore a direct measure of the degree of translational order that
remains in the system. Three possible cases are shown in
figure~\ref{fig1}.
For simple Gaussian fluctuations 
(and isotropic displacements) $C_K(r) = e^{-\frac{K^2}2 \tilde{B}(r)}$
but such a relation holds only qualitatively in general (as a lower 
bound).
Depending on how much cristalline order
remains in the system the structure factor will have
extremely different behaviors as depicted in figure~\ref{structure}.
Quite surprisingly, if one takes into account correctly the {\it periodicity}
of the lattice, a thermodynamic phase {\it without dislocations} 
was predicted to exist in $d=3$ at weak disorder.
\cite{giamarchi_vortex_short,giamarchi_vortex_long}.
This phase, named the Bragg Glass, posses quasi long range order
with Bragg peaks diverging as least as $q^{-(3-A_3)}$ (with 
$A_3 \approx 1$), similar to dashed line in Fig. \ref{structure}
At the same time displacements $B(r)$ grow logarithmically at large scale 
\cite{giamarchi_vortex_short,giamarchi_vortex_long}.
Similar predictions hold for other elastic models
such as random field XY systems,
and a priori also for liquid crystals. The Bragg glass theory
has by now received considerable numerical
\cite{ryu_diagphas_numerics,gingras_dislocations_numerics}
and analytical confirmations 
\cite{carpentier_bglass_layered,kierfeld_bglass_layered,fisher_bragg_proof}.

If disorder is increased above a threshold it is predicted 
that there is a transition at which topological defects proliferate.
They destroy the translational long range order exponentially fast beyond a length
$R_D$ leading to finite height
diffractions peaks. The height of the peak will be inversely proportional
to the scale at which translational order is destroyed.
This transition is thus characterized by the
loss of the divergence in the Bragg peaks.
In type II superconductors it implies that there is
a transition, upon increasing the magnetic field \cite{giamarchi_vortex_long},
from the Bragg glass (at low fields) to another phase.
The high field phase is either 
the putative vortex glass \cite{fisher_vortexglass_short,fisher_vortexglass_long}
or is simply continuously related
to the high temperature phase. These predictions for the phase diagrams
of superconductors has received experimental support
(see Ref. \onlinecite{giamarchi_diagphas_prb} for a review).

\begin{figure}
\centerline{\epsfig{file=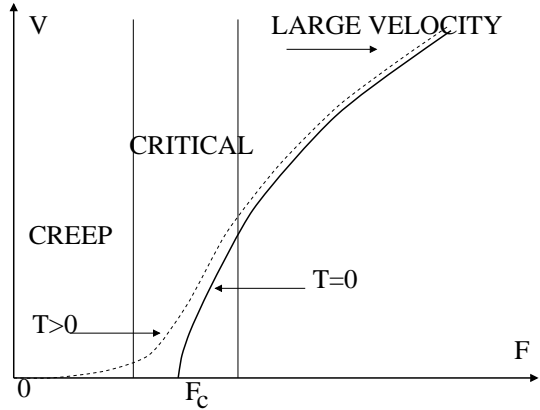,angle=0,width=7cm}}
\caption{A typical $v-f$ characteristics at $T=0$ (full line) and
finite temperatures (dashed line). Three main regimes can be distinguished:
the creep regime for forces well below threshold, the critical regime
around the threshold and the large velocity regime well above threshold.
\label{regimes}}
\end{figure}

What happens when an external force is applied to such a structure ?
One obviously important quantity to determine is the
curve of velocity $v$ versus the applied force $f$.
Through this $v-f$ characterics, three main regimes can be
distinguished and are shown on Figure~\ref{regimes}.
Far below the depinning threshold $f_c$ the system moves
through thermal activation. This is the so called creep regime.
Since the motion is extremely slow in this regime, it has been analyzed based
on the properties of the 
{\it static} system \cite{feigelman_collective,blatter_vortex_review}.
The resulting $v-f$ curve crucially depends on whether
the static system is in a glass state (such as the Bragg Glass)
where the barriers $U(f)$ become very large when $f \to 0$, or a liquid where
barriers remain finite at small $f$, resulting in a linear
resistivity. The general form expected in the creep regime
is:

\begin{equation}
v \sim \rho_0 f e^{-U(f)/T}
\end{equation}

Let us emphasize that this ``longitudinal'' $v-f$ characteristics 
has mainly be used to determine whether the the {\it static} system 
(i.e the limit $f=v=0$) is or not in a glass state. It may not
be enough though, if one wants to probe glassiness of the moving system
itself.

The second regime, near the depinning
transition $f \approx f_c$, has been intensely investigated in similarity with usual
critical phenomena (see e.g
\cite{fisher_depinning_meanfield,narayan_fisher_depinning,%
nattermann_stepanow_depinning})
where the velocity plays the role of an order parameter.
A particularly important question is that regime is to determine
whether plastic rather than elastic motion occurs.
Close to the threshold in low dimensions and in strong disorder situations
the depinning is observed to proceed through ``plastic
channels'' \cite{jensen_plasticflow_short,jensen_plasticflow_long}
between pinned regions. This type of filamentary flow has been found
in \cite{gronbech_jensen_filamentary} in simulations
of 2D (strong disorder) thin film geometry (with $c_{11} \gg c_{66}$)
where depinning proceeds via filamentary channels which
become increasingly denser.

\begin{figure}
\centerline{\epsfig{file=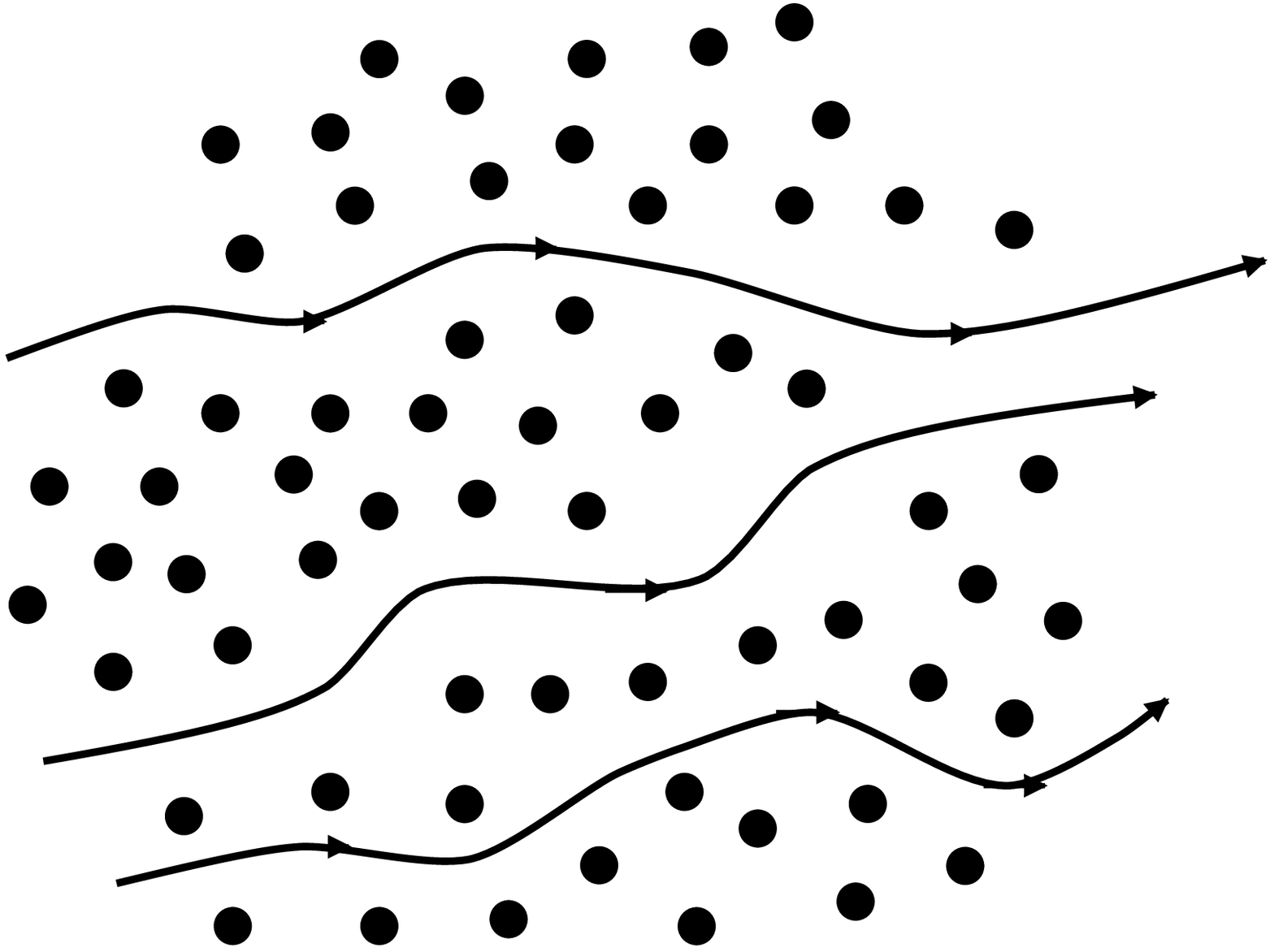,angle=-90,width=7cm}}

\caption{Plastic flow of a 
network of objects submitted to an external force
$F$, shown for simplicity in two dimensions. The motion
occurs through plastic channels around pinned regions. Plastic
flow might occur close to the depining threshold whereas
at large velocities one expects to recover elastic flow
where the whole lattice moves coherently.
\label{plasticchann}}
\end{figure}

The third regime is above the depinning
threshold $f > f_c$  This is situation on which we will focus in this paper
(though some of our considerations will have consequences
in the other regimes as well). An important phenomenon in this regime 
is that of {\it dynamical reordering}.
Indeed, it was observed experimentally, some time
ago in neutron diffraction experiments \cite{thorel_neutrons_vortex},
and in more details recently \cite{yaron_neutrons_vortex},
that at {\it large velocity} the vortex lattice is more translationally ordered
than at low velocity. Intuitively the idea is that at large velocity $v$,
the pinning force on each vortex varies rapidly and disorder should produce
little effect. This phenomenon was
also known in the context of CDW \cite{fisher_depinning_meanfield}.
The tendency to reorder has also been seen in numerical simulations
\cite{shi_berlinsky,pla_nori,koshelev_dynamics_first}.

Since the effect of disorder were expected to vanish at high velocity
perturbation theory in $1/v$ were developped mainly to compute the
$v-f$ characteristics \cite{larkin_largev,schmidt_hauger,sneddon_cross_fisher}.
Recently it was extended by Koshelev and Vinokur
in \cite{koshelev_dynamics} to compute the vortex displacements
$u$ induced by disorder in the moving lattice and in the moving liquid.
The effect of disorder on the moving liquid
was found to be equivalent to {\it heating } to an effective temperature
$T' = T + T_{sh}$ with $T_{sh} \sim 1/v$. Thus the moving liquid was argued to survive
at temperatures lower than the melting temperature $T<T_m$,
and a {\it dynamical melting transition} to occur below $T_m$ from a
moving liquid to a moving solid upon increase of the
velocity \cite{koshelev_dynamics},
when $T'=T_m$. These arguments were then extended to describe the
moving solid itself, and it was argued that there the effect
of pinning could also be described \cite{koshelev_dynamics}
by some effective shaking temperature $T_{sh} \sim 1/v^2$ defined by the
relation $<|u(q)|^2> = T_{sh}/c_{66}q^2$. This suggests bounded displacements
in the solid and that at low $T$ and above a certain
velocity the moving lattice is {\it a perfect crystal}.
As will be discussed in the remainder of this paper
the picture of the moving lattices emerging from the above
bold qualitative arguments \cite{koshelev_dynamics} goes wrong
in several ways.

There are several other important questions to be answered in addition to the 
$v-f$ characteristics. The first one is the question of
the effect of the motion on the spatial correlations
and in particular whether translational order exist in a 
moving system. This is related to the question of
plastic versus elastic flow. 
If plastic flow occurs, the structure factor should 
signal some destruction of lattice.
However because a moving system is inherently anisotropic
new effects may appear and the decay of the structure factor
will not be as isotropic as in the static system (the 
Lorentzian in Figure~\ref{structure}). This question thus remains to
be investigated. A possibility, suggested by the idea of
a shaking temperature \cite{koshelev_dynamics}, 
would be that at large velocity one should observe $\delta$-function
Bragg peaks characteristic of a crystal a finite temperature.
Such questions will be discussed in details in section~\ref{sec:physics}.
Finally determining how motion affects the phase diagram of the statics 
has to be investigated and depends of course on the above issues.
In particular what remains of the glassy properties of the systems
when in motion (slow relaxation, history dependence, 
non-linear behaviors) needs to be addressed.

For moving periodic systems, an equivalent question can be asked 
also about ``temporal order'' and its associated effects
such as noise spectrum. In particular if one looks at a signal 
at a fixed position in space but as a function of time, one expects
a periodic signal with a periodicity of $a/v$, having $\delta$ peaks 
in frequency at the multiples of the washboard frequency $\omega_0 = 2\pi v/a$.
If the lattice becomes imperfect one could naively expect the Fourier peaks in 
frequency to broaden in a way that reflects the loss of translational order. 
Quite surprisingly this is not so. Indeed it can be shown for a single component
displacements field (CDW) \cite{middleton_theorem} that the perfect periodicity
in time remains (in the absence of topological defects). However
this result is not readily applicable to a moving lattice,
and it is thus crucial to determine whether this remarkable property 
holds in that case. 

\subsection{The Moving Glass} \label{sec:themg}

\begin{figure}
\centerline{\epsfig{file=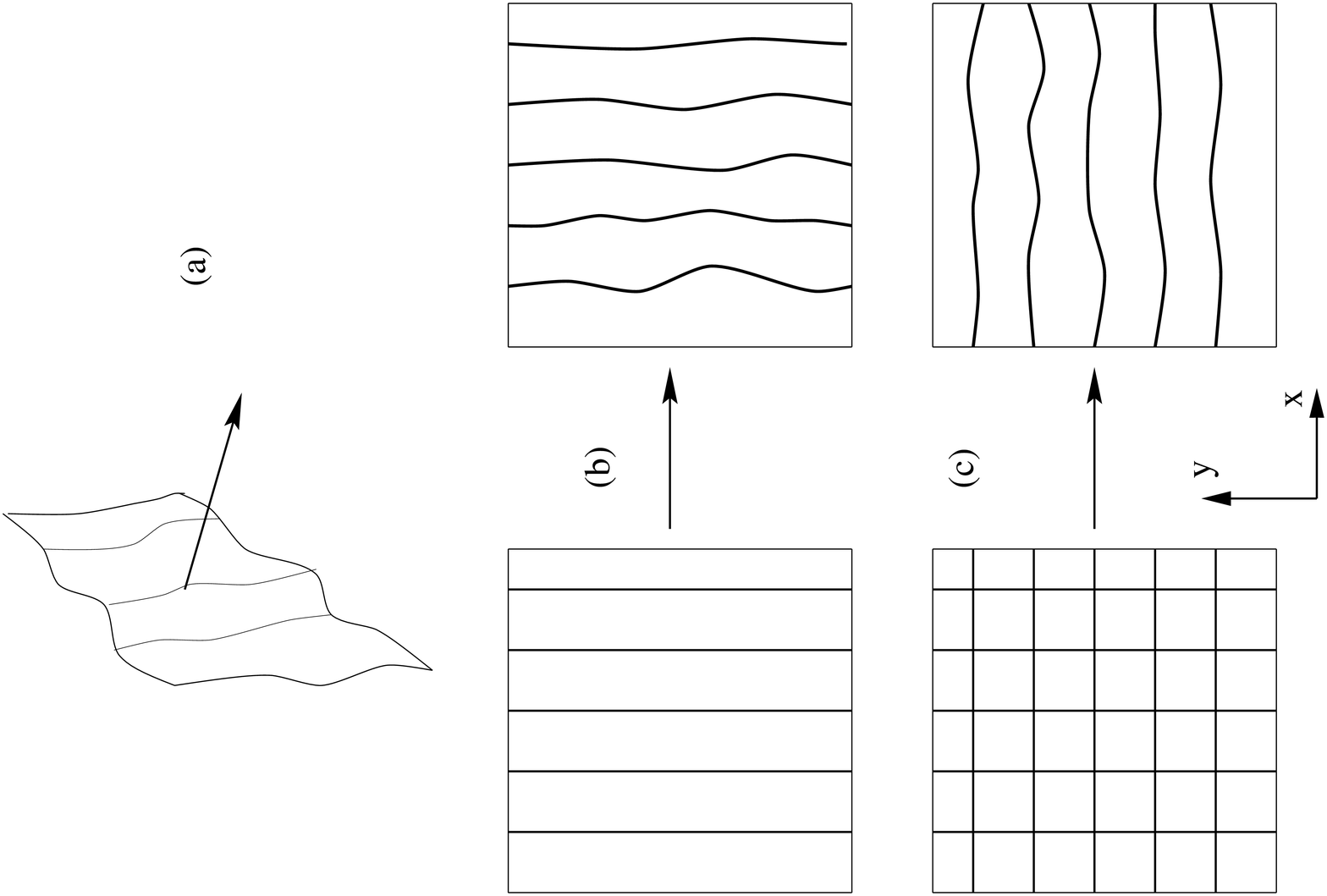,angle=-90,width=10cm}}
\caption{Three types of dynamical systems. (a) a manifold driven
perpendicular to itself. (b) a single Q CDW system. Only
displacements in the directions of motion exist, but periodicity
{\bf along} the direction of motion can play a role. (c) A
periodic system with transverse degrees of freedom driven
along one of its symmetry directions. The correct
description of this last class of systems is the moving glass fixed point
where the {\bf transverse} degrees of freedom are the important one
as represented here.
\label{mancdwgla}}
\end{figure}

To tackle the physics of a structure with a displacement field
with {\it more than one component} ($n>1$), such as a triangular
lattice (by contrast with a single $Q$ CDW), two routes
seem to be possible. The commonly followed one
\cite{balents_dynamics_vortex,koshelev_dynamics,littlewood_wigner_largev}
is to simply borrow from, or extend, the physics of single
component CDW \cite{littlewood_sliding_cdw,sneddon_cross_fisher,narayan_fisher_cdw},
or of elastic manifolds driven {\it perpendicularly}
to their internal direction \cite{kardar_review_lines}.
In this case emphasis is put on
the displacements {\bf along} the direction of motion $u_x$
and on the proper way to model its dynamics. Such a 
problem has turned out to be already quite complicated
in particular due to the generation of KPZ type non linearities
in the equation of motion. Even if degrees of freedom transverse to motion
$u_y$ exist as in the cases depicted in
Figure~\ref{mancdwgla} they constitute an extension \cite{ertas_kardar_anisotropic}
of this ``longitudinal'' physics. Thus in this ``CDW paradigm''
it would seem necessary to understand first completely the physics
of longitudinal modes $u_x$ and then incorporate $u_y$ as an extra complication.
Indeed there were a few attempts to describes the physics of
driven vortex lattices along those lines
\cite{koshelev_dynamics,balents_dynamics_vortex}.

The second approach is based on the realization
that the physics of periodic structures driven along one of their
internal direction is radically different \cite{giamarchi_moving_prl}
from the above descriptions.
This stems from the fact that due to the periodicity in the transverse
direction $u_y$ a {\it static non linear pinning force} $F^{\rm{stat}}$ persists even in a
fast moving system. We want to stress that this is a very general property
of a {\it any} moving structure which contains uniform density modes $K_x=0$ in the
direction of motion (as can be seen on the Fourier decomposition of
the density \cite{giamarchi_moving_prl}). As illustrated in
Fig. \ref{mancdwgla} the substructure formed by these
modes can deform elastically in the $u_y$ direction and sees essentially
a static disorder. As is obvious from the picture
\ref{mancdwgla} (c), this
substructure has generically a liquid crystal type of
(topological) order and
can be termed a ``smectic'' (though when $d_z=0$, e.g for $d=3$ and $n=3$,
it is rather a ``line crystal'' - see below).
In all cases the basic starting point thus
involves the {\bf transverse} degrees of freedom
as shown on figure~\ref{mancdwgla}, and is quite different from the ``CDW
description''. The equation which capture the main ingredients 
of such moving systems
was derived in Ref. \onlinecite{giamarchi_moving_prl}. It leads
to a new and interesting model for transverse components
$u \equiv u_y$, which has the general form in the laboratory frame:
\begin{eqnarray}   \label{movglasseq2}
\eta \partial_t u +\eta v\partial_x u &=& c\nabla^2 u
+ F^{\text{stat}}(r,u(r,t)) + \zeta (r,t) 
\end{eqnarray}
Since this equation captures glassy features of moving systems
we call it {\it the moving glass equation}. Although it 
looks like a standard pinning equation the {\it convection term}
$\eta v \partial_x u$ dissipates even in the static limit
(a reminder that we are looking at a
moving system) and does {\it not} derive from a potential.
Thus we will consider this problem and its generalizations as a prototype for
a new class of physical phenomena which are glassy and do not
derive from a potential (or from a hamiltonian).
The first example is to choose $F^{\text{stat}}(r,u)$
periodic in the $u$ direction:
\begin{eqnarray}  \label{startingdis}
F^{\text{stat}}(r,u_y) &=& V(r) \rho_0 \sum_{Ky \ne 0} K_y \sin{K_y(u_y-y)}
\end{eqnarray}
and corresponds to lattices (or to liquid crystals)
driven in a random potential with a short range correlator
$\overline{V(r) V(r')}= g(r-r')$ of range $r_f$. The study
of this case in Ref. \onlinecite{giamarchi_moving_prl} gave the first
hint that non potential dynamics can indeed exhibit
glassy properties and lead to {\it dissipative glasses}.
This is a rather delicate notion because the
the constant dissipation rate in the system would naturally
tend to generate or increase the effective temperature and
kill the glassy properties. However this type of competition
between glassy behaviour and dissipation arises in other systems
which are a generalization of the above equation. Let us
briefly indicate some of the generalizations that we are proposing
which are being studied here or in related works.

An interesting generalization is the
case of a periodic manifold with {\it correlated disorder}
\cite{ledou_giamarchi_futur}.
This is relevant to describe the {\it moving Bose glass}
state of driven vortex lattices in the presence of correlated
disorder.

Another generalization is to extend the equation (\ref{movglasseq2})
to a $N$ component vector $u_\alpha$.
It is easy to see in that case that a non potential
non linear disorder is generated if $v>0$ (which reduces to the 
``static random force'' for $N=1$). Thus in that model it is natural
to look at a generic non potential disorder
$F_\alpha^{\text{stat}}(r,u)$ from the start.
The mean field dynamical equations for large $N$
and the FRG equations at any $N$ for a large class
of such models are derived in
Section \ref{renormalizationtransverse}
and in Appendix \ref{parametrization}.
A subclass of these models is non periodic models
(manifold). They are relevant to describe the
random manifold crossover regime in the moving glass
(see below).
A further subclass is then obtained by setting $v=0$.
Interestingly the resulting
model describes polymers (and manifolds) in random flows
and can be studied both in the large $N$ limit
\cite{ledou_polymer_longrange} and using RG \cite{ledou_wiese_ranflow}
for any $N$.

Finally, there are other simpler but interesting situations
such as disorder correlated along the direction of motion
or lattices moving in a
periodic tin roof potential. These potentials which
are independent of $x$ have the interesting property that the
steady state measure $P[u(r)]$ is {\it identical}
(at any $T>0$) to the one with $v=0$.

Thus we see that the moving glass equation hides a 
whole class of new interesting dissipative models
whith glassy properties.

\section{Physical results} \label{sec:physics}

In this Section we present all the physical results
on the Moving Glass that we have obtained
in Ref. 
\onlinecite{giamarchi_moving_prl,giamarchi_m2s97_vortex,giamarchi_book_young}
and in the 
present paper. We deliberately avoid technicalities
and refer to the proper Sections for details.

\subsection{Channels}

\label{sec:channels}

One of the most striking property of moving structures
described by (\ref{movglasseq2}) is that the non linear static
force $F^{\rm{stat}}$ results in the pinning of the 
transverse displacements $u_y(r,t)$ into preferred static 
configurations $u_y(r)$ in the laboratory frame.
Thus the resulting flow can be described
in terms of {\it static channels} where the particles follow
each others like beads on a string.
In the laboratory frame these channels are determined by the static disorder
and do not fluctuate in time. They can be visualized in simulations
or experiments by simply superposing images at different times. 
What makes the problem radically new compared to conventional
systems which exhibit pinning is that despite the static nature
of these channels there is constant dissipation in the
steady state. This can be seen in the moving frame
where each particle, being tied to a given channel
(which is then moving) must wiggle along $y$ and dissipate.
In fact the existence of the channels shows in 
a transparent way that the wiggling of different
particles in the moving frame is highly correlated in
space and time, thus leading to a radically different
image as the one embodied in the ``shaking temperature''
based on thermal like incoherent motion \cite{koshelev_dynamics}.

\begin{figure}
\centerline{\epsfig{file=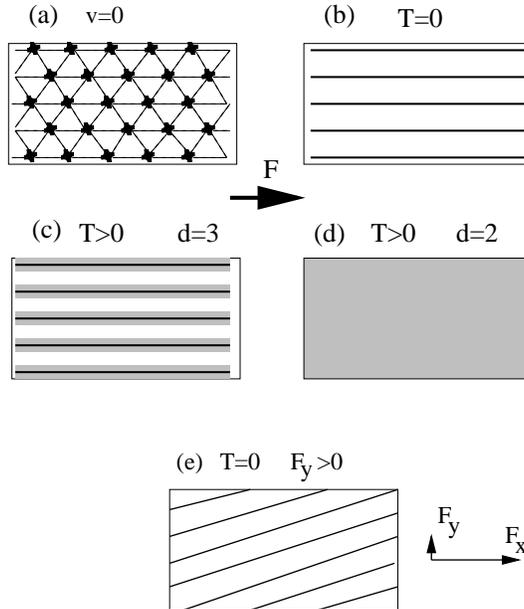,angle=-90,width=7cm}}
\caption{(a) a snapshot of a perfect
(non-disordered) lattice moving along the $x$ direction.
(b) upon superposing images at different times one would see
that at $T=0$ the particles follow perfectly straight lines.
(c) at $0<T<T_m$ in $d=3$ the channels remain perfectly straight with
a finite width due to the thermal fluctuation of the particles.
(d) in $d=2$ since thermal fluctuations are unbounded 
channels are completely blurred and cannot be defined 
even for $T<T_m$. (e) Even in situation (b), (c) applying a
additional small force along $y$ immediately results in tilted channels
with angle $F_y/F_x$. 
\label{fig:perfect}}

\end{figure}

The channels are thus the easiest paths
followed by the particles. One can see that the ``cost''
of deforming a channel along $y$ is that dissipation is increased.
Thus the channels
are determined by a subtle and novel competition
between elastic energy, disorder and dissipation.
As a consequence these channels are {\it rough}.
This is a crucial difference between what would be
observed for a perfect lattice as illustrated in
Figure~\ref{fig:perfect}

\begin{figure}
\centerline{\epsfig{file=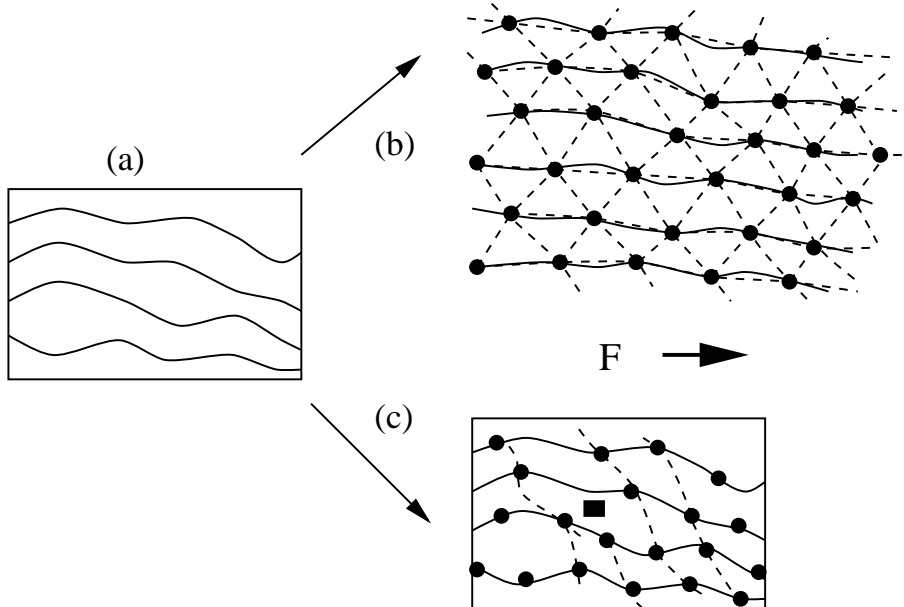,angle=-90,width=12cm}}
\caption{(a) The motion in the moving glass occurs through rough
static channels. Relative deformations grow with distance
and become of order $a$ at distances $R_a^x \sim {R_a^y}^2$.
Only the {\bf channels} themselves are elastically coupled along 
the $y$ direction. Depending on the dimension, velocity and disorder strength
two main cases can occur: (b) an elastic flow where the particles positions
are elastically coupled between channels 
(in $d=3$ and weak disorder or large velocities).
In this regime the lattice is topologically ordered (no dislocations)
and the rows of the lattice follow the channels.
This is a Moving Bragg Glass. (c) In $d=2$ or 
at stronger disorder in $d=3$ the
positions of particles in different channels may decouple.
Dislocations with Burgers vectors along $x$ (indicated by the square)
are then injected between
some channels beyond the length
$R_a$. This situation describes a Moving transverse glass
(with a smectic or a line crystal type of topological order).
\label{chanfig}}
\end{figure}

\begin{figure}
\centerline{\epsfig{file=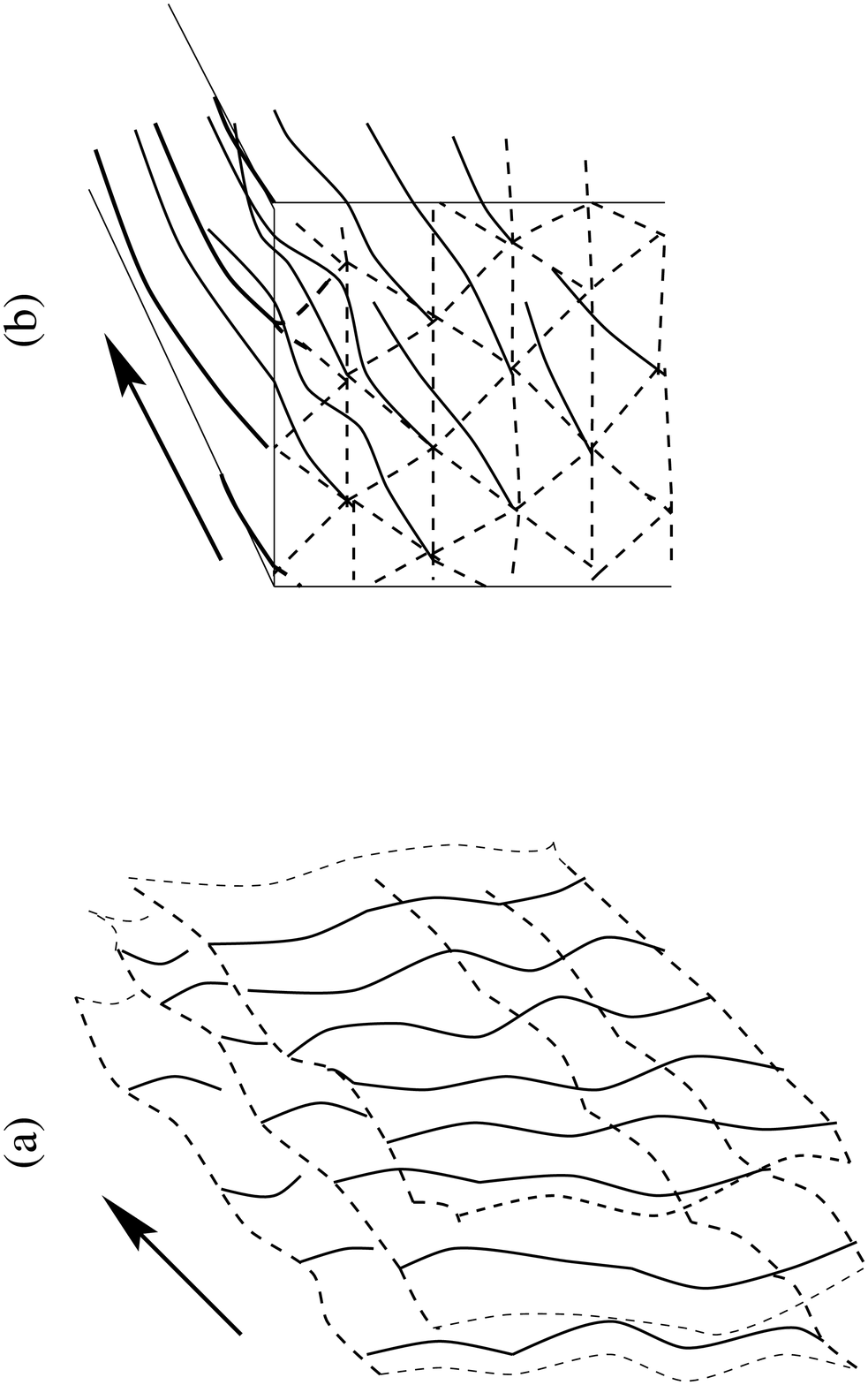,angle=-90,width=10cm}}
\caption{ Different types of topological order for
the manifold of channels in $d=3$ (a) for line lattices
in motion the channels are ``sheets'' and thus form an anisotropic type
of smectic layering (b) for 3d lattices of pointlike objects
or for triply periodic structures (triple $Q$ CDW) 
they have instead the topology of a line crystal \label{fig:chann3d}}
\end{figure}

By contrast the channels which are predicted in the
Moving Glass are illustrated
in Figure~\ref{chanfig} and in Figure~\ref{fig:chann3d}.
It is important to
stress that the Moving Glass equation
(\ref{movglasseq2}) introduced in \cite{giamarchi_moving_prl}
does not assume anything about the
coupling of the particle in {\it different} channels
but only implies that the channels themselves are
elastically coupled along $y$, and thus through compression modes.
Indeed on specific models such as model II one can verify 
explicitely that although coupling between longitudinal and 
transverse degrees of freedom exists a priori, 
the longitudinal degrees of freedom $u_x$ do not feed back
{\bf at all} in the moving glass equation 
(see section~\ref{frgcomplete}).

The existence of channels naturally leads to several a priori
possible regimes for the coupling between particles in different channels.
The first case, represented in Fig. \ref{chanfig} (b),
is a topologically ordered moving structure
corresponding to full elastic coupling between particles in 
different channels.
Since, remarkably, this structure retains perfect topological order
despite the roughness of the channels, it is reminiscent of the properties
of the static Bragg glass, and thus we call it a Moving Bragg glass.
A second case of a Moving Glass corresponds to decoupling between the channels,
by injections of dislocations beyond a certain lengthscale $R_d$
and is called the Moving Transverse Glass.
These two regimes will be discussed in more details in 
section~\ref{sec:decoupling}. Finally note that in $d=3$
channels can be either ``sheets'' (for line lattices) or 
linear (for point lattices) as represented in Fig. \ref{fig:chann3d}.

It is important to note that the channels in the Moving glass
are fundamentally different in nature
from the one introduced previously 
\cite{jensen_plasticflow_short,jensen_plasticflow_long}
to describe slow plastic motion between
pinned islands,
as illustrated in Fig. (\ref{plasticchann}). In the Moving glass they
form a manifold of almost parallel lines (or sheets for
vortex lines in $d=3$), elastically coupled along $y$. 
For that reason we call them generically ``elastic
channels'' (whether or not they are fully coupled or decoupled)
to distinguish them from the ``plastic channels''
(even though some plastic flow may occur when
elastic channels decouple).

Note that in the above discussion we have concentrated on
elastic channels which can {\it spatially decouple}.
It is possible a priori that they may still remain
{\it temporally coupled}, i.e synchronized.
Indeed, examples of synchronization where observed
even in extremely plastic filamentary flow. 

\subsection{Dynamical Larkin length and transverse critical force}

\label{sec:larkin}

Another important property of the moving glass intimately
related to the existence of stable channels is the
existence of ``transverse barriers''. Indeed it is 
natural physically that once the pattern of channels
is established the system does not respond in the transverse
direction along which it is pinned. Thus we have predicted
in Ref. \onlinecite{giamarchi_moving_prl}
that the response to an additional small external force $F_y$
in the direction transverse to motion vanishes at $T=0$.
A true {\it transverse critical force} $F^y_c$ exists 
(and thus a transverse critical current $J^y_c$ in
superconductors) for a lattice driven along a 
principal lattice direction.

The transverse critical force is a rather subtle effect,
more so than the usual longitudinal critical force.
It does {\it not} exist for a single particule at $T=0$
moving in a short range correlated random potential.
By contrast even a single particule 
experiences a non zero longitudinal critical force.
It does not exist either for a single driven vortex line
or any manifold driven perpendicular to itself in a
pointlike disordered environment. It would exist however,
even for a single particle if the disorder is
sufficiently correlated {\it along the direction of motion}
(such as a tin roof potential constant along $x$
and periodic along $y$). Such disorders break the rotational
symmetry in a drastic way. Still, in the case of
a lattice driven in an uncorrelated potential it
does seem to break the rotational symmetry of the problem.
In some sense in the Moving glass the transverse topological
order which persists (and the elasticity of the manifold) provide
the necessary correlations (through a spontaneous breaking
of rotational symmetry). Thus the transverse critical force
is a dynamical effect due to barriers preventing the 
channels to reorient. 

We have investigated the equation \ref{movglasseq2}
numerically in $d=2$ and found that indeed starting from a random
configuration and at zero temperature the field $u(r,t)$
relaxes towards a {\it static} configuration  $u_{stat}(r)$
solution of the static equation. Applying a small force
in the $y$ direction (i.e adding $f_y$ in \ref{movglasseq2})
yields no response. The manifold is indeed {\it pinned}.

Thus we have proposed the Moving Glass as a new 
dynamical phase (a new RG fixed point) and the transverse critical
force as its order parameter at $T=0$. 
The upper critical dimension of this phase is $d=3$ 
instead of $d=4$ for the static Bragg glass. Above
$d=3$ weak disorder is irrelevant and the moving glass
is a moving crystal. For $d \leq 3$ disorder is relevant
in the moving crystal and leads to a breakdown of
the $1/v$ expansion of \cite{koshelev_dynamics}. Divergences
in perturbation theory can be treated using a renormalization
group RG procedure (Section \ref{renormalizationtransverse}).
One indeed finds a new fixed point
which confirms the prediction that the moving glass
is a new dynamical phase. Using RG and the properties
of this new fixed point one can compute various physical
quantities (Section \ref{physicalt0}). We find that the
transverse critical force is given by:
\begin{equation}  \label{transcrit1}
F_c^y \sim A \frac{c~r_f}{(R^y_c)^2}
\end{equation}
with $c=c_{11}$ in $d=2$, $c=\sqrt{c_{11} c_{44}}$ in $d=3$
and $A$ is a nonuniversal constant. The length
scale $R_c^y$ is the {\it dynamical Larkin length}. It is defined
as the length scale along $y$ at which perturbation theory breaks down,
non analyticity appears in the FRG and the (scale dependent)
mobility vanishes. Before we proceed further 
let us define now disorder strength parameters. For
uncorrelated disorder the random potential $V(r)$
which couples to the density of the structure has short range
correlations of range $r_f$,
$\overline{V(r,z) V(r',z)}= g(r-r') \delta^{d_z}(z-z')$
(see Section \ref{model}). As in \onlinecite{giamarchi_moving_prl}
we denote by $\Delta$ (also denoted $\Delta(u=0)$, see
Section \ref{model}) the bare static pinning force correlator
$\Delta = \rho_0^2 \sum_{K_y,K_x=0} K_y^2 g_K$ where
$\rho$ is the average density and $g_K$ the Fourier
transform of $g(r)$ at the reciprocal lattice vectors.
Throughout $\Delta_2$ will denote the second derivative
of the non linear pinning force correlator
$\Delta_2 = - \Delta''(0) \approx \Delta/r_f^2$
(see Section \ref{renormalizationtransverse}).
Our result is that in $d=3$ the dynamical Larkin length
is given by 
\begin{equation}     \label{transcrit2}
R_c^y \sim a \exp{\frac{4 \pi \eta v c}{\Delta_2}}
\end{equation}
while in $d \le 2$ it reads
\begin{equation}   \label{transcrit3}
R_c^y \sim  (1 + \frac{4 \pi \eta v c (3-d)}{\Delta_2})^{1/(3-d)}
\end{equation}
with again $c=c_{11}$ in $d=2$ and $c=\sqrt{c_{11} c_{44}}$ in
$d=3$. These results are valid for large enough velocities
$v \gg v^*_c$ (see below for the definition of $v^*_c$ and results for all
velocities). Note that
for $v>v*_c$ the dynamical Larkin length depends only
on $c_{11}$ (and of $c_{44}$ in $d=3$) as it should since the
physics of the moving glass is controlled by the compression
modes and thus largely independent of the detailed
behaviour along $x$.

Another way to estimate the Larkin length is to compute
the displacements in perturbation theory of the disorder.
At very short distance one can treat the pinning force 
in (\ref{movglasseq2})
to lowest order in $u$. This gives a model where disorder
is described by
a {\it random force} $F^{\text{stat}}(x)$ independent of $u$
whose correlator is $\langle F^{\text{stat}}(r)F^{\text{stat}}(r')\rangle
=\Delta \delta^d(r-r^{\prime })$. This regime is
the equivalent of the short distance Larkin regime for the statics.
In the Moving glass at very large velocity $v \gg v_c^*$ 
the displacements along $y$ grow as $B(r) = B_{RF}(r)$ (at $T=0$) with:
\begin{equation} \label{larkinmove2}
B_{RF}(y)=\int \frac{dq_x
dq_y d^{d_z} q_z}{(2\pi)^d} \frac{\Delta (1-\cos(q_y y))}{(\eta v q_x)^2 +
(c_{11}^2 q_y^2 + c_{44}^2 q_z^2)^2}
\end{equation}
The scale along $y$ at which 
$u_y$ becomes of order $r_f$ defines the 
dynamical Larkin length $R_c^y$, i.e $B_{RF}(y=R_c^y,x=0) \sim r_f^2$.
The resulting expression coincides 
with the one obtained within the RG approach (up to
non universal prefactors).

Similarly one can define a Larkin length for transverse pinning
along the $x$ direction by the condition that
$u_y(x=R_c^x,y=0) - u_y(x=0,y=0) \sim r_f$. Since what
determines this length is only $u_y$ (and not $u_x$)
it is independent of the detailed behaviour along $x$.
It is important to note that
the Moving glass is a very anisotropic object at large scale
with a scaling $x \sim y^2$ of the internal coordinates.
This implies that at large velocity ($v>v_c^*$)
the Larkin length along $x$ is very large
(much larger than $R_c^y$), with $R_x = v (R_c^y)^2 /c_{11}$
(One has also the more conventional
behaviour $R_c^z \sim \sqrt{c_{44}/c_{11}} R_c^y$ in $d=3$ ).
Estimating the random
force acting on a Larkin volume for the transverse
displacements \cite{giamarchi_moving_prl} one recovers
the above estimate for $F_c^y$.

\begin{figure}
\centerline{\epsfig{file=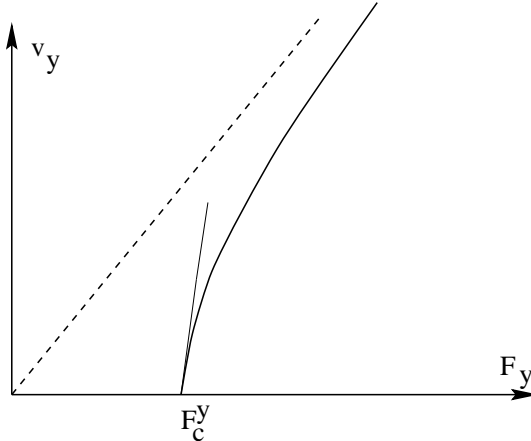,angle=-90,width=7cm}}
\caption{Transverse $v$-$f$ characteristics at $T=0$: transverse 
velocity $v_y$ as a function
of the applied transverse force $F_y$ at a fixed longitudinal
velocity. The behaviour near threshold is found to be
linear with a large slope for $v \le v^*_c$.
\label{fig:ivtrans}}
\end{figure}

The resulting transverse I-V characteristics at $T=0$ is 
depicted in Fig. \ref{fig:ivtrans}. The transverse 
depinning is studied in Section \ref{renormalizationtransverse}
and we find the behaviour near the threshold $v_y \sim |F^y-F_c^y|^\theta$ 
for $F^y>F_c^y$ with $\theta=1$ to lowest order in
$\epsilon$. A reasonable conjecture which would be interesting
to verify is that it remains $\theta=1$ to all orders.
Thus it starts linearly with a slope which depends
on the velocity $v$. It is very large for $v \ll v_c^*$
and diverges in the limit $v \to 0$.

The existence of a transverse critical force in a moving state
raises interesting issues about {\it history dependence}. These
issues are largely open and should be explored in
further numerical, experimental and theoretical work.

Let us for instance consider consider two experiments.
In the first one a force $f_x e_x + f_y e_y$ is applied
to the lattice at time $t=0$ and then wait until a steady state is
reached. The velocity is then $(v_x(f),v_y(f)$.
In the second experiment one first apply a force
$f_x$ along the direction $x$, wait for a steady state
and then apply $f_y$ along $y$. One then measures
the velocities $(v^{w}_x(f),v^{w}_y(f)$. The question
is should one find the same result in the two experiments
or not. Of course there are subtle issues which complicate
the problem and needs to be further investigated 
(such as (i) the order of the limit system size versus waiting
time before deciding a steady state is reached (ii)
whether the lattice will globally rotate or break into 
crystallites, (iii) some non universality of $T=0$ dynamics)
but one should still be able to find an
{\it operational} answer. If it is found that there are
such history dependence effects then that would 
be a strong characteristic of a glassy state
(it should not happen in the liquid where one expects
both answers to be the same, but in the same trivial
sense as for a single particle).
On the other hand, if no clear history dependence is found
it has interesting consequences. We assume in the following
that the global orientation of the lattice is unchanged.
Then the first consequence is that 
there is a well defined history-independent
global $v$-$f$ function. This function however is 
non analytic in a large region of the $f_x,f_y$ plane.
$v_y(f_x,f_y)$ should remain zero at least in 
the region $f_y < F_c^y(v_x(f_x,f_y))$ and similar
regions near each of the principal symmetry axis of the crystal.
This is clearly the result of the FRG calculation
presented here.
But then one may also guess that it may be non analytic too
along other lattice directions (though it is possible that some of
the higher symmetry directions be screened by lower ones).
The transverse mobility as a function of the angle
and the force should exhibit a complex (and rather strange)
behaviour which would be interesting 
to investigate further. 
A second interesting consequence would be that if in the above
described first experiment one chooses a $f_y>0$ smaller than
$F_c^y(v_x)$, the lattice would first glide in the direction of
the applied force (as small time perturbation theory would 
indicate) but would soon change its velocity to lock it 
along a symmetry axis. It is quite possible that this
locking effect exists and be a possible explanation for the behaviour
ubiquitously observed in experiments, namely that lattices
tend to flow along their principle axis directions.
Such a behaviour near depinning was observed in recent decoration experiments
\cite{marchevsky_new}.

Another important question for experiments is to determine
the transverse critical force as a function of 
the longitudinal velocity $v$. As $v$ decreases $F_c^y$ 
increases but it is intuitively clear that
$F_c^y$ cannot become larger than the longitudinal
critical current (strictly speaking in the same direction $y$).
We will neglect for now the dependence of the longitudinal
critical current in the orientation with respect to
the lattice (which gives a numerical factor which can be
incorporated). We will call $F^{iso}_c$ the critical
current for $v=0$. As $v$ is decreased below
$v^*_c$ the transverse critical force saturates at $F^{iso}_c$.
This is depicted in Fig. \ref{gend2} (the large $v$ behaviour was
given in \ref{transcrit1},\ref{transcrit2},\ref{transcrit3}).
There is thus a crossover towards the
static isotropic behaviour (e.g in the Bragg glass)
- assuming no dynamical phase transition as $v$ decreases
which would complicate the analysis.

\begin{figure}

\centerline{\epsfig{file=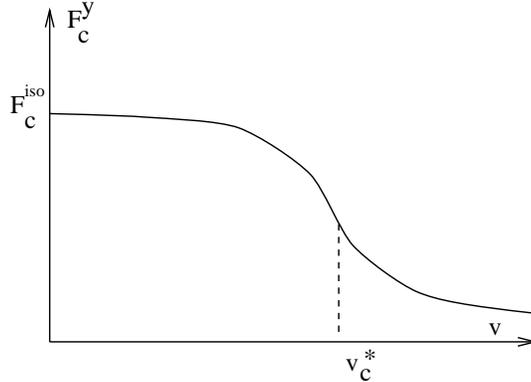,angle=-90,width=7cm}}
\caption{Transverse critical force as a function
of the longitudinal velocity.
For a relation between $\eta v_c^*$ and the longitudinal
critical force see the text.
\label{gend2}}
\end{figure}

\begin{figure}

\centerline{\epsfig{file=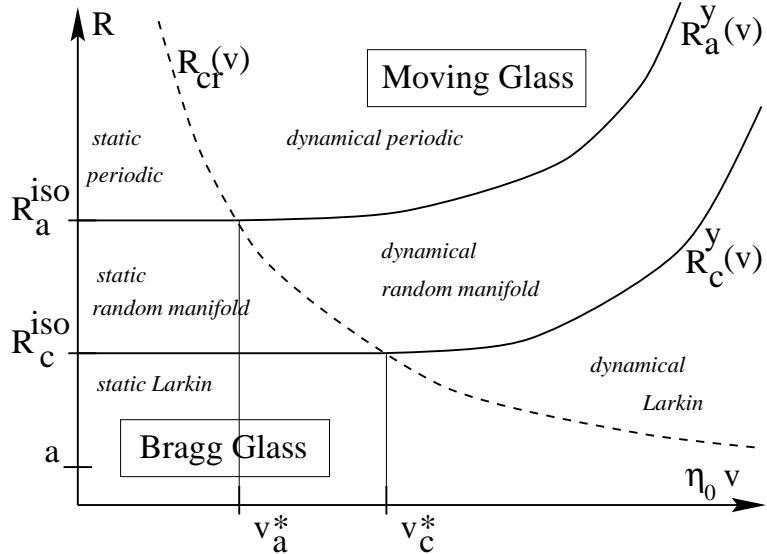,angle=-90,width=10cm}}
\caption{Crossover as a function of the length scale
$R$ and longitudinal velocity $v$ from the static
Bragg glass behaviour (at small $v$) to the
Moving glass behaviour (at large $v$). The dashed line represents
the crossover between these two regimes $R=R_{cross}=c/v$.
The dynamical Larkin length $R_c^y(v)$ as a function of $v$ and the 
transverse translational order length $R_a^y(v)$ are indicated as plain curves.
This is valid in the collective pinning regime $R^{iso}_c>a$
where $R^{iso}_c$ is the static Larkin length.
\label{bragg}}
\end{figure}

This crossover can be explicitly estimated using the
FRG in Sections \ref{renormalizationtransverse},
\ref{frgcomplete} and physical arguments. It is convenient
to discuss it using the Figure \ref{bragg} (also useful for
studying the crossover in the
correlations- see next Section). Let
us first discuss it for simplicity with isotropic
elasticity $c_{11}=c_{66}=c_{44}=c$. There is a crossover
length scale $R_{cr} = c/v$ below which the Moving glass
looks very isotropic and very similar to the Bragg glass.
This length scale is represented in Figure \ref{bragg}
as a dashed line. Increasing the length scale $R$ starting from 
$a$, at fixed $v$, one is first controlled by the static
behaviour until reaching that line ($R < R_{cr}$)
and then one is controlled by the dynamical MG regime
for $R > R_{cr}$. Similarly one can also represent the 
Larkin length  at $v=0$ $R_c^{iso} = (c^2 r_f^2/\Delta)^{1/(4-d)}$ (in $d < 4$).
The crossover velocity $v^*_c$ corresponds to the
velocity at which $R^y_c = R_{cr}$ when one has also
$R^y_c=R_c^{iso}$. One finds that:

\begin{eqnarray}
&& v_c^* = (\Delta / c^2 r_f^2 ) \ln (c^2 r_f^2/a \Delta)  \qquad (d=3) \\
&& v_c^* = c (\Delta / c^2 r_f^2)^{1/(4-d)} \qquad (d \le 3)
\end{eqnarray}

These results are valid when ($R_c^{iso} > a$),
i.e. in the collective pinning regime
for the statics. We denote $r_f =\min (r_f,a)$, i.e
if $r_f>a$ one can simply replace $r_f$ by $a$ in
all the above formulae.

Thus for $v < v_c^*$ the transverse critical current becomes
of order the longitudinal one $F_c^{iso}=c r_f/(R^{iso}_c)^2$.
It is useful for purpose of comparison with experiments to
compare $\eta v_c^*$ with $F_c^{iso}$. One finds the general
relation:

\begin{eqnarray}
\frac{\eta v_c^*}{F_c^{iso}} = \frac{R_c^{iso}}{r_f}
\end{eqnarray}

with logarithmic corrections in $d=3$, 
$\eta v_c^* = F_c^{iso} R_c^{iso}/r_f \ln(R_c^{iso}/a)$. This result
is remarkable. Since for weak disorder one has usually that
$R_c^{iso} \gg r_f$ it shows that for a system with
isotropic elasticity, the transverse critical force should
remain of order the longitudinal one up until very far above the
longitudinal threshold ($F_x \gg F_c$) (very high up in the
$v_x$-$F_x$ curve in Fig. \ref{regimes}). As we see now
this is different when $c_{66} \ll c_{11}$.

Incorporating $c_{66}$, $c_{11}$ and $c_{44}$ one finds 
that $v_c^*$ is indeed smaller, with:
  
\begin{eqnarray}   \label{probe}
&& \frac{\eta v_c^*}{F_c^{iso}} = \frac{c_{66}}{c_{11}} \frac{R_c^{iso}}{r_f} 
\qquad d=2  \\
&& \frac{\eta v_c^*}{F_c^{iso}} = (\frac{c_{66}}{c_{11}})^{1/2}
\frac{R_c^{iso}}{r_f}  \qquad d=3
\end{eqnarray}
up to a logarithmic factor $\ln(R_c^{iso}/a)$ in $d=3$.
we have used $R_c^{iso} = r_f^2 c_{66}^{3/2} c_{44}^{1/2}/\Delta$ in $d=3$ and
$R_c^{iso} = r_f c_{66}/(\Delta)^{1/2}$ in $d=2$ and
$F_c^{iso}=c_{66} r_f/(R^{iso}_c)^2$ (we have assumed 
$c_{66} \ll c_{11}$ and neglected the contribution of
compression modes). Thus the value of $v_c^*$ is then
much smaller. One sees from \ref{probe} that a measure
of the transverse critical current may lead to interesting information
about the elasticity of the lattice.

Finally note that one can make a simple minded argument showing 
directly on the equation (\ref{movglasseq2}) that 
the new convective term should not change pinning much
at small $v$.
Indeed starting from the case $v=0$, where one has a pinned
state $u^{v=0}_{stat}(r)$ and treating the convection term
as a perturbation (which should be OK at small scales)
one sees that this terms acts on the $v=0$ pinned
state as an additional quenched random force. 
Since there is a critical force $f_c(v=0)$ in that case,
it is intuitively clear that this term will not destroy completely
the state $u^{v=0}_{stat}(r)$ until $v r_f/R_c \sim f_c$.
This argument gives back the correct value for $v_c^*$.

\subsection{Displacements and correlation functions}

\label{sec:correlations}

Due to the presence of the static disorder
one expects unbounded growth of displacements in the
Moving glass. The relative displacements induced by disorder
in the moving system can be first computed in naive perturbation theory
using (\ref{larkinmove2}). One finds:
\begin{equation} \label{displ}
B(x,y) \sim \Delta \frac{y^{3-d}}{c \eta v}H(\frac{c x}{\eta v y^2})
\end{equation}
where $H(0)=\text{cst}$ and $H(z)\sim z^{(3-d)/2}$ at large $z$. Thus $x$
scales as $y^2$ and the displacements are very anisotropic.

The above formula, if taken seriously, leads to displacements
growing unboundedly for $d \le 3$. This is similar to the Larkin
calculation for the static problem. As in the statics it indicates
that the crystal is unstable to weak disorder in $d \le 3$ and
that perfect TLRO is destroyed. Note that due to motion the
upper critical dimension is now $d=3$ instead of $d=4$ for the 
statics. As in the statics, the above formula and perturbation
theory breaks down above $R_c^y$ and an RG approach is absolutely 
necessary to compute the displacements.

Using that the RG calculation one finds that the behaviour
of displacements is controlled by a new fixed point characteristic
of the Moving glass phase. One finds that the correlation function
of displacements {\it averaged over disorder} can be rigorously 
separated into two parts:
\begin{equation} 
B(r) = \overline{ \langle [ u(r) - u(0) ]^2 \rangle } =
B_{\rm{RF}}(r) + B_{\rm{NL}}(r)
\end{equation}
where $B_{\rm{NL}}(r)$ comes from the {\it nonlinear part} of
the pinning force. While this part is dominant in the
Bragg glass (and was computed in \cite{giamarchi_vortex_long})
in the Moving glass this contribution is
subdominant and we will neglect it for now.
The main contribution
comes from the {\it static random force}
which is generated both along $y$ and $x$ direction. The generation 
of such a random force, forbidden in a static system, occurs here
because of the non potentiality induced by the motion.
The complete expression of the generated random force is
given in Section \ref{frgcomplete} (see also 
Section \ref{renormalizationtransverse}.

This random force gives a contribution to the displacement 
which at large scale has the same spatial dependence than
the one naively extrapolated from Larkin regime 
formula (\ref{larkinmove2}) and thus (\ref{displ}).
One thus finds:
\begin{eqnarray} 
&& B_{\rm{RF}}(r)  \sim \frac{\Delta_R}{4 \pi c \eta_0 v}~~ ln r  \qquad d=3   \\
&& B_{\rm{RF}}(r)  \sim C_d \frac{\Delta_R}{4 \pi c \eta_0 v}~~ y^{3-d} \sim  
\frac{\Delta_R}{c \eta_0 v} x^{(3-d)/2}  \qquad d<3
\end{eqnarray}
At large scales the random force contribution
to $B(r)$ dominates.
Although the formula resembles the perturbative one,
the amplitude of the random force is given by the 
{\it renormalized} $\Delta_R$ which has been to be extracted
from the RG analysis and is determined by the non linear
pinning force. In general $\Delta_R$ can be different from
the perturbative $\Delta$. In particular $\Delta_R$ must vanish when 
$v \to 0$. 

\begin{figure}

\centerline{\epsfig{file=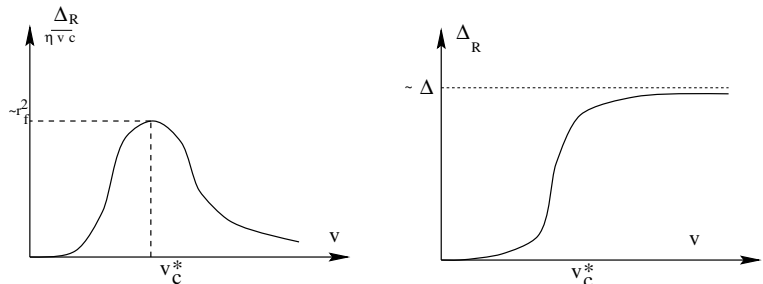,angle=-90,width=10cm}}
\caption{Renormalized random force strength as a function
of the velocity (left) and resulting amplitude in
displacement correlations (see text).
\label{figranforce}}
\end{figure}

$\Delta_R$ is a non universal quantity (contrarily to the behaviour
in the Bragg glass) but one can still
obtain a reliable estimate for $\Delta_R$ by studying
the crossover depicted in Fig. (\ref{bragg}).
If the velocity is smaller than the crossover 
velocity $v^*_c$ the random force will be renormalized 
downwards according to the behaviour in the Bragg glass
phase. Thus $\Delta_R$ will be smaller than the bare 
$\Delta$. This is illustrated in figure \ref{figranforce}

The amplitude of the displacements (e.g the prefactor of
the logarithmic growth in (\ref{displ}) ) generated by the renormalized random
force is maximum around the velocity $v^*_c$. Even at this velocity 
the displacements can be estimated as 
$B_{\rm{RF}}(r) \sim r_f^2 ln (r/R^y_c)$ in
$d=3$ and $B_{\rm{RF}}(r)\sim r_f^2 (y/R^y_c)$ in $d=2$. At all
other velocities the amplitude is much smaller.

Given the form of the displacement correlation function
the Moving glass will have QLRO in $d=3$. One finds
for transverse translational order correlations:
\begin{eqnarray}
C_{K_y,K_x=0}(0,y,z) \sim  (\frac{R^y_a}{\sqrt{y^2 + z^2 \frac{c_{11}}{c_{44}} }})^{A_K}
\qquad A_K= \frac{K_y^2 \Delta_R}{16 \pi v c} 
\end{eqnarray}
In particular $A_{K_0} = \pi a^2 \Delta_R/(4 v c)$. The dependence in the
coordinate $x$ is:
\begin{eqnarray}
C_{K_y,K_x=0}(x,0,0) \sim  (\frac{R^x_a}{x})^{A_K/2}
\end{eqnarray}
and thus one finds an anistropic divergence of the
Bragg peaks corresponding to $K_x=0$ of the form:

\begin{equation} 
S(q) \sim \frac{1}{(c_{11} q_y^2 + c_{44} q_z^2)^{2-A_K/2}} \sim  \frac{1}{q_x^{2-A_K/2}}
\end{equation}

The question of the divergences of peaks associated
to $K_x >0$ is discussed in the next section.

In $d=2$ algebraic growth of displacements imply 
a stretched exponential decay of $C_K(r)$ and thus
that the peaks in the structure factor are rounded
(as the dotted line in figure \ref{structure})

Note that in each configuration of the disorder
the random force along $y$ and the transverse critical force
will compete. The physics of the moving glass will be determined
by this competition. 

The roughness of the channels define an additional lengthcale at
which the wandering becomes similar to the lattice spacing.
As in the statics (Bragg glass) it is possible to 
estimate these lengths. At large velocity these lengths are
large and at these scale the system
is very anisotropic. A simple argument 
a la Fukuyama-Lee similar to the one in 
Ref. \onlinecite{giamarchi_vortex_long} give:
\begin{equation} \label{fukulee2}
R_a^y \sim (a^2 v c/\Delta)^{1/(3-d)}, \qquad
R_a^x = v (R_a^y)^2/c
\end{equation}
At large $v$ one can also obtain these lengths by
looking at the displacements generated by the
random force. For small $v < v^*_a$ there is a long
crossover since at small scales the system looks
more like the Bragg glass. As a consequence the
estimates for $R_a$ change. This illustrated in
Fig. \ref{bragg}. $v^*_a$ is 
determined roughly by $R_{cr}=R^y_a$.

Let us summarize the main regimes as a funtion
of the velocity of the moving glass, as can be seen
on Fig. \ref{bragg}. At large velocity $v > v^*_c$
the system is already anisotropic at the scale
$R_c$ and pinning and correlations are determined
directly by the asymptotic moving glass behaviour.
For $v^*_a < v < v^*_c$ the system is isotropic
at the Larkin length and pinning is similar
to the static, but the system is still very anisotropic
at scales $R^y_a$. Finally for $v>v^*_a$ the system is 
almost static like up to $R^y_a$ and isotropic.
The randon force is enormously reduced, and transverse
barriers are very large

\subsection{Decoupling of channels and dislocations}

\label{sec:decoupling}

Most of the novel properties of a moving structure
discussed in the previous Section were obtained from
the moving glass equation (\ref{movglasseq2}), which contains only
the transverse displacements $u_y$. They thus rest
only on the channel structure itself and not on the
precise motion of the individual particles along these
channels. Let us now discuss the problem of the coupling
of particles between different channels which is important
for the issue of topological, translational order and
structure factor. 

An oustanding problem in the statics is whether or
not topological defects will be generated by disorder 
in an elastic structure. Using energy arguments 
it was predicted that due to the periodicity a
lattice is stable to dislocations at weak disorder in
$d=3$ giving rise to the Bragg glass. The similar 
question of whether disorder will generate dislocations
arise also for moving structures. At first sight the
situation looks even more complicated to tackle
analytically and furthermore precise energy arguments
cannot be used because the system is out of equilibrium.
However, as is becoming clear from the discussion in the
previous Section the problem of dislocation can now be discussed
here in term of decoupling of channels. Even in the presence 
of dislocations our picture of pinned channels should remain valid as 
long as periodicity along $y$ is maintained. Before the channel 
structure was identified in \onlinecite{giamarchi_moving_prl} it was 
unclear how dislocations could affect a moving structure. The existence 
of channels then {\bf naturally} suggests a scenario by which dislocations
will appear. In fact Ref. \onlinecite{giamarchi_moving_prl}
naturally suggests that transitions from elastic to plastic
flow may now be studied as {\it ordering transitions} in the structure
of channels. We thus discuss the problem of the coupling
of particles between different channels which is important
for the issue of topological, translational order and
the structure factor.

The peaks at vectors with a non zero $K_x$ will
distinguish between the Moving Bragg glass and the
Moving Transverse glass. Indeed $P^{T}_{xx}$ now determines
the large scale behaviour, and thus $c_{66}$.

Thus we call
this system the Moving Bragg glass. The question of the
decoupling is examined in the next section.

Let us examine first whether dislocations will appear in the
Moving Bragg glass in $d=3$. The relative deformations due
to disorder grow only logarithmically with distance, resulting
in quasi long range order. At weak disorder or large 
velocity (since the relevant parameter is $\Delta/v$)
the prefactor of the logarithmic displacements is
very small. This suggests, by analogy with the statics, that
dislocations {\it will not appear}, leading to a stable
Moving Bragg glass at weak disorder or large velocity.
In that case
the structure factor will exhibit Bragg glass type peaks
(at all the small reciprocal lattice vectors).
Note however that due to the anisotropy
inherent to the motion the {\it shape} of each peak will be highly anisotropic
the length $R_a^x$ being much larger than $R_a^y$.

Upon increase of disorder the first likely transition
corresponds to a decoupling of the channels, while
the periodicity along $y$ is maintained. This corresponds
to the loss of divergent Bragg peaks at reciprocal lattice
vectors with non zero component along the direction of
motion. The peaks at reciprocal lattice
vectors along $y$ will still exhibit divergences
(computed in the last Section). This particular case of a Moving glass was observed
numerically \cite{moon_moving_numerics} and called the 
Moving Transverse Glass (see next Section). This 
phase has also a smectic type of order. One
question is whether particles can hop between the channels
in this phase. This however seems unlikely at zero temperature
provided the channels are well defined. 
In the absence of such hops this decoupled phase can still
be described by equation (\ref{movglasseq2}) it has non zero transverse
critical current. Increasing further the disorder 
should destroy even the channel structure leading to
a fully plastic flow.

An estimate of the locus of the transition between the
Moving Bragg Glass and the Moving Transverse Glass
is given by a Lindemann criterion.

\begin{equation}
\overline{\langle[u_x(y=a)-u_x(0)]^2\rangle} = c_L^2 a^2
\end{equation}

This indicates that edge dislocations appear first 
and decoupling of channels come from displacements along
the direction of motion.

In $d=2$ displacements grow algebraically. It is thus much more
likely that dislocations will appear at large scale. Presumably this
scale will be when the displacements will be of order
$a$. One then easily see that dislocations will appear between the 
layers. Indeed 

\begin{equation}
\overline{\langle[u_x(L^y_a)-u_x(0)]^2\rangle}  \sim a^2
\end{equation}

is controlled by the random force along $x$ and by $c_{66}$
(the displacements $u_y$ are down by a factor $R^y_a/R^x_a$ - see below).
In this regime blocks of channels of variable
transverse size (depending on the strength of the disorder) will be
separated by dislocations.

The peaks at vectors with a non zero $K_x$ thus allow to
distinguish between the Moving Bragg glass and the
Moving Transverse glass. In systems with a small ratio $c_{66}/c_{11}$
and stronger disorder the peaks with $K_x>0$ have a
tendency to be smaller (and decoupling becomes easier as $u_x$
becomes larger than $u_y$). Indeed the displacements at large scales
(and thus the decay of TLRO) are controlled by the
random forces along $y$, $\Delta_{yy}$
and along $x$, $\Delta_{xx}$ (they remain
uncoupled -see Section \ref{frgcomplete}). They act
differently on $u_y$ and $u_x$. Indeed,
only $P^T_{xx}$ (shear, $c_{66}$) and
$P^L_{yy}$ (compression $c_{11}$) lead to unbounded
displacements (e.g. in $d=3$). The $y$ random force
will thus act mainly on $u_y$ via compression,
and the $x$ random force
mainly on $u_x$ via shear. Though generally
one has $\Delta_{xx} < \Delta_{yy}$ at weak disorder
if the ratio $c_{66}/c_{11}$ is small this could
strongly favor the weakening of the $K_x>0$ peaks 
and channel decoupling.

Thus, the problem of the behavior of dislocations in the moving glass system
is of course still open, and constitutes as for the statics one of the
most important issues to understand. It is noteworthy, however, that
although these issues are of course important to obtain the structure
factors and such, they {\bf do not} affect the physics of the moving
glass.

\subsubsection{the Moving Bose glass}

In presence of correlated disorder along the $z$ 
direction we predict the existence of a ``moving Bose glass''
in $d=3$. Indeed the same calculation as above:

\begin{equation} \label{larkinmove2}
B_{RF}(y)=\Delta \int \frac{dq_x
dq_y d q_z}{(2\pi)} \delta(q_z) \frac{(1-\cos(q_y y))}{(\eta v q_x)^2 +
(c_{11}^2 q_y^2 + c_{44}^2 q_z^2)^2}
\end{equation}

now yields a fast growing displacement. Thus the
disorder effects are stronger and one can expect thermal
effects to be weaker for correlated disorder. The situation
resembles the $d=2$ case at $T=0$. One can still predict 
a transverse critical current. Full translational LRO
is unlikely but a Moving Transverse glass type of order
along $y$ is likely. This should be enough to guarantee
a localization effect of the layers and thus a transverse
Meissner effect along $y$. A detailed study will be given
in \cite{ledou_giamarchi_futur}.
The effect of correlated disorder
on another dissipative glass system (nonpotential)
were found to be quite strong in \cite{ledou_wiese_ranflow}.

\subsection{Moving glass at finite temperature}

\label{sec:temperature}

Thus the moving glasses, in its different forms, described by equation
(\ref{movglasseq2}) is a new disordered fixed point at $T=0$. An
important question is to understand what is the effect of thermal fluctuations.
 
Indeed in moving systems, as can be seen by perturbation theory, the 
fluctuation dissipation theorem is violated and a generation of 
temperature by motion will occur. This corresponds to the physical 
effect of heating by motion. Note however that this effect is different 
from the ``shaking temperature'' effect \cite{koshelev_dynamics} which
would occur even at $T=0$. Indeed at $T=0$ a system in the absence 
of thermal fluctuations retains perfect time order which implies that no 
temperature can be generated !

Although the temperature will grow due to motion, this effect competes
with the fact that naive power counting in glassy systems suggests
that the temperature is an irrelevant variable flowing to zero.
The competition between these two effects is highly non trivial 
and leads to new physics which needs to be investigated for the whole class
of moving glasses of section~\ref{sec:themg}. 

Remarkably, for this class of systems, new finite temperature fixed points
exists. In the case of driven lattices, we find a fixed point at finite temperature
in a $d=3-\epsilon$ expansion and $d=2+\epsilon$ expansion. Similarly 
for randomly driven polymers very similar fixed points are obtained
\cite{ledou_wiese_ranflow}. Thus a large class of moving glasses seems to 
exist at non zero temperature.

In $d=3$, the fixed point is slightly peculiar since both temperature 
and disorder flow to zero but can be analyzed along the same lines.
The properties of the finite temperature phase are continously related to the 
$T=0$ one. In particular the finite temperature moving glass exhibits the same
type of rough channel structure. Channels are slightly broadened due to bounded 
thermal displacements around the average channel position. Thus the asymptotic
behavior the displacements, and structure factor, still remains 
similar to the ones at $T=0$ discussed in previous Sections.
There is in addition a contribution of thermal displacements.
In $d=3$ they are small, and one sees that the RG methods developped
here allow to estimate more precisely the thermal heating effect, and
to distinguish it clearly from the disorder effects. This will
be important for determining the dynamical phase diagrams. In
$d=2$ the thermal displacements are large (see Section \ref{dynamicalco}).

The main effect of temperature is to modify the $v-f$ characteristics.
One finds (Section \ref{renormalizationtransverse}) that the
asymptotic mobility $\mu_R$ is non zero. However
at low temperatures or at velocities not too large the 
$v-f$ characteristics remains highly non linear. 
There is still an ``effective'' (or apparent) transverse critical force $F_c^y(T)$
as shown on figure~\ref{gend2}. At low temperature
the mobility $\mu_R$ is very small. If the velocity is 
not too large $v<v_c^*$ there are several regions 
in the $v$-$f$ curve. Below the transverse pinning force slow motion 
due to effective barriers exist. They are a growing 
function of $1/f$ until one reaches the 
finite temperature moving glass fixed point. Indeed reducing the transverse 
force probes larger and larger lengthscales. As depicted in figure~\ref{bragg}
one is dominated until the scale $R_{\rm cr}$ by the Bragg glass
fixed point for which temperature is strongly renormalized downwards.
In this regime the $v-f$ curve are nearly similar to the one in the 
static Bragg glass and thus highly non-linear. This corresponds to a creep 
regime. For smaller forces (i.e. 
when probing scales larger than $R_{\rm cr}$) one crosses over to the 
moving bragg glass fixed point. At that point the FRG calculation 
shows that the barriers saturate. Thus below the scale $f^*$ one recovers a 
linear $v-f$ charactersitics, with an extremely small mobility. 
Note that the scale $f^*$ which corresponds to the crossover scale 
$R_{\rm cr}$ can be much smaller than the critical transverse pinning
force if $v<v_c^*$.

These properties shows that even at finite temperature the 
moving Bragg glass remains different from a perfect crystal.
The definition of what is ``glassiness'' in a moving structure
is a new concept which has to be defined. In that respect too 
close analogies with the statics can be misleading.
A first obvious glassy characteristics is 
the loss of translational order, contrarily to the crystal.
Note however that a similar effect could be obtained by adding a random
force by hand to a perfect crystal. However the response of 
such a structure to an externaly applied force would be {\bf identical}
to the one of a perfect crystal. Thus 
the glassy properties of the moving Bragg glass are necessarily
stronger than such a state. The same question of
history dependence as discussed above at $T=0$ can be
asked. If these effects exist the question of the finiteness of the
barriers might not be as an important issue as in the static case. 
Note however that in some other examples of non potential
dynamics barriers can indeed be infinite \cite{ledou_wiese_ranflow}

Since the finite temperature Moving Bragg glass is described by
a new fixed point which still contains non-linear disorder $\Delta(u)$
the system remains obviously in a glassy regime. Some 
correlation functions of the system will necessary depend on the existence of
this finite $\Delta(u)$. This is reminiscent of what happens in the 
statics where the order parameter is the fluctuation of the 
susceptibility \cite{hwa_fisher_flux} whereas the averaged susceptibility
itself remain inocuous.

Note finally that for driven lattice (and for experimetal purposes), the
predicted existence of high (even if asymptotically finite)
barriers in some regime of velocities is a totally un-anticipated
property of disordered moving systems.

\begin{figure}

\centerline{\epsfig{file=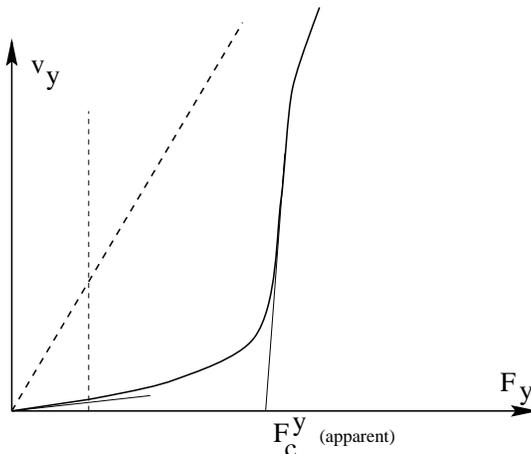,angle=-90,width=7cm}}
\caption{Transverse critical force as a function
of the velocity.$v-f$ characteristics at finite temperature.
\label{gend2}}
\end{figure}

\subsection{Phase diagrams}

\label{sec:diagrams}

Having established the existence of the moving Bragg Glass 
in $d=3$ and of the moving Transverse Glass in $d=3$ and $d=2$ and discussed 
their properties we now indicate in which region of the phase diagram 
these phases are expected to exist. We study the phase diagram as a function
of disorder, temperature and applied force (or velocity).

\begin{figure}
\centerline{\epsfig{file=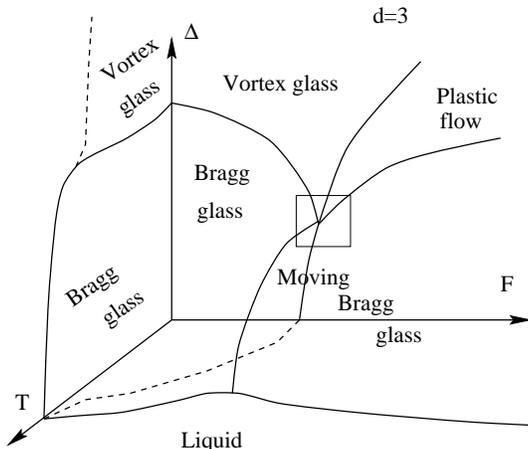,angle=-90,width=7cm}}
\caption{Phase diagram in the temperature $T$, disorder $\Delta$,
applied force $f$ variables. In $d=3$ for very weak disorder
since the moving glass is likely to be topologically ordered, the
possibility of a depining without a plastic regime exists.
Note that the Moving Bragg glass then should extend all
the way down to small $f$. We have not represented intermediate
phases (Hexatic, Moving Transverse Glass) for clarity.
Note also that the lower plane corresponds to a small but
non zero $\Delta$.
\label{gend3}}
\end{figure}

Let us first discuss the case $d=3$. We have represented in Figure~\ref{gend3}
a schematic expected phase diagram in the three axis. For clarity
we have not represented intermediate phases. The static phase diagram
at $F=0$ was dicussed in \onlinecite{}. There is a transition at finite
disorder strenght between the Bragg glass to an amorphous glass where 
dislocations proliferate. Upon applying a force the bragg glass phase 
becomes the Moving Bragg glass in the low velocity regime (creep regime)
and continuously extend to the Moving Bragg glass at higher drives.
At weak enough disorder the continuity between the two phases suggests 
that depinning should be elastic without an intermediate plastic region.
Upon raising the temperature the Moving Bragg glass melts to a liquid,
presumably through a first order dynamical melting transition.

The $T=0$ plane contains a pinned region for $F < F_c(\Delta)$
it is natural to expect the Bragg glass to still exist even for a finite 
force $F < F_c$ until the depinning transition. At higher disorder dislocations 
appear and the Bragg glass is replaced by an amorphous glass. The depinning of 
this amorphous glass should be via a highly disordered filamentary plastic 
flow. Upon increasing the force and thus the velocity, the system should 
reverse back to the moving bragg glass. 
At strong disorder and finite drive the liquid extends to zero temperature.

\begin{figure}

\centerline{\epsfig{file=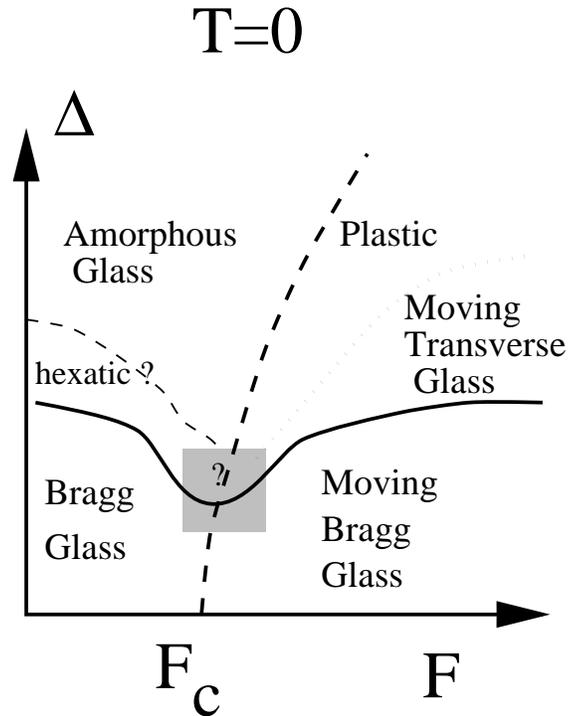,angle=-90,width=7cm}}
\caption{Phase diagram in the force $f$, disorder $\Delta$
in $d=3$ at $T=0$. The behaviour in the square region is
unclear. An interesting possibility would be a direct depinning
of an hexatic into the Moving Transverse glass, but other
scenario are possible.
\label{fig:delf}}
\end{figure}

\begin{figure}

\centerline{\epsfig{file=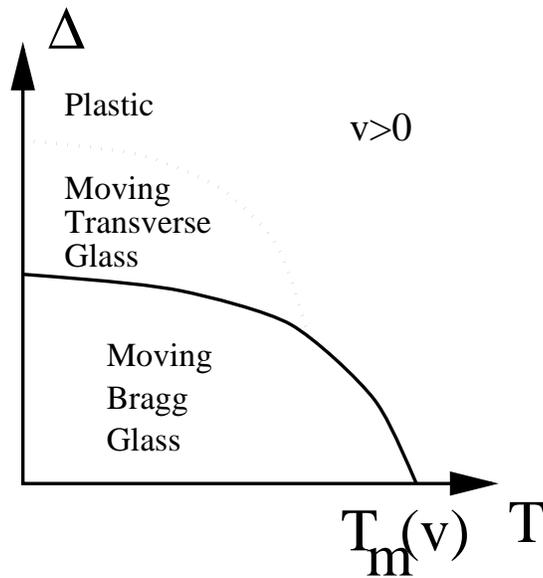,angle=-90,width=7cm}}
\caption{Phase diagram in the temperature $T$, disorder $\Delta$
in $d=3$ for a fixed velocity (not too small).
This phase diagram is the dynamical version to
the static one (containing the Bragg glass, the
vortex glass, and the field driven transition).
The MBG can either thermally melt (via a first order
transition) or decouple because of disorder.
\label{fig:delt}}
\end{figure}

\begin{figure}

\centerline{\epsfig{file=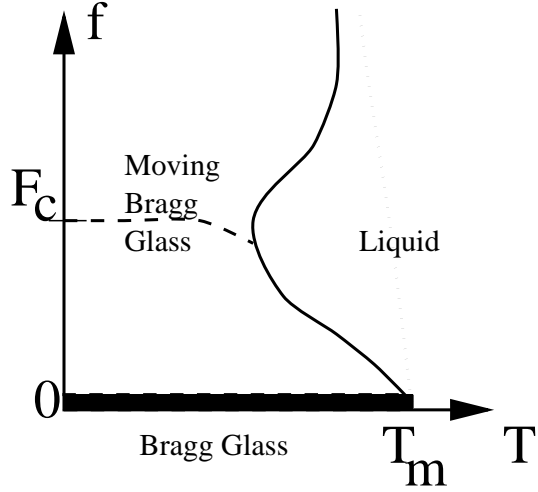,angle=-90,width=7cm}}
\caption{Phase diagram in the force $f$, disorder $T$
in $d=3$. The Bragg glass phase also exist at $T=0$ for
$f<F_c$.
\label{fig:ft}}
\end{figure}

This different behaviors are also represented along each plane in 
Figure~\ref{fig:delf},\ref{fig:delt},\ref{fig:ft}, where we have also 
indicated intermediate phases such as the Moving transverse glass.

Determining the exact shape of the various boundaries is still an
open and challenging problem, in particular in the square regions
in Fig. \ref{gend3}

One of the strong features that emerges from this phase diagrams 
is the fact that the Bragg glass is able to survive motion by 
turning into the moving bragg glass. On the other hand
other, more disordered phases such as the amorphous glass (vortex glass)
are likely to be immediately destroyed at finite drive (and finite temperature)
and to be continuously related to the liquid.

In d=2, most of the transitions reduce to simple crossover. 
At $F=0$ and finite disorder dislocations are expected to be present.
The resulting phase should thus be continuously connected to the liquid,
although it can retain good short distance translational order.
At $T=0$ there is a pinned phase until $F_c$, which should depin 
by a plastic flow. At larger drive disorder effects become smaller 
and one expects the system to revert to a moving glass state. As discussed
earlier, due to the presence of disorder induced dislocations, this 
state is a Moving transverse glass (if $d=2$ is above its lower
critical dimension).
At any finite temperature, one can use the RG flow of section~\ref{dynamicalco}.
Since the temperature renormalizes {\it above} the melting temperature and 
disorder flows to zero the resulting phase should be a driven liquid.

\begin{figure}

\centerline{\epsfig{file=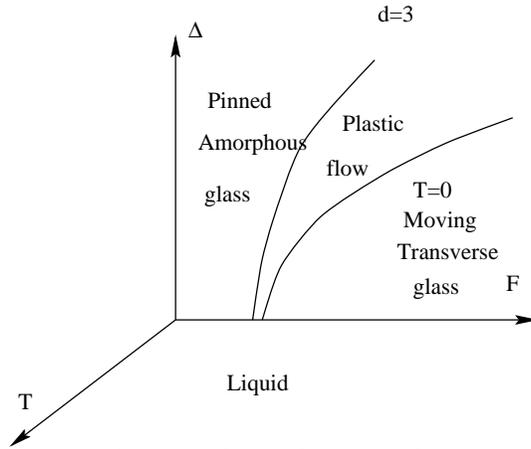,angle=-90,width=7cm}}
\caption{Phase diagram in the force $f$, disorder $T$
in $d=3$. There will be very long crossover not represented.
\label{fig:2d}}
\end{figure}

\subsection{comparison with numerical simulations}

\label{sec:numerics}

Some of the predictions of the Moving Glass theory
contained in the short account of our work
\cite{giamarchi_moving_prl} have been later verified in several
numerical simulations in $d=2$ and $d=3$.

The static channels were clearly observed at $T=0$ in $d=2$
\cite{moon_moving_numerics}.
They were also observed in Ref. \onlinecite{holmlund_mglass_numerics}.
The transverse critical force at $T=0$ in $d=2$ was observed very clearly in
\cite{moon_moving_numerics} and in \cite{ryu_mglass_numerics}
(see also \cite{holmlund_mglass_numerics})
and found to be a fraction of the longitudinal critical force
which is a reasonable order of magnitude.
The effect of a non zero temperature $T>0$ was observed to
weaken the effect of transverse barriers in \cite{ryu_mglass_numerics}
in $d=2$. Note that some non linear effects were observed to persist
for low enough transverse force and temperature.
Such an observation is in agreement in the discussion of Section ()
and can be interpreted as a long crossover.

Sharp Bragg peaks were observed in the direction transverse to motion
in \cite{moon_moving_numerics} at $T=0$. However the order along
the $x$ direction was found to have fast decay. This is consistent
with a decoupling of the channels, and the resulting state was termed
the ``moving transverse glass''. This is consistent with
the expectations from the theory presented here, as illustrated on
the phase diagram (\ref{fig:2d}). This phase is presumably the
$T=0$ moving transverse glass fixed point analyzed in Section ()
which does have a non zero transverse critical force.
This is in fact confirmed by the observation of 
\cite{ryu_mglass_numerics} of a smectic type of order where
well separated dislocations exist between the channels consistent
with the expectations discussed in Section ().
The absence of long range order was also observed in
\cite{spencer_mglass_numerics} in a stronger disorder 
situation.

In $d=3$, a simulation of a driven discrete superconducting XY gauge model
\cite{dominguez_mglass_numerics} finds not perfect but still
well defined Bragg peaks at $T>0$ (near the melting),
a result which indicates that the driven lattice
is in a quasi ordered moving Bragg glass state.
In another study on the simpler $d=3$ driven XY model at $T=0$
\cite{huse_mglass_numerics},
it was found that indeed there is a phase without
topological defects at large enough drive. If it carries to the
lattice problem it would indicate that indeed there is a
$d=3$ moving Bragg glass state.

Finally note that there are also very recent simulations
of a lattice with a periodic substrate
\cite{braun_periodic_simulation,reichhardt_nori_periodic}.
This is a simpler case where
a transverse critical current exist (it does exist for
a single particule). It may be worthile
to investigate this case in all details.

All the above numerically observed effects seem to be in
qualitative agreement with the predictions in
\cite{giamarchi_moving_prl}. However, it would be
very useful to be able to make a more {\it quantitative}
comparison. This should now be possible, as we will give here
more detailed predictions than the short account
\cite{giamarchi_moving_prl}. Among the various 
interetsing topics to check are the algebraic decay of
translational order in $d=3$, a detailed study of the
dependence of the transverse critical force on the
velocity, the exponent $\theta$ of the transverse
depinning, a measure of the barriers at low temperatures,
a characterization of the history dependence, and
zero and at low temperature. 

\subsection{comparison with experiments}

\label{sec:experiments}

The moving glass picture has also been 
confronted with experiments. Since these experiments 
need the characterization of a moving structure
they are challenging. The transverse critical current can in principle
be observed in transport experiments (and will show up
as hysteretic effects). These are difficult
though because of dissipation in the longitudinal direction.

Decoration experiments on the {\it moving} vortex lattice have
been performed recently by Marchevsky et al. \cite{marchevsky_decoration_channels}.
In these experiments the external field is slowly varied and vortices
are decorated while they move. The decoration particles thus accumulate
on the regions where vortices are flowing preferentially.
The lattice is observed to move in the symmetry axis
direction and relatively large regions of highly correlated static
channels are observed. These channels do not look like
``plastic channels'' but rather like the channels predicted
in \cite{giamarchi_moving_prl}. Note however that 
some dislocations along $y$ appear (defects in the layered
structure). This may be due to strong disorder effects
or since the advancing front geometry is in the shape of a droplet 
some dislocations are unavoidable.
Another set of experiments in NBSe2 was also reported
in \cite{pardo_decoration_motion}. Note that there has
also been several decoration experiments performed {\it just after} the
current is turned off. These can in principle probe
the defect structure of the flowing lattice (though one may worry about
transient effects) but cannot show the channel structure.

In \cite{giamarchi_moving_prl}
we have suggested that the transverse barriers may explain the
anomalies recently observed in the Hall effect in a Wigner crystal
in a constant magnetic field \cite{perruchot_thesis,perruchot_prl}. 

The qualitative analysis suggested by
the moving glass theory is as follows.
An electric field $E_x$ is applied in along the $x$ direction.
The Wigner crystal starts moving along $x$ when the applied field is
larger than a ``longitudinal'' threshold $E_x > E_c$. It produces
a current along $x$, $I_x = q v_x$ which is directly measured. Below
the longitudinal threshold a highly non linear regime is observed
where activated motion dominate.
Since it is moving in a high magnetic field,
the moving Wigner crystal is submitted to a transverse Lorentz force
$F^L_y = q v_x B$. The geometry of the experiment is such, however, that
no transverse motion is permitted in the stationary state (because
of zero current boundary conditions), and thus $v_y=0$.
Thus the transverse Lorentz force must be balanced by a
transverse electric field, which is thus generated, and
is measured as the Hall voltage $V_y$. In the absence of transverse
pinning the Hall voltage is $V_y = L B I_x$. Remarkably,
it is found in the experiment that the actual measured Hall voltage
is indeed $V_y = L B I_x$ for small $I_x$, then experiences a plateau,
and finally starts again growing linearly with 
a slope $d V_y/d I_x \approx L B $.
We have interpreted the different behaviours upon increase of
$I_x$ as follows. For small $I_x$ one is near the longitudinal
depinning and it is probably a plastic flow regime with little transverse
barriers. Then upon motion, a transition
to a moving solid occurs, which is presumably a moving glass.
The existence of a
non zero transverse critical force $F^c_y > 0$ then immediately
implies that there are sliding states with $v_y = 0$
as long as $F^L_y < F^c_y$ and no Hall voltage necessary.

Finally note a recent experiment
\cite{gurevich_periodic_dynamic}
on superconducting multilayers where it was found that
the flux flow resistivity exhibit quasi periodic oscillations
as a function of the field. This was interpreted
\cite{gurevich_periodic_dynamic} in terms
of dynamics matching of the moving vortex lattice with the periodic
substrate. This is compatible with the presence 
of a quasi ordered structure in motion.

It would be interesting to probe further the channel
structure by direct imaging techniques. In particular
one may investigate the degree of reproducibility of the
channel pattern. It is predicted that upon sudden reversal
of the velocity the channels should be {\it different}.
The question of order and quasi order can be probed
in experiments such as neutron scattering, flux lattice imaging
magnetic noise experiments, NMR experiments and more indirectly
in transport measurements. Other imaging techniques such
as $\mu$-SR NMR electron holography can also be used.
Finally it would be interesting to check for similar effects
in the presence of columnar defects since, as discussed
in this paper we predict the formation of a moving Bose glass.

\section{The Model and Physical Content}
\label{model}

\subsection{Derivation of the equations of motion} \label{sec:demotion}

Let us first derive the equation of motion for a lattice submitted
to external force $f$. We work in the {\it laboratory frame}.
This offers several advantages that will become obvious later.
We denote by $R_i(t)$ the true position of an individual vortex
in the laboratory frame. The lattice as a whole moves
with a velocity $v$.
We thus introduce the displacements $R_i(t)=R_i^0+v t+u_i(t)$
where the $R_i^0$ denote the equilibrium positions in
the perfect lattice with no disorder. $u_i$ represent the displacements
compared to a moving perfect lattice (and corresponds to the
position of the i-th particle in the moving frame). The definition of
$v$ imposes $\sum_i u_i(t) = 0$ at all times. We furthermore assume that
the motion is overdamped. The exact equation of motion can then be
obtained from the Hamiltonian $H$ by
\begin{eqnarray} \label{labase}
\eta \frac{d u_i(t)}{dt} = - \frac{\delta H}{\delta u_i} + f - \eta v +
\zeta_i(t)
\end{eqnarray}
where $\eta$ is the friction coefficient and the thermal noise
satisfies $\overline{\zeta_i(t) \zeta_j(t)}
= 2 T \eta \delta_{ij} \delta(t-t^{\prime})$.
The Hamiltonian is the standard Hamiltonian for periodic structure in a
random potential  $H = H_{\text{el}} + H_{\text{dis}}$. $H_{\text{el}}$
is the standard elastic Hamiltonian, and $H_{\text{dis}}$ describes the
interaction with the random potential
\begin{equation}
H_{\text{dis}} = \int_r V(r) \rho(r) = \sum_i \int_r V(r) \delta(r -
R_i^0+vt+u_i(t) )
\end{equation}
where the random potential has correlations $\langle
V(r)V(r^{\prime})\rangle = g(r-r^{\prime})$
of range $r_f$.

In order to use the standard field description of the
displacement
$u$ instead of focussing on the equation for one particle, one rewrites
(\ref{labase}) as
\begin{eqnarray} \label{labasem}
\eta \frac{d u_i(t)}{dt} = - \frac{\delta H_{el}}{\delta u_i}
+ \int_r \partial V(r) \delta(r - R_i^0+vt+u_i(t) ) +
f - \eta v + \zeta(R_i(t),t)
\end{eqnarray}
In doing so one would get the same thermal noise for two particles being
at the same place at the same time, instead of the two independent
noises of equation (\ref{labase}). Since such a configuration cannot
happen, going from (\ref{labase}) to (\ref{labasem}) is essentially
exact.

As for the static case
\cite{giamarchi_vortex_short,giamarchi_vortex_long} the difficulty is
to take the continuum limit of
(\ref{labasem}) since the disorder can vary at a much shorter scale than
the lattice spacing $a$. To proceed one follows the same steps than for
the static case, suitably modified to take into account the time
dependence of the displacements. One first
introduces a smooth interpolating displacement field $u(r,t)$ such that
$u(R_i^0 + v t,t) = u_i(t)$ (see formula $A2$ of
\onlinecite{giamarchi_vortex_short,giamarchi_vortex_long}). The field
$u(r,t)$
is the smoothest field interpolating between the actual positions
$u_i(t)$. All coordinates $r$ are expressed in the laboratory frame.
The field $u(r,t)$,
whose components we denote by $u_\alpha(r,t)$ thus expresses the
displacement in the moving frame, as a function of the coordinates of
the laboratory frame.

As for the static, if one assumes the absence of dislocations at all
times the particles can be labeled in a unique way. One then introduces
the continuous labelling field:
\begin{eqnarray} \label{label}
\phi(r,t) = r - v t - u[\phi(r,t) + v t,t]
\end{eqnarray}
Thus $\phi(R_i(t),t)=\phi(R_i^0 + v t + u_i(t) ,t)=R_i^0$
by definition, and $\phi$ numbers the particles by their initial
positions. In the absence of dislocations
the field $\phi(r,t)$ is uniquely defined.

To obtain the continuum limit of the equation (\ref{labasem}) one
first performs the continuum limit in the Hamiltonian
as in \onlinecite{giamarchi_vortex_long}, to obtain for the disorder term
\begin{equation} \label{couplage}
H_{\text{pin}} = \int dx V(x) \rho(x) = -\rho_0 \int dx V(x)
\partial_\alpha u_\alpha
 + \rho_0 \int dx \sum_{K\ne 0} V(x) e^{iK\cdot(r-vt-u(r,t)) }
\end{equation}
where $K$ spans the reciprocal lattice and $\rho_0$ is the
average density.

In (\ref{couplage}) we have made the approximation
\begin{equation}
u(\phi(r,t)+ vt,t) \sim u(r,t)
\end{equation}
Such an approximation is exact up to higher powers of $\partial u$,
negligeable in the elastic limit, as for the static case
\cite{giamarchi_vortex_long}. However the dynamic case is more subtle
since such terms could generate when combined with a non-zero velocity
relevant terms. This is the case for example of the so-called KPZ terms
generated through cutoff effects. Since it is hopeless to try to tackle
from first principle all such additional terms the only safe procedure
is to assume that every term allowed by symetry will be generated, and
has to be examined. We will proceed with such a program in
section~\ref{beyond}. For the moment we only retain the dominant terms
of (\ref{couplage}). If one then takes the derivative with
respect to the smooth field
$u(r,t)$ one obtains for
the equation of motion in the laboratory frame
\begin{equation} \label{eqmotion}
\eta \partial_t u_{rt}^\alpha + \eta v \cdot \nabla u_{r t}^\alpha =
\int_{r'} \Phi_{\alpha \beta}(r-r')  u_{r t'}^\beta
+ F_{\text{pin}}^{\alpha}(r,t)
+ f_{\alpha}-\eta v_{\alpha} + \zeta_{\alpha}
\end{equation}
where $\Phi_{\alpha \beta}(r-r')$ is the elastic matrix. The term
$\eta v \cdot \nabla u_\alpha$ comes from the standard Euler
representation when expressing the displacement field in the laboratory
frame. $-\eta v_{\alpha}$ is the average friction and in the continuum
$v$ is determined by the condition that the average of $u$ is zero.
The thermal noise satisfies in the continuum limit
\begin{equation}
\overline{\zeta_{\alpha} (r,t)\zeta_{\beta} (r^{\prime},t^{\prime})}
= 2 T \eta \delta_{\alpha \beta}
\delta^d(r-r^{\prime})\delta(t-t^{\prime})
\end{equation}
and
\begin{equation} \label{pinham}
F^{\text{pin}}_{\alpha}(r,t) = - \delta {H_{\text{pin}}}/\delta
u_{\alpha}(r,t) = V(r)\rho_0\sum_{K} iK_\alpha \exp
(iK\cdot(r-vt-u(r,t))) - \rho_0 \nabla_\alpha V(r)
\end{equation}
is the pinning force.
Note the difference between our equation (\ref{eqmotion}) and the one
derived in \onlinecite{larkin_largev}, which does not contain the
convective term. This difference comes simply from a different
definition of the displacement fields. They consider displacements
fields labelled by the original position of the particle (i.e. the
actual position of the particle is $r + u$) whereas for us $r$ denotes
the actual position of the vortex considered (i.e. in the presence of an
external potential $V$ the potential acting on the vortex at point $r$
is $V(r)$ instead of $V(r+u)$ for \onlinecite{larkin_largev}).

In Appendix~\ref{nonham} we give a more general derivation of (\ref{pinham})
valid even for cases where the equation of motion is {\it not} the
derivative of a potential.

\subsection{Models and Symmetries} \label{sec:models}

Before we even attempt to solve (\ref{eqmotion}), let us examine the
various symmetries of the problem and define several models
which approximate the physical problem at various levels.

The physical symmetry of the original equation
of motion (\ref{labase}) is the global inversion symmetry
($r \to -r$, $u \to -u$, $v \to - v$ $f \to -f$).
When the force (and thus $v$) is along a principal lattice direction,
one has then two independent inversion symmetries
$I_x =$ ($x \to -x$, $u_x \to - u_x$, $v \to - v$ $f \to -f$)
and $I_y =$ ($y \to -y$, $u_y \to - u_y$). These symmetries
are exact and hold in all cases. They are the only symmetries
of the original model (\ref{labase}). The proper continuum limit of
equation (\ref{labase}) must thus include all terms which
are relevant and consistent with these exact symmetries.
We define such a model as Model I, which will be
studied in more details in Section (\ref{beyond}).
The additional terms can originate from e.g anharmonic
elasticity, cutoff effects or higher order terms
in $\nabla u$, as will be dicussed in Section~\ref{beyond}.

If one drop in Model I the terms which are small
in the elastic limit $\nabla u \ll 1$, one obtains
another model that we call Model II. It corresponds
to the continuum
limit of the equation of motion to obtain equation
(\ref{eqmotion}) i.e. (\ref{eqmotion}) in the elastic limit.
Although model II is slightly simpler than model I, it only
misses terms which are small in the bare equation but would
be allowed by the above symmetries. Even if some of them are
relevant, the would only be able to change the physics compared
to model II at very large length scales. One thus expects
model II to give in practice an extremely accurate description of
the physics.

The Model II posesses a higher symmetry than Model I:
let us examine the
symmetries of the pinning force (\ref{pinham}).
Using the correlator of
the random potential $V$, the correlator of the pinning force is:
\begin{equation} \label{eq1}
\Delta_{\alpha \beta} =
\overline{F^{\text{pin}}_\alpha(r,t,u_{rt})
F^{\text{pin}}_\beta(r',t',u_{r't'})} =
\rho_0^2 g(r-r')  \sum_{K,K' \neq 0} iK_\alpha iK'_\beta
e^{iK\cdot(r-vt-u_{rt})+iK'\cdot(r'-vt'-u_{r't'})}
\end{equation}
Since $u$ is a smooth field it has no rapidly oscillating components and
thus  in (\ref{eq1}) the terms that are rapidly
oscillating in $r+r'$ can be discarded. Setting  $K'=-K$ in (\ref{eq1}),
one is left with
\begin{equation} \label{copinfor}
\Delta_{\alpha \beta} =  \rho_0^2 \sum_{K \neq 0} K_\alpha K_\beta
g(r-r') \exp(iK\cdot(r-r') - iK\cdot(u_{rt}-u_{r't'}+ v(t-t'))
\end{equation}
The symmetries of (\ref{copinfor}) thus a priori
depends on the precise form of
the correlator $g(r)$. However in the elastic limit it
is legitimate to replace $u_{r't'}$ by $u_{r t'}$ in
the above expression. Integrating then over $r'$ one obtains:
\begin{equation} \label{copinfor2}
\Delta_{\alpha \beta} =  \rho_0^2 \sum_{K \neq 0} K_\alpha K_\beta
g_K \exp(- iK\cdot(u_{rt}-u_{rt'}+ v(t-t'))
\end{equation}
where $g_K$ is the Fourier coefficient of the correlator $g(r)$.
Since $g_K$ is essentially zero for $ K \gg 1/r_f$, the
error made in the above approximation is itself of order
$\nabla u$ and thus consistent with the elastic limit
approximation. This justifies the choice of (\ref{copinfor2})
as the random force correlator in Model II. The disorder term
then posseses the statistical tilt symmetry (STS)
$u_{rt} \to u_{rt} + f(r)$ where $f(r)$ is an arbitrary function.
In this case one can absorb any {\bf static}
change in $u$ without affecting the correlations of the pinning force.
Furthermore in the case of isotropic elasticity, the additional inversion
symmetry $y \to -y$ holds.

Though we will study the complete Model II in Sections
\ref{sec:pert} and \ref{beyond} its main physics can be
understood\cite{giamarchi_moving_prl} by noticing that the
pinning force $F^{\text{pin}}_{\alpha}(r,t)$ in (\ref{pinham})
naturally splits into a {\it static} and a time-dependent part:
\begin{eqnarray}
F_\alpha ^{\text{stat}}(r,u) &=& V(r)\rho_0\sum_{K.v=0} iK_\alpha \exp
(iK\cdot(r-u))  - \rho_0 \nabla_\alpha V(r)  \label{eqmotion2} \\
F_\alpha^{\text{dyn}}(r,t,u) &=& V(r)\rho_0\sum_{K.v\neq 0} iK_\alpha \exp
(iK\cdot(r-vt-u)) \nonumber
\end{eqnarray}
The static part of the random force comes from the modes
such that $K.v=0$ which exist for any
direction of the velocity commensurate with the lattice.
The maximum effect is obtained for $v$ parallel to one
principal lattice direction, the situation we study now.
This force originates {\bf only} from the periodicity along $y$
and the uniform density modes along $x$, i.e the smectic-like modes.
Since this static pinning force $F_\alpha^{\text{stat}}(r,u)$
is along the $y$ direction, it is useful
consider only the transverse part (along $y$) of the equation
of motion (\ref{eqmotion}) dropping
$F_\alpha ^{\text{dyn}}$. This leads to introduce Model III,
defined by the following equation of motion in the laboratory frame:
\begin{eqnarray}
\eta \partial_t u_y +\eta v\partial_x u_y &=& c\nabla^2 u_y
+ F^{\text{stat}}(r,u_y(r,t)) + \zeta_y (r,t) \nonumber \\
F^{\text{stat}}(x,y,u_y) &=& V(x,y) \rho_0 \sum_{Ky \ne 0} K_y
\sin{K_y(u_y-y)} - \rho_0 \partial_y V(r)
\label{staticequ}
\end{eqnarray}
Although for non isotropic elasticity $u_x$ also appears in
the equation of motion for $u_y$, and can in principle lead
to additional static effects, it is not included in Model
III  for reasons which will be explained below. Thus
Model III only involves
the {\it transverse} displacements $u_y$. It posesses the same
symmetries as Model II with the three additional independent
symmetries $y \to -y$, $u_y \to -u_y$ and ($x \to -x$, $v \to -v$)
and is also defined in the elastic limit.

It is to be emphasized that although the derivation of model III was given
here systematically starting from an elastic description. the
only serious hypothesis behind Model III is the existence of transverse
periodicity \cite{giamarchi_reply_movglass}. As discussed in
section~\ref{sec:themg} (\ref{staticequ}) will be the correct starting
point to describe {\bf any} kind of structure have such transverse periodicity
properties. One thus expects model III to be the generic equation
containing the physics necessary to describe these structures.

\section{Perturbation theory for the complete time-dependent equation}
\label{sec:pert}

Let us start by a simple perturbation analysis of the equation of
motion Model II. Such a large velocity or weak disorder expansion has a
long history
in various contexts such as vortex lattices \cite{schmidt_hauger,larkin_largev}
and charge density waves \cite{sneddon_cross_fisher,russes_cdw}.
The natural idea is that at large velocity the disorder term oscillates
rapidly and averages to a small value and that $1/v$ is a good
expansion parameter. As we will see such an idea is in fact incorrect,
since previously unnoticed divergences appear in the perturbation theory.

\subsection{Analysis to first order}

We start from the initial equation
(\ref{eqmotion}) defining Model II that we rewrite as:
\begin{equation} \label{startexp}
{(R^{-1})}^{\alpha \beta}_{rt r't'} u^{\beta}_{r't'} = f_{\alpha} -
 \eta_{\alpha \beta} v_{\beta} + f_{\alpha} (r,t,u_{rt})
\end{equation}
where from now on we drop the $\text{pin}$ subscript on $f$.
The response kernel
$R$ is defined in Fourier space:
\begin{equation} \label{defr}
{(R^{-1})}^{\alpha \beta}_{qt, q't'} = \delta_{t t'} \delta_{q',-q}(
\eta_{\alpha \beta} \partial_t  + i \eta_0 v_{\gamma}
q_{\gamma} \delta_{\alpha \beta} +
C_T(q) P^T_{\alpha \beta}(q)
+ C_L(q) P^L_{\alpha \beta}(q) )
\end{equation}
where the elastic matrix is : $C_T(q) = c_{66} q^2$, $C_L(q) = c_{11} q^2$
for a two dimensional problem and
$C_T(q) = c_{66} q^2 + c_{44} q_z^2 $, $C_T(q) = c_{11} q^2 + c_{44} q_z^2$
for a three dimensional problem.  The bare value of the
friction coefficient $\eta_0$ is defined as $\eta_{\alpha \beta} = \delta_{\alpha \beta} \eta_0$.
In (\ref{startexp}) the velocity is fixed by the constraint that
$\langle u^{\beta}\rangle=0$ to all orders in perturbation theory.
This is equivalent to enforce that the linear term
in the effective action \cite{zinnjustin_norenorm} is exactly zero.

Instead of working directly with the equation of motion it is more
convenient to use the Martin-Siggia-Rose formalism
\cite{martin_siggia_rose}. The generic de Dominicis-Janssen MSR functional is given by
\begin{equation}
Z[h,\hat{h}] = \int Du D\hat{u} e^{- S[u,\hat{u}] + \hat{h} u + i h \hat{u}}
\end{equation}
where $\hat{h},h$ are source fields. The MSR action corresponding to the
equation of motion (\ref{startexp}) and the disorder correlator
(\ref{copinfor2}) is
\begin{equation} \label{totalmsg}
S[u,\hat{u}]  =   S_0[u,\hat{u}] + S_{int}[u,\hat{u}]
\end{equation}
with
\begin{eqnarray}
S_0[u,\hat{u}] & = & \int_{rtr't'} ~ i \hat{u}^{\alpha}_{rt}
(R^{-1})^{\alpha \beta}_{rt,r't'}  u^{\beta}_{r't'}
- i \hat{u}^{\alpha} ( f_{\alpha} - \eta_{\alpha \beta} v_{\beta} )
- \eta T \int_{r,t} (i \hat{u}^{\alpha}_{rt}) (i \hat{u}^{\alpha}_{rt})
\label{freemsg} \\
S_{int}[u,\hat{u}]  & = & - \frac{1}{2} \int dr dt dt' (i \hat{u}^{\alpha}_{rt})
(i \hat{u}^{\beta}_{r t'})
\Delta^{\alpha \beta}(u_{rt} - u_{rt'} + v (t - t'))
\label{disordermsg}
\end{eqnarray}
Note that (\ref{totalmsg}) corresponds to the action derived in
Appendix~\ref{nonham}.

The fundamental functions to compute
are the
disorder averaged displacements correlation function
$C^{\alpha,\beta}_{rt,r't'} = \overline{ \langle u^{\alpha}_{rt}
u^{\beta}_{r't'} \rangle }$
and the response function
$R^{\alpha,\beta}_{rt,r't'} = \delta \overline{ \langle u^{\alpha}_{rt}
\rangle }/
\delta h^{\beta}_{r't'} $ which measures the linear response to a
perturbation
applied at a previous time. They are obtained from the
above functional as
$C^{\alpha \beta}_{rt,r't'} = \langle u^{\alpha}_{rt} u^{\beta}_{r't'}
\rangle_S $
and $R^{\alpha \beta}_{rt,r't'} = \langle u^{\alpha}_{rt}
i \hat{u}^{\beta}_{r't'} \rangle_S$
respectively. Causality imposes that $R_{rt,r't'}=0$ for $t'>t$
and we use the Ito prescription for time discretization
which imposes that
$R_{rt,r't}=0$.
We assume here time and space translational invariance
and denote indifferently $C_{rt,r't'} =C_{r-r',t-t'}$ and
$R_{rt,r't'} =R_{r-r',t-t'}$ by the same symbol,
as well as their Fourier transforms when no confusion is possible.
Note that in this problem $C_{-r,t} \neq C_{r,t}$ when $v$ is non zero.
In the absence of disorder the action is simply quadratic
$S=S_0$. The response and correlation fucntion in the absence of disorder
are thus (for $t>0$) and introducing the mobility $\mu = 1/\eta$:
\begin{eqnarray}
&& R^{\alpha \beta}_{q,t} = P^L_{\alpha \beta}(q) \mu
e^{- ( c_L(q) + i v q_x ) \mu t } \theta(t)
+ P^T_{\alpha \beta}(q) \mu e^{- ( c_T(q) + i v q_x) \mu t }  \theta(t) \\
&& C^{\alpha \beta}_{q,t} = P^L_{\alpha \beta}(q) \frac{T}{c_L(q)}
e^{- ( c_L(q) \mu |t| + i v q_x \mu t ) }
+ P^T_{\alpha \beta}(q)
\frac{T}{c_T(q)} e^{- (c_T(q) \mu |t| + i v q_x \mu t})
\end{eqnarray}
Note that the fluctuation dissipation theorem (FDT)
\begin{equation} \label{fdtth}
T R^{\alpha \beta}_{r,t} = - \theta(t) \partial_t
C^{\alpha \beta}_{r,t}
\end{equation}
does {\it not} hold here (it holds only
for $v=0$) since we are studying a moving system which does not
derive from a Hamiltonian.

It is easy to show that the disorder does not produce any
correction to the part $i \hat{u_t} ( c q^ 2 + i v q_x) u_t$
of the action, and thus that the parameters $c$ ($c_{11}$ and $c_{66}$ )
and $\eta_0 v$
are not renormalized (we consider here for simplicity the isotropic version $c=c_{11}=c_{66}$ but
this property holds in general). This is similar to the property of
non renormalization of connected correlations in the
statics (for $v=0$) due to the statistical tilt symmetry. Indeed here
one has the exact relation:
\begin{equation} \label{tiltsymmetry}
\overline{\log Z[h_t,\hat{h}]}
= - \int dt ~ h_t (c q^2 + i v q_x)^{-1} \hat{h} + \overline{\log Z[h_t,0]}
\end{equation}
where $\hat{h}$ is an arbitrary {\it static} field. This relation implies
that the static response function $\int dt' R_{q,t,t'} =
(c q^2 + i v q_x)^{-1}$
is not renormalized.
The property (\ref{tiltsymmetry}) is demonstrated
by performing the change of variable
$u_{t,r} = u'_{t,r} + f_r$ with $f_q= - (c q^2 + i v q_x)^{-1} \hat{h}_{q}$
in the action and noticing that the change in the random
force $F(r,u) \to F'(r,u)=F(r,u+f)$ is thus that all averages over $F$ and
$F'$ are identical since they have the same correlator from
(\ref{copinfor2}).

Let us now study the perturbation theory in the disorder and
compute the effective
action $\Gamma[u,\hat{u}]$ to lowest order in the
interacting part $S_{int}$, using a standard cumulant expansion
\begin{equation} \label{cumulant}
\Gamma[u,\hat{u}] = S_0[u,\hat{u}] + \langle S_{int}[u+\delta u,
\hat{u} + \delta  \hat{u}]
\rangle_{\delta u, \delta \hat{u}}
\end{equation}
where the averages in (\ref{cumulant}) over $\delta u$, $\delta \hat{u}$
are taken with respect to $S_0$.

The calculations are performed in Appendix \ref{appendixb}. One finds
that the effective action has the same form as the bare action,
up to irrelevant higher order derivative terms, with the following
modifications. First the full non linear form of the
correlator of the pinning force is corrected
by thermal fluctuations $\Delta^{\alpha \beta}_K \to
\tilde{\Delta}^{\alpha \beta}_K$.
In $d>2$ it reads:
\begin{eqnarray}
\tilde{\Delta}^{\alpha \beta}_K = \Delta^{\alpha \beta}_K
e^{- \frac{1}{2} K^2 B_{\infty}}
\end{eqnarray}
or equivalently $\tilde{g}_K = g_K e^{- \frac{1}{2} K^2 B_{\infty}}$ where
$B_{\infty} = \langle u^2\rangle_{th}$. We have defined
$B^{\alpha \beta}_{r,t} = 2 (C^{\alpha \beta}_{0,0} -
C^{\alpha \beta}_{r,t})$.
This amounts to
a smoothing out of the disorder by thermal fluctuations.
Secondly, the friction coefficient matrix is corrected
by $\delta \eta_{\alpha \beta}$, and the temperature
by $\delta T$. Finally, the driving force is
corrected by $\delta f$ (we are working at
fixed velocity, enforcing order by order that
$f + \delta f = \eta v$).

Let us start by the corrections to the driving force.
We find:
\begin{eqnarray} \label{hauger}
\delta f_\alpha(v) = \sum_K  \sum_{I=L,T} \int_{BZ}  dq
K_\alpha (K.P^I(q).K) g_{K} \frac{ v.(K+q) }{ c_I(q)^2 +
(\eta_0 v.(K+q))^2 }
\end{eqnarray}
To derive this formula from
(\ref{correction1}) we have used the symmetry $K \to -K$.
This formula gives
the lowest correction to the driving force at fixed velocity or,
equivalently
to the velocity at fixed driving force. It is identical to the
formula (22) of
Schmidt and Hauger \cite{schmidt_hauger}. There are small differences,
unimportant in
the elastic limit, which come from the different definitions of
the continuum
limit of the model (see discussion in section~\ref{sec:demotion}).

A salient feature of the above formula was noticed by Schmidt and Hauger,
i.e the velocity and the force are not, in general, aligned.
There are aligned
however when the velocity is along one of the principal lattice directions,
i.e $K_0 . v=0$ where $K_0$ is one of the principal reciprocal
lattice vector
(note that this is also the case for the median direction $\pi/6$). Such
a feature is reasonable on physical grounds and can be confirmed
by higher order analysis of the
perturbation theory (see section~\ref{renormalizationtransverse}).

Furthermore, using the approximation $v (K+q) \sim v K$ Schmidt and
Hauger found that, in $d=2$, the transverse pinning force versus the angle
$\alpha$ between the velocity and one principal direction of the
lattice has a discontinuity at $\alpha=0$. One could naively think
that such a discontinuity could
be interpreted as the existence of a transverse critical
current. Indeed a natural interpretation of figure~1 of
\onlinecite{schmidt_hauger} would be
that one need to apply a finite force to the lattice (opposite to the
transverse pinning force) to tilt sightly its velocity from the
principal axis direction. Notable confusion on this subject existed at
that time in the litterature \cite{schmidt_hauger,larkin_largev}.
Such an interpretation is in fact incorrect.
First, as Schmidt and Hauger correctly pointed out such a discontinuity
is an {\bf artefact} of the approximation $v (K+q) \sim v K$,
and diseappears
if the correct expression (\ref{hauger}) is used. Furthermore it is easy
to check that even with the above approximation the discontinuity exists
only in $d=2$ and the function is continuous for $d>2$. Thus the first
order perturbation does not exhibit any divergence and {\em does not}
give rise to a transverse critical current.

In order to have divergences in the perturbation theory (and the
associated effects) it is thus necessary to examine the perturbation
theory to second. We will perform such a calculation in
section~\ref{renormalizationtransverse}.

Before we do so it is interesting to examine the first order
corrections to the friction coefficient and the temperature.
\begin{eqnarray} \label{corrfrict}
\delta \eta_{\alpha \beta}(v) =
\int_{rt} t R^{\gamma \delta}_{r=0,t} K_\gamma K_\beta
\Delta^{\alpha \delta}_K e^{- i K.v t} e^{- \frac{1}{2} K.B_{0,t}.K}
\end{eqnarray}
At zero temperature it reads:
\begin{equation}
\delta \eta^{\alpha \beta}(v) = - \int d\tau dr \tau
R^{\gamma \delta}(r=0,\tau) \Delta^{\alpha \delta ; \gamma \beta}(v \tau)
\end{equation}
Still at $T=0$ and using the bare form of the
disorder one finds:
\begin{eqnarray}
\delta \eta_{\alpha \beta} = \sum_K \sum_{I=L,T} \int_{BZ}  dq
K_\alpha K_\beta (K.P^I(q).K) g_{K}
\frac{1}{ ( c_I(q) + i \eta_0 v.(K+q))^2 }
\end{eqnarray}
When the velocity is along a principal lattice
direction one finds that $\delta \eta_{xy}=0$ and thus the
friction matrix remains diagonal. However the corrections
to the friction is clearly not the same along $x$ and $y$.

Next we give the corrections to temperature.
They are:
\begin{eqnarray}  \label{corrtemp}
\delta (\eta T)_{\alpha \beta} = \frac{1}{2} \sum_K \int_t
\Delta^{\alpha \delta}_K e^{- i K.v t} ( e^{- \frac{1}{2} K.B_{0,t}.K}
- e^{- \frac{1}{2} K.B_\infty.K} )
\end{eqnarray}
Contrarily ot the velocity corrections, corrections to the temperature
(\ref{corrtemp}) exhibit divergences for any $v$ already at the first
order in the disorder. However these divergences are well hidden and can
{\bf only} appear if one looks at the {\it non zero temperature}
perturbation theory, which was not done in
\cite{larkin_largev,schmidt_hauger}. At
$T=0$ one finds trivially that
$\delta T=0$, showing that disorder alone cannot generate
a finite temperature.
Such corrections are thus non trivial and there is in general no simple
relation between $\delta(\eta T)_{\alpha \beta}$ and
$\delta(\eta)_{\alpha \beta}$, due to the absence of FDT theorem. Only
in the particular case where $v=0$ and of potential random forces
$\Delta^{\alpha \delta}_K = K_\alpha K_\beta g_K$ the FDT
theorem enforces $\delta T=0$ (see, e.g (\ref{fdtth}) and
\cite{carpentier_ledou_triangco}). The way to treat these divergences
will be examined together with the second order in perturbation in
section~\ref{renormalizationtransverse}.

\subsection{Correlation functions}

The last physical information that can be extracted from the
perturbation theory is about the correlation functions.
The calculation is performed in appendix~\ref{appendixb}, and
this gives (see (\ref{corrtfin})) in the static limit
(see (\ref{corrtfin}) for the full expresion)
\begin{eqnarray} \label{corrfonc}
\overline{\langle u^\alpha_{-q} u^\beta_{q} \rangle}
&=& \sum_{K,K.v=0}  \sum_{I=L,T, I'=L,T}
\int_{BZ}  dq (2 \pi) g_K
\frac{P^I_{\alpha \gamma}(q)}{c_I(q) + i \eta_0 v.q}
\frac{P^{I'}_{\beta \delta}(q)}{c_I'(q) - i \eta_0 v.q}
D_{\gamma \delta}
\end{eqnarray}
where the disorder is renormalized by the temperature
\begin{eqnarray}
g_K D_{\gamma \delta}(\omega) =  g_K K_\gamma K_\delta
e^{- \frac{1}{2} K.B_{0,\infty}.K}
\end{eqnarray}
Let us first focuss on the $T=0$ limit of (\ref{corrfonc}).
The length $R_c^{x,y}$ can be estimated by looking
at $A= \overline{\langle u^x_{-q} u^x_{q} + u^y_{-q} u^y_{q}\rangle}$.
This gives
\begin{equation} \label{compr}
A = \sum_{K,K.v=0} \int_{BZ}  dq (2 \pi) K_y^2 \left[
\frac{q_x^2}{q_\perp^2(v^2 q_x^2 + (c_{66}(q_x^2+q_y^2) + c_{44}
q_z^2)^2)}
+
\frac{q_y^2}{q_\perp^2(v^2 q_x^2 + (c_{11}(q_x^2+q_y^2) + c_{44}
q_z^2)^2)}
\right]
\end{equation}
where $q_\perp^2 = q_x^2 + q_y^2$.
Since $q_x \sim q_y^2$ one immediately sees from (\ref{compr}) that the
compression modes are the one responsible for the divergences.
Other modes are not divergent around $d=3$. Lengthscqles $R_c$ qre theus
controled by $c_{11}$.

On the other hand in order to obtain a decoupling of the channels one
can use a simple Lindermann criterion.
\begin{equation}
\overline{\langle[u_x(y=a)-u_x(0)]^2\rangle} = C_L^2 a^2
\end{equation}
Using (\ref{corrfonc}) one gets
\begin{equation} \label{linder}
B_{xx}(y=a) \approx \Delta_{yy} \int_{q,BZ}
\frac{q_x^2}{q_\perp^2(v^2 q_x^2 + (c_{66}(q_x^2+q_y^2) + c_{44}
q_z^2))}
\end{equation}
we have neglected all terms containing $c_{11}$.
One can thus neglect the compression
mode. The decoupling between the channels is thus controled by $c_{66}$
whereas the roughness of the channels and the characteristics
lengthcales of the moving glass will directly depend on the compression
mode $c_{11}$. Estimating the integral one finds in $d=3$:

\begin{equation} \label{linder}
B_{xx}(y=a) \sim \frac{\Delta_{yy}}{a^2 c_{44}^{1/2}}
\min( (\frac{a^2}{c_{66}})^{3/2} , (a/v)^{3/2} )
\end{equation}

One recovers the Bragg glass estimate (in the simpler case
$a=r_f$).

The above perturbation expansion for the Lindemann criterion
implicitely supposes that the random force is directed along the $y$
direction. In fact, under renormalization a random force along $x$ is
also generated as discussed in section~\ref{sec:physics}.
One can use again the Lindemann criterion with a random force 
along $x$. In that case only the transverse modes remain are 
important (fo small enough $q_x$).

\newpage

\section{Renormalization group study of the transverse physics (moving glass)}

\label{renormalizationtransverse}

Up know we have studied the perturbation expansion of the full continuous
model II, keeping both the $x$ and $y$ directions.
Doing the second order perturbation on that full model is tedious.
Since one knows on physical grounds that the singularities in the
perturbation theory will come from the {\it static components}
of the disorder \cite{giamarchi_moving_prl} , which as was discussed in 
Section \ref{sec:themg} originate from
the transverse degrees of freedom, we now study the
perturbation theory of the simplified {\it transverse equation of motion},
Model III. If, as we indeed find, this perturbation theory is singular, this
implies divergences in the full model II as well. Model III will
thus give a good description of the physics. We thus study it here
and come back to the full Model II in the next Section ~\ref{frgcomplete}.

\subsection{zero temperature perturbation theory to second order}

To avoid cumbersome expressions, and since in this whole section we only
discuss transverse degrees of freedom we skip the index $y$ for $u_y$.
We also discuss here for simplicity a $n=1$ component model for
the transverse displacement $u$ (which is appropriate for flux lines in $d=3$
and point vortices in $d=2$). Generalizations to $n>1$ will be briefly
mentionned in Section \ref{extensions}.

We thus study the dynamical equation for Model III \cite{giamarchi_moving_prl}.
Since we will be dealing with an anisotropic fixed point it is useful to
distinguish $c_x$ and $c_y$.
\begin{equation}   
\eta \partial_t u_{rt} = ( c_x \nabla_x^2 + c_y \nabla_y^2  - v \partial_x ) u_{rt}
+ F(r,u_{rt}) + \zeta (r,t) \label{dyneq}
\end{equation}
The bare value of the friction coefficient along $y$ is
denoted by $\eta_0$, and for simplicity we
denote by $v$ the quantity $\eta_0 v$ (which remains
uncorrected to all orders in this model). The correlator of the static transverse
pinning force is (\ref{staticequ}):
\begin{equation}
\overline{F(r,u) F(r',u')} =  \Delta(u-u') \delta^d(r-r')
\end{equation}
Averages over solutions of (\ref{dyneq}) can be
performed using the Martin-Siggia-Rose (MSR) action:
\begin{eqnarray} \label{action}
Z[h,\hat{h}] &=& \int Du D\hat{u} e^{- S_0[u,\hat{u}] - S_{int}[u,\hat{u}]
+ h u + i \hat{h} \hat{u}} \\
S_0[u,\hat{u}] & = & \int_{rt}  ~ i \hat{u}_{rt} (  \eta \partial_t + v \partial_x
- c_x \nabla_x^2 - c_y \nabla_y^2) u_{rt}
- \eta T  (i \hat{u}_{rt}) (i \hat{u}_{rt}) \\
S_{int} &=& - \frac{1}{2} \int_{rtt'}
(i \hat{u}_{rt}) (i \hat{u}_{rt'})
\Delta(u_{rt}-u_{rt'})
\end{eqnarray}
In this section we will restrict ourselves to $T=0$.
The corrections coming from
correlation functions $\langle \delta u_t \delta u_{t'} \rangle \sim T$
then vanish, which simplify the analysis. This can be used to show
that to all orders the temperature remains zero. The first
order corrections where computed in the Section \ref{timedependentpt}.
At $T=0$ there is no correction to the disorder term (to this order). There is a
non trivial correction to the kinetic term, which gives the
following correction to the friction coefficient $\eta$:
\begin{eqnarray}
\delta \eta = - \Delta''(0) \int_q \int_0^{+\infty} dt t R(q,t)
\end{eqnarray}
This leads to:
\begin{eqnarray}
&& \frac{\delta \eta}{\eta} = - \Delta''(0) \int \frac{d^{d-1}q_y}{(2 \pi)^{d-1}}
\frac{2 c_x}{(4 c_x c_y q_y^2 + v^2)^{3/2}}
\end{eqnarray}
(check)
This is {\it not} a divergent integral, except when $v=0$. To find
divergences in the perturbation theory at $T=0$ one has to go
to second order.

The second order corrections to the effective MSR action, and thus to
the coarse grained equation of motion, are computed in
Appendix~\ref{appendixc}. To second order a correction to
the full nonlinear disorder correlator appears and reads:
\begin{eqnarray} \label{corr2nd}
\delta \Delta(u) = \Delta''(u)
( \Delta(0)  - \Delta(u) )
\int_r  G(r) G(r)  - \Delta'(u)^2 \int_r  G(r) G(-r)
\end{eqnarray}
where $G(r)$ is the static response function
\begin{eqnarray}
G(r) = \int_0^\infty dt R(r,t) \qquad,\qquad G(q) = 
\frac{1}{c_x q_x^2 + c_y q_y^2 + i v q_x}
\end{eqnarray}
At zero velocity both terms in (\ref{corr2nd}) are infrared divergent
for $d \leq 4$, as is well known leading to the glassy effects in the
statics. The key novelty with respect to the problem at $v=0$ is
that due to the assymetry introduced by motion,
$G(r)$ is different from $G(-r)$. As a consequence the second term
in (\ref{corr2nd}) is now {\em convergent} for $v > 0$. Indeed
the integral
\begin{equation}  \label{irrel}
\int_r  G(r)  G(-r) = \int_q  G(q)^2
= \int \frac{d^{d-1}q_y}{(2 \pi)^{d-1}}
\frac{2 c_x }{ (4 c_x c_y q_y^2 + v^2)^{\frac{3}{2}}}
\end{equation}
is {\it convergent} in all dimensions for $v>0$.  On the other hand, one
divergence remains from the first term:
\begin{equation} \label{laclef}
\int_r  G(r)  G(r) = \int_q  G(q)  G(-q)
= \int \frac{d^{d-1}q_y}{(2 \pi)^{d-1}}
\frac{1}{2 c_y q_y^2 \sqrt{4 c_x c_y q_y^2 + v^2} } \sim
\int \frac{d^{d-1}q_y}{(2 \pi)^{d-1}}  \frac{1}{2 v c_y q_y^2}
\end{equation}
The integral (\ref{laclef}) is divergent for $d\leq 3$, even
for $v>0$.
Thus, contrarily to general belief 
\cite{larkin_largev,schmidt_hauger,koshelev_dynamics}
originating mainly form the study of the
first order perturbation in $1/v$, analysis to second order
confirms the surprising conclusion
that even at {\bf large} velocity infrared divergences occur in the
perturbation theory \cite{giamarchi_moving_prl,footnote_hauger}.
Such divergences indicates the instability of the
zero disorder fixed point and the breakdown of the large $v$ expansion.
They lead the system to a novel fixed point where the disorder is
plays a crucial role. The above divergence is the key to
the physics of the moving glass.

\subsection{Renormalization group study at zero temperature}

In order to handle these new divergences, and to find the new fixed
point which describes the large scale physics, we use a dynamical
functional renormalization group (DFRG) procedure on the effective action
using a Wilson scheme.
This allows to keep track of the full function $\Delta(u)$, which is
necessary since the full function is marginal at the upper critical
dimension.
This is equivalent to decompose the fields into fast and slow components 
$u \to u + \delta u$
and $\hat{u} \to \hat{u} + \delta \hat{u} $ and
to integrate the fast fields $\delta u$ and $\delta \hat{u}$
over a momentum shell.
This method is very similar to the method introduced in Ref.
\onlinecite{nattermann_stepanow_depinning,narayan_fisher_depinning}
for the $v=0$ case, though differs in details.

\subsubsection{derivation of the RG equations}

The first task is to perform a dimensional analysis of the MSR action
and to determine the appropriate rescaling transformation. Since we want
to describe both the $v=0$ and $v>0$ fixed points, we perform the
following redefinition of space, time and the fields
(keeping also an arbitrary $T$):
\begin{eqnarray}  \label{rescaling0}
&& y = y' e^{l} ~~~~ x = x' e^{\sigma l} ~~~ t= t' e^{z l} \\
&& \hat{u} = \hat{u}' e^{\alpha l} ~~~~ u = u' e^{\zeta l} \nonumber
\end{eqnarray}
We now impose that the action $S$ in (\ref{action}) is unchanged, which yields
redefinitions of the coefficients. Since $c_y=c'_y$ since this
quantity remains uncorrected to all orders, this fixes 
$\alpha=3 - d - \sigma - z - \zeta$. One finds the rescaling:
\begin{eqnarray}
&& \eta \to \eta'=\eta e^{(2-z)l}~~~~~ v \to v' = v e^{(2-\sigma)l}\\
&& c'_x = c_x e^{(2-2\sigma)l} \qquad T \to T'= T e^{(3-d-\sigma-2 \zeta)l} \\
&& \Delta \to \Delta' = \Delta e^{(5-d-\sigma-2 \zeta)l}
\end{eqnarray}
In the case $v=0$ the natural choice is $\sigma=1$, which
yields $\Delta' = \Delta e^{(4-d-2 \zeta)l}$. Power counting
at the gaussian fixed point ($z=2$, $\zeta=0$) yields
the upper critical dimension $d_{uc}=4$ below which disorder is
relevant (and a $d=4-\epsilon$ expansion can be performed
\cite{nattermann_stepanow_depinning,narayan_fisher_depinning}.

For $v > 0$ since $v$ is uncorrected to all orders a natural choice is
$\sigma = 2$. Power counting near the Gaussian fixed point
($z=2$, $\zeta=0$)
indicates that now the upper critical dimension is thus $d_{uc}=3$
( with $\Delta \to \Delta' = \Delta e^{(3-d-2 \zeta)l}$ ).
As a consequence disorder terms are
relevant for dimensions $d \leq 3$, whereas the temperature appears
to be formally irrelevant (see however Section \ref{rgfinitetemp} ).
The elasticity term along $x$ ($c_x$) now corresponds to an irrelevant
operator at the anisotropic fixed point. Note that the above
rescaling (\ref{rescaling0}) indicates that 
the proper dimensionless disorder parameter is $\Delta/v \Lambda^{d-3}$.
Here we are mostly interested in periodic systems for which,
as in the statics \cite{giamarchi_vortex_long}, one must set 
$\zeta=0$. For completeness we give however the equations 
for non periodic systems ($\zeta \ge 0$).

The standard RG method consist in two steps. Firstly, one integrates the
modes between $a<y<a e^l$ or equivalently $\Lambda_0 > q_y > \Lambda_0 e^{-l}$
with $\Lambda_0 \sim \pi/a$,
which yields corrections to the bare quantities. 
The cutoff procedure we choose 
here for convenience is to integrate over the following momentum shell:
\begin{equation}
\int_{sh} dq = \int_{\Lambda e^{-l}}^{\Lambda} \frac{d^{d-1} q_y}{(2 \pi)^{d-1}}
\int_{-\infty}^{+\infty} \frac{dq_x}{2 \pi}
\end{equation}
This results in 
the same theory but with a different cutoff and corrected parameter.
Secondly,one perform the length, time, and field rescaling 
(\ref{rescaling0}), as well as
the corresponding change of quantities (\ref{rescaling0}), so as to leave the
effective action invariant. The cutoff has thus been brought back
to its original value. Scale invariant theories will thus correspond
to fixed points of this combined transformation.

The RG equation for the disorder can be also established.
The shell contribution of the integral (\ref{laclef})
is asymptotically:

\begin{eqnarray}  \label{laclef2}
\int_r \delta G(r) \delta G(r)  \sim
\int_{sh} dq_y
\frac{1}{2 v c_y q_y^2} \sim \frac{1}{2 v c_y} A_{d-1}
\Lambda^{d-3}
\end{eqnarray}

where $A_d=\frac{S_{d}}{(2 \pi)^{d}}$ and 
in $d=3$ we need $A_{2} = 1/(2 \pi)$. Using
(\ref{corr2nd},\ref{irrel},\ref{laclef2}) we obtain
after rescaling we obtain the following FRG equation for the disorder
correlator:
\begin{eqnarray}  \label{frgsmallscale}
&& \frac{ d \Delta(u)}{dl}
= (3 - d - 2 \zeta) \Delta(u) + \zeta u \Delta'(u)
- \Delta'(u)^2 \frac{1}{2 \pi}
\frac{2 c_x(l)  \Lambda_0^2 }{ (4 c_x(l) c_y \Lambda_0^2 + v^2)^{\frac{3}{2}}}
+ \\
&& \frac{1}{4 \pi v c_y
\sqrt{1 + 4 c_x(l) c_y \Lambda_0^2/v^2} }
\Delta''(u) ( \Delta(0)  - \Delta(u) )
\end{eqnarray}
where $c_x(l)=c_x e^{-2l}$. In the large scale limit it reduces to
\cite{giamarchi_m2s97_vortex,giamarchi_book_young}:
\begin{equation} \label{frglargescale}
\frac{ d \Delta(u)}{dl}
= (3 - d - 2 \zeta) \Delta(u) + \zeta u \Delta'(u)
+  \frac{1}{4 \pi v c_y}
\Delta''(u) ( \Delta(0)  - \Delta(u) )
\end{equation}

The equation $\ref{frgsmallscale}$ allows in principle to
examine the intermediate scales crossover 
when $v$ is not very large. Indeed there is a characteristic
crossover length scale:

\begin{eqnarray}
L_{cross} = 2 \sqrt{c_x c_y}/(\eta_0 v)
\end{eqnarray}

such that 
(\ref{frglargescale}) becomes valid for $e^l \gg L_{cross}/a$.
Note that setting $v=0$ in the above equation
(\ref{frgsmallscale})
leads back the FRG equation for the usual
manifold depinning
\cite{nattermann_stepanow_depinning,narayan_fisher_depinning}
(up to numerical factors originating from
choice of short distance cutoff, and different choices for
rescalings).

We also give the RG equation for the friction coefficient
for the case of the periodic problem ($\zeta$ =0).
Before rescaling it reads:

\begin{eqnarray}
\frac{d \eta}{\eta dl} = - \Delta''(0) A_{d-1}
\frac{ 2 c_x \Lambda^{d-1} }{(4 c_x c_y \Lambda^2 + v^2)^{3/2}}
\end{eqnarray}

Taking into account that $\Lambda = \Lambda_0 e^{-l}$ and
$c_x(l)=c_x e^{-2l}$, and $\Delta_l''(0) = \Delta''(0) e^{(3-d)l}$
one find the RG equation, after rescaling:

\begin{eqnarray}  \label{etaren}
\frac{d \eta}{\eta dl} = 2 - z - \Delta_l''(0) A_{d-1}
\frac{ 2 c_x(l) \Lambda_0^{d-1} }{(4 c_x(l) c_y \Lambda_0^2 + v^2)^{3/2}}
\end{eqnarray}

thus except for $v=0$, $\eta$ is corrected only by a finite
amount (as long as $\Delta_l''(0)$ is finite, see below).

\subsubsection{study of the FRG equation}

We now study the FRG equation (\ref{frglargescale})
for the periodic problem ($\zeta=0$). Thus we 
impose $\Delta(u)$ to be periodic of period
$1$ and study the interval $[0,1]$. Let us
look for a perturbative fixed point in $d=3-\epsilon$.
Absorbing the factor $\frac{1}{4 \pi v c_y \epsilon}$
in $\Delta(u)$ and redefining temporarily $\epsilon l \to l$.
the FRG equation reads:
\begin{equation}  \label{frg1}
\frac{ d \Delta(u)}{dl}
= \Delta(u) +  \Delta''(u) ( \Delta(0)  - \Delta(u) )
\end{equation}
No continuous solutions such that $d\Delta(u)/dl=0$
exist \cite{footnote_solutions}. This is due to the fact that
the average value of $\Delta(u)$ on the
interval [0,1] must increase unboundedly. Indeed integrating inside
the interval one finds:

\begin{equation}
\frac{d}{dl} \int \Delta(u)
= \int \Delta(u) +  \int \Delta'(u)^2 + [\Delta'(u) ( \Delta(0)  - \Delta(u) )]_{0^+}^{1^-}
= \int \Delta(u) +  \int \Delta'(u)^2
\end{equation}

It is thus natural to define $\overline{\Delta}(u)=\Delta(0) - \Delta(u)$
which satisfies:
\begin{equation}
\frac{ \overline{\Delta}(u)}{dl} = \overline{\Delta}(u) (1 + \overline{\Delta}(u))
\end{equation}
Note that physically one expects $\overline{\Delta}(u) \ge 0$.
This equation has a fixed point $\overline{\Delta}^*(u)=u(1-u)/2$.
It is shown in Appendix \ref{stability} that this fixed
point is stable (locally attractive). Since the flow
equation for $\Delta(0)(l)$ is simply:
\begin{equation}
\frac{ d \Delta(0)(l)}{dl}
= \Delta(0)(l)
\end{equation}
the value of $\Delta(0)(l)$ thus grows unboundedly
as $\Delta(0)(l) = \Delta(0) e^{\epsilon l}$ (restoring the $\epsilon$ factor).

\begin{figure}
\centerline{\epsfig{file=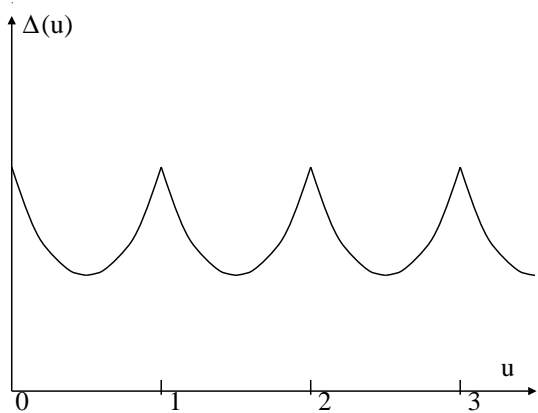,angle=0,width=7cm}}

\caption{Solution of the FRG equation. Note the non analyticity
at all integers
\label{FRGsolution} }

\end{figure}

Thus the full fixed point solution in a $d=3-\epsilon$ expansion 
is \cite{giamarchi_m2s97_vortex,giamarchi_book_young}:

\begin{equation}
\Delta_l(u) = \Delta^*(u) - \Delta^*(0) + C e^{\epsilon l} ~~~~  \Delta^*(u)''=1
\end{equation}

where $C$ is an arbitrary constant and:

\begin{equation}  \label{fixedpoint}
\Delta^{*}(u) = C^* + (\epsilon 4 \pi v c_y) (\frac{1}{2} u^2 - \frac{1}{2} u) \qquad (0 \le u \le 1)
\end{equation}
and the solution repeats periodically as shown in Fig. \ref{FRGsolution}
We have restored the factor $1/(4 \pi v c_y)$ and
$\epsilon=3-d$.

In $K$-space the fixed point solution can be written
$\Delta_K=1/K^2$ for $K \ne 0$ ($K=2 \pi k$ with $k$ integers)
and $\Delta_{K=0}(l)=\Delta_l(u=0)+(1/12)=\Delta_0(u=0) e^{l}+(1/12)$.

Thus there is an ever growing average to the correlator.
Remarkably, {\it does not spoil} the above fixed point,
since one can always separates $\Delta(0)$ and  $\Delta(u) - \Delta(0)$ in
the starting MSG action. In perturbation
theory one sees that $\Delta(0)$ has no feedback at all
into the non linear part. It simply means that there is an
unrenormalized random force which simply
adds to a non linear pinning force, which is described
by $\Delta^*(u)$.

Note that this solution has cusp non analyticity at all integer
$u$. This was to be expected since one has exactly:
\begin{equation}  \label{grow}
\frac{ d \Delta''(0)}{dl}
= \Delta''(0)  -  \Delta''(0)^2
\end{equation}
with $C=1$ in the asymptotic regime.
At the start $\Delta''(0)$ is negative (since 
$\Delta(u)$ is an analytic function with a maximum at
$u=0$). One easily sees that $\Delta_l''(0)$
becomes infinite at a finite length scale (interpreted
as the Larkin length, see next section), and the
function becomes non analytic (also $\Delta''(0+)$
becomes positive).

Once the solution is known in $d=3 -\epsilon$ it is
straightforward to obtain it in the physically
relevant dimension $d=3$. In $d=3$:

\begin{equation}
\frac{ d \Delta(u)}{dl}
= \Delta''(u) ( \Delta(0)  - \Delta(u) )
\end{equation}

Defining $\Delta_l(u)=\frac{1}{l} \hat{\Delta_l}(u)$
and introducing $l'=\ln l$ one finds that:
\begin{equation}
l' \frac{ d \hat{\Delta}(u)}{dl'}
= \hat{\Delta}(u) +  \hat{\Delta}''(u) ( \hat{\Delta}(0)  - \hat{\Delta}(u) )
\end{equation}
which is identical to (\ref{frg1}). Thus the
physics will be controlled by the
slow decrease to zero of
disorder at large scale
with the following stable fixed point behaviour:

\begin{equation}
\Delta_l(u)  \sim \Delta(0) + \frac{4 \pi v c_y}{2 l}
 (u^2 - u)
\end{equation}

( note that the random force term does not grow 
by rescaling in $d=3$).

Finally, note that the equation (\ref{frglargescale}) presents several
differences and some remarkable similarities with the
one describing the statics FRG and the dynamical FRG
for $v=0$ in a $d=4 -\epsilon$ expansion.
The statics FRG equation is:
\begin{equation}
\frac{ d R(u)}{dl}
= R(u) +  \frac{1}{2} R''(u)^2 - R''(u) R''(0)
\end{equation}
where $R(u)$ is the correlator of the random
potential and the dynamic FRG equation for ($v=0$)
is
\begin{equation}   \label{vnul}
\frac{ d \Delta(u)}{dl}
= \Delta(u) +  \Delta''(u) ( \Delta(0)  - \Delta(u) ) - \Delta'(u)^2
\end{equation}
Since one has $\Delta(u)=- R''(u)$ it yields the 
solution periodic in [0,1] \cite{giamarchi_vortex_long}:

\begin{equation}
\Delta^*(u) = \frac{1}{36} (1 - 6 u + 6 u^2)
\end{equation}

Remarkably both the solution for $v=0$ and for $v >0$
are non analytic at integer $u$, though the detailed
form of these solutions is different. Since this non analyticity
is related to glassiness and pinning 
one can expect a certain continuity of properties
between the moving and non moving case.

The main difference 
however is that in (\ref{vnul}) $\Delta(0)=\Delta(0^+)$ 
starts growing at initial RG stages as for $v>0$
but is stopped to its fixed point
value $\frac{1}{36}$ beyond the scale at 
which a nonanalycity develops (Larkin length).
This effect is due to the term $- \Delta'(0+)^2$
and physically means that in the case $v=0$
displacements grow much more slowly at larger scales.
The system remembers that it is a potential system
and thus $\int \Delta(u)$ remains zero if it is 
zero at the start (at least formally, see however
\cite{narayan_fisher_depinning}). By contrast for the moving system 
one has asymptotically
$\Delta_l(0) \sim \Delta_{\infty} e^{\epsilon l}$.
As will be discussed later this corresponds to
the generation of a random force (which cannot exist
in the statics). 

Note that for $v$ not very large one can see
on equation \ref{frgsmallscale}) that there will be
a long crossover during which the term $- \Delta'(0+)^2$
will act (random manifold regime, see below).
Thus the actual value of $\Delta_{\infty}$ should be
decreased compared to the value naively suggested by
perturbation theory, an effect studied in the next 
Section.

\subsubsection{Physical results at $T=0$}

\label{physicalt0}

We now extracts some of the physics of the moving glass
from the FRG analysis.
>From the equation for the second derivative of the
force correlator:

\begin{equation}  \label{grow}
\frac{ d \Delta''(0)}{dl}
= (3-d) \Delta''(0)  -  C(l) \Delta''(0)^2
\end{equation}
with  $1/C(l)=4 \pi v c_y
\sqrt{1 + 4 c_x e^{-2 l} c_y \Lambda_0^2/v^2}$
it is possible to extract the length scale
$R_c^y$ at which $\Delta''(0)$ becomes infinite.
We first estimate it in the large velocity regime
$L_{cross} \ll a$ where one can set $C(l)=1/(4 \pi v c_y)$.
In $d=3$ one has
\begin{equation}  \label{grow2}
\Delta_l''(0) = - \frac{\Delta_2}{1 - 
 \frac{\Delta_2 l}{4 \pi v c_y} }
\end{equation}

where $\Delta_l''(0) = - \Delta_2$ is the bare value.
Thus:

\begin{equation}
R_c^y = a e^{\frac{4 \pi \eta_0 v c_y}{\Delta_2}}
\end{equation}

\begin{figure}
\centerline{\epsfig{file=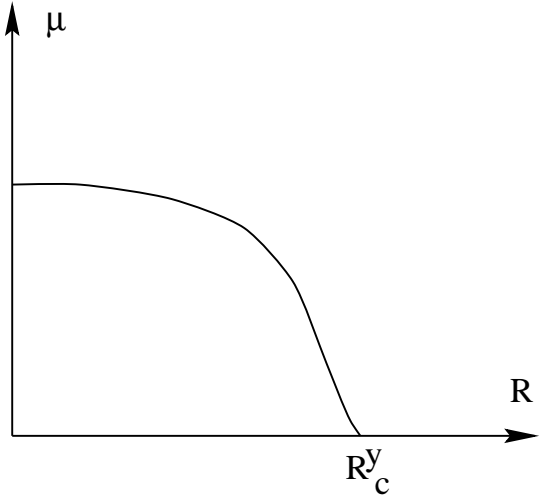,angle=-90,width=7cm}}

\caption{Scale dependent mobility. It vanishes beyond the dynamical
Larkin length it vanishes (at $T=0$).
\label{muren} }

\end{figure}

This length scale, introduced in
\cite{giamarchi_moving_prl,giamarchi_book_young},
is the analogous of the Larkin length
for the statics. Indeed $R_c^y$ coincide with
the scale at which the scale dependent mobility
$\mu(L)$ vanishes as depicted in Fig. \ref{muren}. This can be
seen from (\ref{etaren}). The divergence of
$\Delta_l''(0)$ drives at $L=e^l=R_c^y$ drives 
$\mu(L)=1/\eta(L)$ to zero for all larger scales.
Beyond that scale pinning starts to play a role. 

In $d<3$ one has:
\begin{equation}  \label{grow2}
\Delta_l''(0) = - \frac{\Delta_2 e^{\epsilon l}}{1 - 
 \frac{\Delta_2 (e^{\epsilon l}-1)}{4 \pi v c_y \epsilon} }
\end{equation}

Thus the dynamical Larkin length is:

\begin{equation}   \label{larkin2}
R_c^y =  (1 + \frac{4 \pi v c_y (3-d)}{\Delta_2})^{1/(3-d)}
\end{equation}

Note that it is the second derivative $\Delta_2$ of the
force correlator which appears in the Larkin
length. For a realistic disorder with a 
correlation length $r_f$ one has $\Delta_2=\Delta(0)/r^2_f$.
Using this relation, one checks that 
 (\ref{larkin2}) is the one obtained in 
\cite{giamarchi_moving_prl} by estimating the
length scale at which $u_{dis} \sim r_f$.

One can also determine the dynamical Larkin length 
when the velocity is not very large. Restoring the
proper dependence of $C(l)$ in $l$ in (\ref{grow})
one gets that $1/\Delta_2 = \int_0^{l_c} e^{\epsilon l} C(l)$
with $l_c=\ln (R^y_c/a)$.
This yields after some algebra in $d=3$:

\begin{equation}   \label{larkin3}
R_c^y = e^{\frac{4 \pi \eta_0 v c_y}{\Delta_2}}
\frac{1}{2} (
1+e^{\frac{-8 \pi \eta_0 v c_y}{\Delta_2}}
+(1-e^{\frac{-8 \pi \eta_0 v c_y}{\Delta_2}})
\sqrt{1 + \frac{4 c_x c_y \Lambda_0^2}{(\eta_0 v)^2}})
\end{equation}

and in $d=2$ one finds:

\begin{equation}   \label{larkin3}
(R_c^y/a)^2  = 1 + (\frac{4 \pi \eta_0 v c_y}{\Delta_2})^2 +
\frac{8 \pi \eta_0 v c_y}{\Delta_2}
\sqrt{1 + \frac{4 c_x c_y \Lambda_0^2}{(\eta_0 v)^2}}
\end{equation}

where we recall $\lambda_0 \sim \pi/a$.
These formulae interpolate smoothly between the
Bragg glass and Moving glass results.

Finally, note that since the above equations \ref{grow2}
 are exact
(within the $3-\epsilon$ FRG approach) the calculations 
of the Larkin lengths are independent of whether there is
an intermediate random manifold regime, i.e it holds both for
$a>r_f$ and $a<r_f$. Non universal irrelevant operators will
of course change the numerical values of the prefactors
but the above expressions should be correct when all
the Larkin lengths are large.

One of the remarkable properties of the moving state 
is the existence of transverse pinning \cite{giamarchi_moving_prl}.
This is demonstrated from the FRG fixed point, due to the
{\it non analyticity} of the fixed point (\ref{fixedpoint}).
Adding an external force $f_y$ along $y$ (i.e in the
l.h.s of (\ref{dyneq})) generates a velocity $v_y$.
The naive perturbation theory result , formula 
(\ref{corrcont}), for $\delta f_y(v_y)$ 
(the correction to the applied force at fixed $v_y$) reads:

\begin{eqnarray}
\delta f(v_y) = \int_t R_{r=0,t} \Delta'(v_y t)
\end{eqnarray}

In the limit of vanishingly small $v_y$ one gets
a non zero limit $\delta f_y(0^{+}) = - F_c^y$, i.e a transverse critical force
only if the function is nonanalytic with
$\Delta'(0^+) <0$. The critical force is thus given by
summing up the contributions of all the successive shells

\begin{eqnarray}
F_c \approx - \int_{\ln R^y_c}^{+\infty} dl
\Delta_l'(0^+) A_{d-1}
\frac{\Lambda_0^{d-1} e^{-(d-1)l}}{\sqrt{4 c_x(l) c_y \Lambda_0^2 + v^2}}
e^{-(3-d)l}
\end{eqnarray}

where quite logically only scales beyond the Larkin length
give a non vanishing contribution. 
Using the asymptotic value for $\Delta'(0^+)^* = \epsilon 4 \pi \eta_0 v c_y$
one finds:

\begin{eqnarray} \label{crit10}
F_c \approx C c_y a (R_c^y)^{-2} 
\end{eqnarray}

where $C=A_{d-1} (3-d) \pi \Lambda_0^{d-1}/a$.

In $d=3$ one finds:

\begin{eqnarray}   \label{crit11}
F_c \approx C' \frac{ c_y a }{(R_c^y)^2 \ln (R_c^y/a)}
\end{eqnarray}
with $C'=A_{d-1} \pi \Lambda_0^{d-1}/a$.

Note that the prefactors are not universal 
(and in $d=3$ affected by the highr orders). In the above formulae
we have assumed a direct passage from the Larkin scale regime
to the asymptotic periodic regime, thus $r_f \sim a$.
If $r_f \ll a$ an intermediate random manifold regime
will be first reached were the typical value for 
$\Delta_l'(0^+)$ will be rather $a/r_f$. This will yield
to the replacement of $a$ by $r_f$ in the numerators of
(\ref{crit10},\ref{crit11}).

Remarkably this coincide with the estimate 
given in \cite{giamarchi_moving_prl} obtained
by balancing forces. Note that this result can be obtained here
{\it without any reference} to a Larkin length along
the $x$ direction. This illustrate that the physics 
of the moving glass depends only on the periodicity 
along the $y$ direction. 

Note that a more
general expression of $\delta f_y(v_y)(l)$ can be given
and reads:

\begin{eqnarray}
\delta f_y(v_y)(l) = \int_0^l dl \int_{t>0} \frac{1}{2}
(\frac{\mu}{c_x \pi t})^{1/2} e^{-\mu t ( c_y e^{-2 l} \Lambda_0^2
+ v^2/(4 c_x) ) } \Lambda_0^{d-1} e^{- (d-1) l} \Delta'_l(v_y t)
e^{- \epsilon l}
\end{eqnarray}

It is interesting also to study the behaviour near the 
transverse depinning threshold. It is easy to see that
because of the absence of IR divergence in the
integral for $\eta$ the exponent at the threshold
remains uncorrected, i.e $v_y \sim (f_y - F^y_c)^{\theta}$
with $\theta=1$ (to first order in $\epsilon$).
The slope can easily be estimated from 
$\Delta''_l(0)$ the above results and becomes
large at small velocities.

We now study the displacement correlations.
The growth of the average of $\int_0^1 \Delta(u)$ 
implies that there is a static random force generated. 
However, unlike in the $v=0$ case, the critical force
does not kill the random force in the FRG equation.
In fact the moving glass will be dominated by the
{\it competition} between the random force and 
the critical force. 
Although the existence of such a random force has 
no effect on pinning it will affect strongly the 
positional correlation functions. In particular the 
relative displacements correlation function
\begin{equation}
B(r) = \overline{\langle [u(r) - u(0)]^2 \rangle}
\end{equation}
becomes, in the presence of the random force
at large scale:
\begin{equation} \label{larkinmove}
B_{rf}(x,y) \sim \Delta_{\text{ren}}(0) \int \frac{dq_x
d^{d-1} q_y}{(2\pi)^d} \frac{ (1-\cos(q_x x+q_y y))}{(\eta v q_x)^2 +
c^2(q_x^2+q_y^2)^2} \sim \Delta_{\text{ren}}(0)
\frac{y^{3-d}}{c \eta v}H(\frac{c x}{\eta v y^2})
\end{equation}
where $H(0)=\text{cst}$ and $H(z)\sim z^{(3-d)/2}$ at large $z$.
Thus $x$ scales as $y^2$ and the displacements are very anisotropic.
This behaviour results in an algebraic decay of 
translational order in $d=3$ ($B(x,y) \sim \log|y|$)
and exponential decay in $d=2$.

It is thus important to estimate 
$\Delta_{\text{ren}}(0)/v$ which sets the growth of displacements.
At large velocity one finds that 
$\Delta_{\text{ren}}(0)/v = \Delta(0)/v$ the bare disorder,
which at large $v$ is very small. When decreasing $v$ the
displacements due to the random force
at a given scale increase but eventually saturate and
decrease again. This is because in the intermediate
velocity regime the random force will
be reduced strongly, from its original value.

An estimate, from \ref{frgsmallscale} is that:

\begin{eqnarray}
\Delta_{\text{ren}}(0) = 
\Delta(0) - \int_0^{+\infty} dl e^{- \epsilon l} 
\Delta_l'(0)^2 
\frac{1}{2 \pi}
\frac{2 c_x e^{-2 l} \Lambda_0^2 }{ (4 c_x e^{-2 l} c_y \Lambda_0^2 + v^2)^{\frac{3}{2}}}
\end{eqnarray}

and one can simply set $\epsilon=0$ to get the result in $d=3$.

We have given here mainly the results at large $v$. These
calculations can extended to any $v$ by studying the crossover from 
the Bragg glass to the Moving glass, as well as the intermediate 
random manifold regimes. This crossover is discussed
in the Section  \ref{sec:physics}.

\subsection{RG study at finite temperature $T>0$}

\label{rgfinitetemp}

We now extend the analysis to finite temperature. In principle
the FRG equations can also be written for any temperature.
We will study both the case $v=0$ and $v>0$ (since no such derivation
exist in the litterature).

In the $v=0$ case the temperature is formally irrelevant. In fact it
is dangerously so (see below) as it will cutoff the properties of
the fixed point (the non analyticity) and thus modify some observables
leading to barrier determination.

Here as we will see temperature for $v>0$ is even more so
- very - dangerously irrelevant by power counting. The dimensional 
rescaling (\ref{rescaling0}) yields
$T \to T'= T e^{(1-d-2 \zeta)l}$ and thus that $T$ is
irrelevant. This turns out to be incorrect: if one add a small
$T>0$ onto the $T=0$ moving glass fixed point, it will
flow upwards very fast (while if $T=0$ to start with it remains
so).  This is at least what the FRG seems to indicate.

\subsubsection{derivation of the RG equations}

Letting a $T>0$ leads to several important modifications
in the perturbation theory. The general idea is that
the disorder will be changed everywhere roughly as
$\Delta_K e^{- \frac{1}{2} K.B_{0,t}.K}$ (with $t \to \infty$).
Of course this has to be checked carefully which is done in detail
in the appendices and we give here only the main results.
Since near the upper critical dimension ($d_u=4$ for $v=0$ and $d_u=3$
for $v>0$) thermal displacements are bounded:

\begin{eqnarray}
&& \lim_{t \to \infty} B_{0,t} = B_{\infty}
= 2 T \int_q \frac{1}{c q^2} = \frac{2 T}{c} A_d \Lambda^{d-2} ~~~~
(v=0) \\
&& = 2 T \int_q \frac{1}{c_x q_x^2 + c_y q_y^2 + i v q_x }~~~~ (v>0)
\end{eqnarray}

We will denote rescale $T$ by a non universal quantity and
define $\tilde{T}$ the ``dimensionless temperature''"

\begin{eqnarray}
\frac{1}{2} B_{\infty} = \tilde{T}
\end{eqnarray}

One will need to replace everywhere $\Delta_K$ by the smoothed
disorder:

\begin{eqnarray}
\tilde{\Delta}_K = \Delta_K e^{- \tilde{T} K^2}  \qquad
\tilde{\Delta}(u) = \sum_K \Delta_K e^{i K u} e^{- \tilde{T} K^2}
\end{eqnarray}

The divergent part of the correction to the friction coefficient
is now (\ref{resultapp}):
\begin{eqnarray}
\delta \eta = \int_{t} t R_{r=0,t} K^2 \tilde{\Delta}_K
\end{eqnarray}
thus the same as before, except one must use the
smoother disorder.

Let us now compute the renormalization of the
temperature by disorder (for a related calculation
see also \cite{ledou_wiese_ranflow}). 
As was mentionned earlier there is a new and non trivial divergence
in the correction to the temperature. Using (\ref{corrfrict},\ref{corrtemp})
one finds:

\begin{eqnarray}  \label{corrtemp2}
\eta \delta T = \sum_K \int_{t>0}
\Delta_K ( e^{- \frac{1}{2} K.B_{0,t}.K}
- e^{- \frac{1}{2} K.B_\infty.K}
- K^2 t T R_{r=0,t} e^{- \frac{1}{2} K.B_{0,t}.K} )
\end{eqnarray}

When $v=0$ this integral can be simplified using the FDT relation
$ 2 T R_{r=0,t} = \theta(t) d/dt B_{0,t}$ 
which gives $\delta T = 0$. This yields the RG equation, after rescaling:

\begin{eqnarray}
\frac{d T}{dl} = (2-d-2 \zeta) T  ~~~~~~(v=0)
\end{eqnarray}

For $v>0$ the second part does not diverge anymore,
and the divergence of the first one can be extracted
as follows:

\begin{eqnarray}
\eta \delta T \sim \sum_K \int_{t>0}
\tilde{\Delta}_K ( e^{- \frac{1}{2} K.(B_{0,t}-B_\infty).K} -1 )
\sim  \frac{1}{2} \sum_K \int_0^{+\infty} dt
\tilde{\Delta}_K K^2 (B_\infty-B_{0,t})
\end{eqnarray}

with:

\begin{eqnarray}
B_\infty - B_{0,t} = \int_q \frac{2 T}{c_x q_x^2 + c_y q_y^2}
e^{- (c_x q_x^2 + c_y q_y^2) \mu |t|  + i v q_x \mu t}
\end{eqnarray}

Thus:

\begin{eqnarray}
\delta T \sim \sum_K K^2 \tilde{\Delta}_K \int_q
\frac{T}{(c_x q_x^2 + c_y q_y^2)(c_x q_x^2 + c_y q_y^2 + i v q_x)}
= T \sum_K K^2 \tilde{\Delta}_K
\int \frac{d^{d-1} q_y}{(2 \pi)^{d-1}}
\frac{1}{2 c_y q_y^2 \sqrt{v^2 + 4 c_x c_y q_y^2}}
\end{eqnarray}

Extracting the IR divergent part this yields, after
rescaling the following RG equation
near $d=3$:

\begin{eqnarray}
\frac{d T}{T dl} = 2-d-2 \zeta
- \frac{1}{4 \pi c_y v} \tilde{\Delta}''(0)  ~~~~~ (v>0)
\end{eqnarray}

Since we will also discuss the crossover from
$v=0$ to $v>0$ we also give the following more precise
estimate of (\ref{corrtemp2})
(obtained from the large time behaviour):

\begin{eqnarray}  \label{corrtemp3}
\frac{\delta T}{T} \sim \sum_K  K^2
\tilde{\Delta}_K \int_q  \frac{v^2}{2 c_y q_y^2 (4 c_x c_y q_y^2 + v^2)^{3/2}}
\end{eqnarray}

Note that it vanishes, as it should when $v \to 0$.
This yields the following RG equation for the temperature:

\begin{eqnarray}  \label{rgtemp2}
\frac{d T}{T dl} = 2-d-2 \zeta 
- \frac{\tilde{\Delta}''(0)}{(4 \pi c_y v)(1 + \frac{4 c_x c_y \Lambda_0^2 e^{-2 l}}{v^2})^{3/2}} 
\end{eqnarray}

We now look at the corrections to disorder. The calculation is
done in the Appendix \ref{appendixc}. The divergent
contributions to the disorder correlator are, adding first and second order
in perturbation:

\begin{eqnarray}
&& \Delta^R_{P} = e^{- \tilde{T} P^2} [  \Delta_{P}
+ e^{\tilde{T} P^2}
\sum_{K,K'=P-K} \Delta_K e^{- \tilde{T} K^2 } \Delta_{K'} e^{- \tilde{T} K'^2 }
(K^2 \int_r G(r) G(r) + K' K \int_r G(r) G(-r) ) \\
&& - P^2 \Delta_P \sum_{K'}
\Delta_{K'} e^{- \tilde{T} K'^2} \int_r G(r) G(r) ]
\end{eqnarray}

The key point is that using $K.K'=(P^2 - K^2 -K'^2)/2$
all exponential factors rearrange and at the end everything can be
written only using the smoothed function 
$\tilde{\Delta}_K$. Using this smoothed function can write:

\begin{eqnarray} \label{corr2nd2}
\delta \tilde{\Delta}(u) = \tilde{\Delta}''(u)
( \tilde{\Delta}(0)  - \tilde{\Delta}(u) )
\int_r  G(r) G(r)  - \tilde{\Delta}'(u)^2 \int_r  G(r) G(-r)
\end{eqnarray}

This yields the RG equation for the disorder:

\begin{eqnarray}
\frac{d \tilde{\Delta}(u)}{dl} = \tilde{T} \tilde{\Delta}''(u)
+ (3 - d - 2 \zeta) \tilde{\Delta}(u) + \zeta u \tilde{\Delta}'(u)
+ f_1(l)\tilde{ \Delta}''(u) ( \tilde{\Delta}(0)  - \tilde{\Delta}(u) )
- f_2(l)    \tilde{\Delta}'(u)^2
\end{eqnarray}

where $f_1$ and $f_2$ are the same coefficients
as in \ref{frgsmallscale}. We have used 
that the smoothed function 
$\tilde{\Delta}(u)$ itself has an explicit 
cutoff dependence. Note that this equation is correct
for any $T$ and to second order in the disorder.
Note that it can be obtained also by a small $T$
expansion, assuming $T$ small (and expanding the
first order correction to $\Delta$ in $T$).
This equation adds to the renormalization
equation for the friction coefficient:

\begin{eqnarray}
\frac{d \eta}{\eta dl} = 2 - z + \tilde{\Delta}''(0) A_{d-1}
\frac{ 2 c_x(l) \Lambda_0^{d-1} }{(4 c_x(l) c_y \Lambda_0^2 + v^2)^{3/2}}
\end{eqnarray}

with here $z=2$. 

\subsubsection{analysis of FRG equations at $T>0$}

Let us now analyze the FRG equations at $T>0$ 
for the periodic case in an $\epsilon=3-d$ expansion
and in $d=3$. We write $\Delta$ instead of $\tilde{\Delta}$ 
and $T$ instead of $\tilde{T}$ everywhere for convenience.
One has:

\begin{eqnarray}
&& \frac{ d \Delta(u)}{dl}
= \epsilon \Delta(u) + T  \Delta''(u) +
  \Delta''(u) ( \Delta(0)  - \Delta(u) )  \\
&& \frac{d T}{T dl} = - 1 + \epsilon 
- \Delta''(0)
\end{eqnarray}

We have absorbed the factor $\frac{1}{4 \pi v c_y}$
in $\Delta(u)$. 
Let us first search for a fixed point. We thus assume 
that $dT/dl=0$ with $T=T^*$ which implies that
$\Delta''(0) = -1 + \epsilon$.
We search for a fixed point for
$\Delta(u) - \Delta(0)$ (as we did 
for the $T=0$ fixed point).
Let us set $\Delta(u)= \Delta(0) - T^* g(u)$
with $g(u) >0$ and periodic. One gets:
\begin{eqnarray}
g''= - \frac{\epsilon}{T^*} + \frac{\epsilon- \Delta''(0)}{T^*} \frac{1}{1+g}
= - V'(g)
\end{eqnarray}
This is the motion in the potential $V(g)$. It always has
a periodic solution starting from $g=0$. Thus the solution is:
\begin{eqnarray}
u = \int_0^g \frac{dg}{\sqrt{-2 V(g)}} ~~~~ V(g)=\frac{\epsilon}{T^*} g -
\frac{\epsilon- \Delta''(0)}{T^*} Log(1+g)
\end{eqnarray}

This yield a condition since we have fixed the period to be
$u=1$:

\begin{eqnarray}
\frac{1}{2} = \int_0^{g_{max}} \frac{dg}{\sqrt{-2 V(g)}} \qquad V(g_{max})=0
\end{eqnarray}

The other condition being:

\begin{eqnarray}
\Delta''(0) = -1+\epsilon
\end{eqnarray}

Both conditions determine $T^*$ and $\Delta''(0)$. 
From this we get the fixed point temperature:

\begin{eqnarray}
T^* = \frac{\epsilon^2}{4 (\int_0^{y_{max}}
 \frac{dy}{\sqrt{2 (\ln (1+\frac{y}{\epsilon}) - y) }} )^2 }
\sim \frac{\epsilon^2}{8 \ln (1/\epsilon)}
\end{eqnarray}

Thus we find that there is a finite temperature fixed point.
This is the moving glass $T>0$ fixed point.
Though we have not investigated in details the stability of this
fixed point it is likely to be attractive. Indeed one sees
clearly on the equation of renormalization of temperature
that at high T one expects $\Delta''(0)$ to be small (since
$\Delta$ is smoothed by temperature), while at low $T$ 
$\Delta''(0)$ grows very fast thereby $T$ increases.
Note that similar finite $T$ fixed points were found for
other non potential systems \cite{ledou_wiese_ranflow}.

Note that
one has also:

\begin{eqnarray}
\frac{d \Delta(0)}{dl} = \epsilon \Delta(0) - T^* (1-\epsilon)
\end{eqnarray}

and thus the random force is still generated 
(though it grows slower).

Form this we get the temperature:

\begin{eqnarray}
T^* = \frac{\epsilon^2}{4 (\int_0^{y_{max}}
 \frac{dy}{\sqrt{2 (\ln (1+\frac{y}{\epsilon}) - y) }} )^2 }
\sim \frac{\epsilon^2}{8 \ln (1/\epsilon)}
\end{eqnarray}

Thus we find that there is a finite temperature fixed point.
Though we have not investigated in details the stability of this
fixed point it is likely to be attractive. Indeed one sees
clearly on the equation of renormalization of temperature
that at high T one expects $\Delta''(0)$ to be small (since
$\Delta$ is smoothed by temperature), while at low $T$ 
$\Delta''(0)$ grows very fast thereby $T$ increases.

Note that there will be an interesting crossover 
at low $T$ where $\Delta''(0)$ will first start to increase violently
before it finally decreases again to its fixed point value.
This will be discussed in following section.

The case of $d=3$ can be studied similarly. As in the case
$T=0$ one looks again
for a solution decaying as $1/l$ (as we noticed before
$\epsilon$ and $1/l$ play the same role).

Here the solution in $d=3$ is as follows.
One has:

\begin{eqnarray}
\tilde{\Delta}(0) - \tilde{\Delta}(u) \sim \frac{1}{l} g(u)
\end{eqnarray}

This ansatz gives the same equation as in $d=3-\epsilon$
with $\epsilon \to 1/l$. Thus $g=g^*(u, \epsilon=1/l)$ is still the
asymptotic solution. Also:

\begin{eqnarray}
T(l) \sim \frac{1}{8 l^2 \ln (l)}
\end{eqnarray}

Thus at large scale temperature decays back to $T=0$. 
Indeed the fixed point function is very similar to the $T=0$ fixed
point function except in small layer around integer $u$.
Near the origin the term $T \Delta''(0)$ is of same 
order as $\epsilon \Delta(0)$. Thus the main effect
of temperature is to round the non analyticity.

\subsubsection{Physical results at finite temperature $T>0$}

We now discuss the behaviour of the
mobility $\mu_R = 1/\eta_R$ and of the
I-V characteristics. 

One can compute the mobility from the RG 
equation by integrating the $l$ dependent
solution over all scales:

\begin{eqnarray}
\ln(\frac{\mu(l)}{\mu_0}) = \int_0^l \Delta_l''(0) A_{d-1}
\frac{ 2 c_x e^{-2 l} \Lambda_0^{d-1} }{(4 c_x e^{-2 l} c_y \Lambda_0^2 
+ v^2)^{3/2}}
\end{eqnarray}

Since asymptotically there are only finite corrections to $\eta$
as soon as $v>0$ this integral converges. Thus there is a
non zero asymptotic mobility $\mu_R$ in the $T>0$ moving glass
(by contrast with what one would have in the static
Bragg glass at $T>0$). 

However the renormalized mobility $\mu_R$ will be very small
(for experimental purposes) in several cases (i)
at low temperature (ii) for velocities not very
large $v \le v_c^*$.

\begin{figure}
\centerline{\epsfig{file=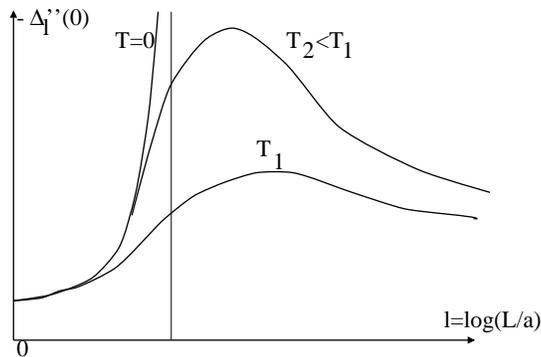,angle=0,width=7cm}}

\caption{Behaviour of the second derivative $- \Delta''(0)$ of the disorder
correlator as a function of the scale
around the $T=0$ dynamical Larkin length
$R_c^y$. At $T=0$ there is a divergence at $R_c^y$ which is rounded
at $T>0$. However $- \Delta''(0)$ still passes through very large
values before eventually decaying slowly towards
its fixed point value. This results in high barriers at low temperature
as discussed in the text.
\label{2deriv} }

\end{figure}

A complete calculation of $\mu_R$ in all regimes
can be made by examining the RG equations derived above.
Here we give an estimate for the low
temperature behaviour. A key point is that at low temperature $\mu_R$ will
be determined by the short scale contributions.
Indeed there must be some continuity with the 
$T=0$ flow, where $-\Delta_l''(0)$ diverges after
a finite length scale, the Larkin length $R^y_c$
(as discussed above). Thus at low temperature  $-\Delta_l''(0)$ will 
first shoot up near the Larkin length, strongly renormalizing
the mobility downwards, before the temperature catches on and reduces
it back to its fixed point value ${-\Delta^*}''(0) \sim 1$.
Note that this fixed point value corresponds to 
values of disorder {\it much larger} than the original 
disorder. Indeed restoring the factors
one sees that (in $d=3$) that the original disorder
dimensionless parameter is 
$\Delta_2/(4 \pi v c) \sim \ln(a/R^y_c) \ll 1$ at weak
disorder while asymptotically one has 
$\Delta^R_2/(4 \pi v c)=1$. The global behaviour
with length scale is illustrated in 
the Fig. (\ref{2deriv}).

The small $T$ behaviour in $d=3$ can be estimated as follows.
let us denote $T_0$ the bare value.
One has the exact equations:

\begin{eqnarray}
&& \frac{d \Delta''(0)}{dl} = T \Delta''''(0) - \Delta''(0)^2  \\
&& \frac{d \Delta''''(0)}{dl} = - 7 \Delta''(0) \Delta''''(0) + T \Delta^{(6)}(0) \\
&& \frac{d T}{T dl} = -1 - \Delta''(0)
\end{eqnarray}

We will roughly estimate the scale $R_c(T_0)=ae^{l^*}$ at which the term $T \Delta''''(0)$ starts
slowing down the growth of $\Delta''(0)$. For this we will 
drop the term $T \Delta^{(6)}(0)$ (thus assume that for $R<R_c(T_0)$
the flow is identical to $T=0$). Integrating the equations
then yields $(1-\Delta_2 l^*)^6 = T_0 e^{-l^*} \Delta_4/(\Delta_2)^2$
denoting the bare values of the derivatives of the disorder correlator
by $\Delta_2 = -\Delta''(0)$ and $\Delta_4 = \Delta''''(0)$.
This length scale is very close to the Larkin length
and the end result is that:

\begin{eqnarray}
- \Delta''(0,l=l^*) = \frac{\Delta_2}{(T_0/T^*)^{1/6}}
\qquad T^* = \frac{\Delta_2}{4 \pi v c} \frac{R_c^y}{a} 
\end{eqnarray}

Thus this argument indicates that as $T \to 0$ the renormalized mobility
will vanish as:

\begin{eqnarray}
\mu_R \sim \mu_0 e^{ - (1/T)^\alpha }
\end{eqnarray}

\subsection{extensions}

\label{extensions}

As was mentionned in the Section \ref{sec:themg}
there are many possible generalization of the moving glass
equation to a larger class of non potential model.
Let us indicate here the general FRG equation for these
models. We consider a dynamical equation
with a $N$ component field $u_{\alpha}$ and static
disorder described by some force correlator $\Delta_{\alpha \beta}(u)$.
Then, as shown in the Appendix, the corrections to disorder
will be:

\begin{eqnarray}
&&\delta \Delta_{\alpha \beta}(u) = \Delta_{\alpha \beta ; \gamma \delta}(u)
( \Delta_{\alpha' \beta'}(0)  - \Delta_{\alpha' \beta'}(u) )
\int_r G_{\gamma \alpha'}(r) G_{\delta \beta'}(r)
\nonumber \\
&& - \Delta_{\alpha \alpha' ; \delta}(u)
\Delta_{\beta \beta' ; \gamma}(u)
\int_r G_{\gamma \alpha'}(r) G_{\delta \beta'}(-r)
\end{eqnarray}

where we have defined the static response
$G(r) = \int_0^{\infty} d\tau R(\tau,r)$. Note that this
formula is valid for a large class of models with arbitrary response
function. It does {\it not}
suppose for instance that the random force correlator is
the second derivative of a random potential.

>From this a generalized FRG equation can be derived,
which depends on the divergences contained in the
response function. A special case is when $v>0$
and isotropy is assumed
(the simplest $N$ component generalization of
the moving glass equation (\ref{movglasseq2}).
Then one finds:

\begin{eqnarray}
\frac{d \Delta_{\alpha \beta}(u)}{dl} = \epsilon  \Delta_{\alpha \beta}(u)
+ \zeta u_\gamma \Delta_{\alpha \beta,\gamma}(u) + C
\Delta_{\alpha \beta ; \gamma \delta}(u)
( \Delta_{\gamma \delta}(0)  - \Delta_{\gamma \delta}(u) )
\end{eqnarray}

where $C$ is a numerical constant.

The temperature can be added. In the isotropic
case it simply produces an extra term $-T \Delta_{\alpha \beta,\gamma \gamma}(u)$
in the above equation.

Finally let us close this section by a remark on the
nature of the MG fixed point. The role played
by $c_x$ (apart from a regulator) is a bit mysterious.
Indeed one can wonder whether these results could be
obtained directly from the following {\it static} equation
for $u_{xy}$:

\begin{eqnarray}  \label{toosimplestatic}
v \partial_x u = c_y^2 \partial_y^2 + F(x,y,u)
\end{eqnarray}

which contains
only the operators which are relevant at the fixed point.
Indeed the final results for the fixed point of the moving
glass involve only (and presumably to all orders in perturbation theory)
zero frequency (i.e static) propagators (at zero temperature and
at zero applied external force). If one is too naive and uses an
Ito scheme in $x$ this equation \ref{toosimplestatic} becomes trivial
and leads to only a random force. It is probable that when the full
non linearity of $F(x,y,u)$ is kept, and the $x$ discretization
treated properly (which may be subtle) the proper result
is found. It probably then lies entirely in
a proper regularization of the corresponding functional determinant
(in a sense the dynamics in $t$ that we use here can serve as
such a regularization). We leave this for future investigations.

\section{Moving glass equation in $d=2$ and $d=2+\epsilon$}

\label{dynamicalco}

As stressed in the Introduction, it is important to
first study the elastic theory as a function of the dimension $d$,
before attempting to include topological defects.
Up to now we have studied the Moving Glass equation 
in an $d=3-\epsilon$ expansion. This study is of course mostly
relevant for the physical dimension $d=3$.
To study the other physically interesting dimension $d=2$
another RG calcluation can be performed.
For the {\it statics}
$v=0$ the RG approach was constructed by Cardy and Ostlund CO
\cite{cardy_desordre_rg}. It was later extended
to study equilibrium dynamics \cite{goldschmidt_dynamics_flux}
and, with some additional assumptions, to study the
problem in $d=2+\epsilon$. In the case $v=0$ it yields a marginal glass phase
in $d=2$ for $T < T_g$ described by a line of perturbative fixed points.
Extensions to models with $n>1$ components necessary to describe a lattice
\cite{carpentier_ledou_triangco} and to far from equilibrium 
dynamics \cite{tsai_shapir_offeq} were also studied.
((((((((rsb quelque part ))))))))))))

In this Section we first show that in $d=2$ the CO fixed line is {\it unstable}
to a finite $v$ on the simplest case of $n=1$ component Moving Glass equation.
We derive the RG equations for the case $v>0$. We stress that this is
a toy model since it is clear that in $d=2$ additional instabilities
to dislocations will occur at the temperatures where we can control
the behaviour of the model. However it is instructive, as we will see,
and provide the first necessary step to introduce the other
instabilities.

\subsection{$d=2$}

\subsubsection{RG equations in $d=2$}

The equation that we study is:
\begin{eqnarray}  \label{eqd2}
( \eta \partial_t - c_x \partial_x^2 - c_y \partial_y^2 + v \partial_x ) u_{rt}
= f_1(r,u_{rt}) + f_2(x)
\end{eqnarray}
where $f_1(r,u_{rt})$ is the random nonlinear pinning force
with $\overline{ f_1(r,u_{rt}) f_1(r',u_{r't'}) } =
\Delta(u_{rt} - u_{r't'})$ and $f_2(x)$ is
the disorder originating from long wavelength disorder (\ref{couplage}) with:
\begin{eqnarray}
\overline{ f_2(q) f_2(- q) } = \Delta q^2 + \Delta_0
\end{eqnarray}
In addition such terms are generated in perturbation theory (at least
for $v=0$) and should thus be added from the start.

We use everywhere the shorthand notation $v \equiv \eta_0 v$ where
is the bare value of the friction coefficient.
Since we are looking at a periodic system one has
$\Delta(u)= \sum_{K} \Delta_K e^{i K u}$. However in $d=2$
where temperature is {\it marginal}
the harmonics are relevant at different temperatures. This remains
true at $v>0$. It is thus enough to consider the lowest harmonic
\begin{equation}
\Delta(u - u') = g \cos(u -u')
\end{equation}
Perturbation theory is carried in Appendix~\ref{appendix2d} using the
MSR formalism. Note that the random forces $f_2$ can be
eliminated by a shift and do not feedback in the RG
(see \ref{appendix2d}). In addition due to the tilt symmetry (galilean
invariance) $c_x$, $c_y$ and $v$ have no corrections.

One finds to first order in $g$ the following
corrections to the friction coefficient, the temperature,
and the disorder:
\begin{eqnarray} \label{corrections}
\delta \eta &=&
g \int_0^{+\infty} d \tau \tau R(0,\tau) e^{-\frac{1}{2} B(0,\tau)}
\nonumber \\
\delta (\eta T) &=& g
\int_0^{+\infty} d\tau
( e^{ - \frac{1}{2} B(0,\tau) } - e^{ - \frac{1}{2} B(0,\tau=\infty) })
\nonumber \\
\delta g &=& g e^{ - \frac{1}{2} B(0,\tau=\infty) }
\end{eqnarray}
where $B(r,t)=\langle[(u_{rt} - u_{00}]^2\rangle_0$ and
$R(r,t) = \langle \delta u_{rt}/\delta h_{00} \rangle_0$ are respectively
the correlation and response functions in the theory without disorder.
In the case $v=0$ the fluctuation-dissipation theorem (FDT) holds:
$2 T R(\tau) = d B/d\tau$. This ensures that
the correction of the friction coefficient and of the
thermal noise are exactly proportional and can be both
absorbed into a single correction of the friction coefficient
(leaving $T$ unchanged $\delta T=0$). This can be seen
immediately by integrating by part (\ref{corrections}) using
(\ref{lescorrel}). This property does not hold any more when $v\neq 0$
and $T$ will renormalize upward in $d=2$. 
As we will discuss in more details later, the generation of 
a temperature due ot disorder occuring here is quite different 
from the notion of shaking temperature \cite{koshelev_dynamics}

Similarly than for the statics, disorder will be relevant below 
a certain temperature $T_g$. To determine $T_g$ one computes
the mean square displacements 
in the absence of disorder
\begin{eqnarray} \label{bregul}
&& B(r,t,a) = \int  \frac{d^2 q}{(2 \pi)^2}
\frac{2 T}{c q^2}
(1 - e^{- c q^2 \mu |t|} e^{i q r + i v q_x \mu t} ) e^{-a^2 q^2}  \\
&& B(r=0,t,a) =  \frac{T}{2 \pi c}
( Log[\frac{c \mu |t| +a^2}{a^2}] + C + Log[\frac{(v \mu t)^2}{4(c \mu |t| 
+a^2)}] - Ei[\frac{- (v \mu t)^2 }{4(c \mu |t| +a^2)}] )
\end{eqnarray}
where the mobility $\mu = 1/\eta$ has been introduced. 
In (\ref{regul}) we have used the 
following regularizations:
(i) an infrared regulator by defining a large time
$t_{max}$ but no infrared regulator in momentum $q$
(ii) an ultraviolet cutoff is enforced via a gaussian
cutoff in momentum. Such a regularization procedure  
turns out to be extremely conveninient to establish the RG
equations.
Thus according to whether $v=0$ or $v>0$ one has the
two different large time behaviours:
\begin{eqnarray} \label{bcut}
B(0,t,a) &=& \frac{4 T}{T_c} ( \ln[v \mu t/a] + C/2 )  \qquad (v>0)
\\
B(0,t,a) &=& \frac{2 T}{T_c} ( \ln[c \mu t/a^2] + C/2 ) \qquad (v=0)
\nonumber
\end{eqnarray}
where $T_c = 4 \pi c$ is the transition
temperature of the static system. 
Remarkably, as can be seen from (\ref{bcut}), $T_g$ is {\it half} of 
the Cardy Ostlund glass temperature $T_c$ of the statics !

Thus the CO line is unstable and both disorder and temperature 
will be generated. To obtain the RG equations we restrict ourselves 
to the case when the
starting cutoff is large enough $a^2 v^2/(4 c^2) \gg 1 $
(or the velocity large enough) so that one is already
in the asymptotic regime. Of course at small velocity there will be
a complicated crossover where the short
distance properties are dominated by the static solution, but the large
distance properties will be again given by the present RG equations.
Introducing the dimensionless coupling constant $\tilde{g} = g a/(v T_c)$
(\ref{corrections}) and (\ref{bcut})
allow to obtain the correction to $\eta T$ as:
\begin{eqnarray}
\frac{ \delta (\eta T) }{\eta T_c} =
\tilde{g} e^{- \frac{T}{T_c} C} \int_{\eta a/v}^{t_{max}}
\frac{ v dt }{\eta a} ( \frac{ v t }{\eta a} )^{- 2 T/T_c}
= \tilde{g} e^{- \frac{T}{T_c} C} dl
+ \tilde{g} e^{dl (1 - 2 T/T_c)} e^{- \frac{T}{T_c} C}
\int_{\eta a'/v}^{t_{max}}
\frac{ v dt }{\eta a'} ( \frac{ v t }{\eta a'} )^{- 2 T/T_c}
\end{eqnarray}
which yield the RG equations upon a change
of cutoff $a' = a e^{dl}$:
\begin{eqnarray} \label{rgdebut}
\frac{d \tilde{g}(l)}{d l} &=& (1  - \frac{2 T}{T_c} ) \tilde{g}(l) + O(g^2) \\
\frac{1}{\eta(l) T_c} \frac{d (\eta(l) T(l))}{d l} &=&
\tilde{g}(l) e^{- C/2} \nonumber
\end{eqnarray}
the last equality being valid near the transition
at $T=T_c/2$. (\ref{rgdebut})
can be compared to the
Cardy-Ostlund RG equation \cite{cardy_desordre_rg} for $v=0$:
\begin{equation}
\frac{d g}{d l} = 2(1  - \frac{T}{T_c} ) g + O(g^2) \qquad [v=0]
\end{equation}
Finally, because of the exponential decay of
the response function 
\begin{equation}
R(0,\tau) = \frac{\mu}{4\pi(c\mu \tau + a^2)} e^{-\frac{v^2 \mu^2
\tau^2}{4(a^2+c\mu\tau)}}
\end{equation}
(as in Section~\ref{}) which
cuts all divergences, the correction to $\eta$
in (\ref{corrections}) is only a finite
number. There is thus no renormalization of $\eta$ at large 
scales.
The final RG equations in $d=2$ read:
\begin{eqnarray}  \label{rgfull}
\frac{d \tilde{g}}{d l} &=& (1  - \frac{2 T}{T_c} ) \tilde{g} + O(\tilde{g}^2)
\nonumber\\
\frac{d T}{T d l} &=& 2 C_1 \tilde{g} + O(\tilde{g}^2) \\
\frac{d \eta}{d l} &=& 0 \nonumber
\end{eqnarray}
with $C_1 = e^{- C/2 }$ is a nonuniversal
constant.

\subsection{Analysis of RG equations in $d=2$}

\begin{figure}

\centerline{\epsfig{file=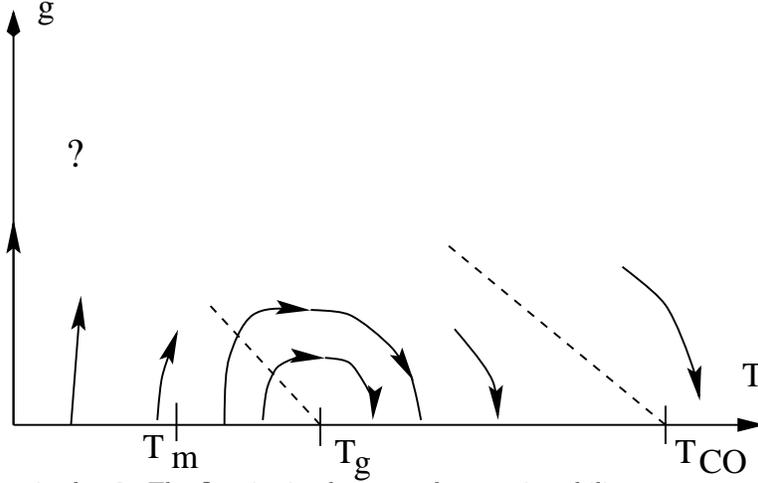,angle=-90,width=10cm}}
\caption{RG flow diagram in $d=2$. The flow is circular around a new
instability temperature at $T=T_g=T_{CO}/2$. The Cardy Ostlund
line of fixed point of the statics which start at $T=T_{CO}$
in unstable when $v>0$. For $d=2+\epsilon$ there is a new 
finite $T$ moving glass fixed point (presumably attractive)
at $g \sim \epsilon$ (the resulting spiraling flow is not shown).
Continuity with the FRG result suggests
that this fixed point moves upward to the $T=0$ axis as
$d$ goes from $d=2$ to $d=3$. In $d=2$ a zero temperature 
moving glass fixed point is expected at infinite 
$g=-\Delta''(0)$ (if the lower critical dimension $d_{lc}$ for
the $T=0$ moving glass is $d_{lc} \le 2$).
\label{fig:co}}
\end{figure}

Let us analyze the RG flows. We introduce the reduced
temperature $\tau = (2 T - T_c)/T_c$ and
$\overline{g}=2 C_1 \tilde{g}$. The RG equations are:
\begin{eqnarray}
\frac{d \overline{g}}{d l} &=& - \tau \overline{g} \\
\frac{d \tau}{d l} &=& \overline{g}
\end{eqnarray}
The trajectories are the arches of parabolaes represented
in Figure~\ref{fig:co} centered around $T_c/2$, of equation:
\begin{eqnarray}
\overline{g} - \overline{g}_0 = \frac{1}{2} (\tau_0^2 - \tau^2)
\end{eqnarray}
(note that close to $T=T_c/2$ these trajectories are
not modified by the higher order terms). 

As can be seen from Figure~\ref{fig:co} if one starts at small disorder
with temperature $T_c/2 - \Delta T$, both
disorder and temperature first increase pushing the system in a region where
the disorder is irrelevant, ending up
with a disorder free system at about $T_c/2 + \Delta T$.
This has several physical consequences:

(i) at finite velocity the effect of disorder is
weaker than in the statics, which manifest itself
in the RG equation since weak disorder becomes irrelevant
for $T > T_c/2$ a region which is already deep in the
glass phase in the statics. This effect is analogous to
the dimensional shift from $d_{uc}=4$ in the statics
to $d_{uc}=3$.

(ii) However we still find a transition at $T=T_c/2$
below which disorder is relevant and grows under RG.
That such a region where disorder is relevant exist
in $d=2$ is compatible with the FRG findings in $d=3-\epsilon$ 
and clearly shows that
even in motion one still has to consider the effect
of the random potential. However due to the importance of the 
thermal effects in $d=2$, at large enough scales
the disorder will stop being relevant since the temperature 
also increases.  The lenght scale
$\xi$ at which disorder becomes again negligible can be estimated
from the RG and reads:
\begin{eqnarray}
\xi \sim (\frac{\tau_0^2}{g})^{1/(4 \tau_0)}
\end{eqnarray}
$\xi$ cna become extremely large when the disorder is weak or when 
one starts at low enough temperatures. 

(iii) Finally, we find that disorder generates an additional
temperature. This renormalization of temperature is physically very different
that the ``shaking' temperature''  of \onlinecite{koshelev_dynamics}. 
In particular the value of the generated temperature in our case does
not depend on the strength
of the disorder but on the temperature itself and
the distance to $T_c$. In particular, and in similar way than 
for the FRG, if one had started at $T=0$ no temperature is
generated as can be seen from (\ref{rgfull}).

Using the RG flow one can compute the displacements.
For the connected correlations one finds:
\begin{equation}
\overline{\langle [u(x)-u(0)]^2 \rangle - \langle u(x)-u(0) \rangle^2} 
\sim T^*(\tau,g) \ln x
\end{equation}

We have used the exponent at the fixed point, which is a correct
procedure because the fixed point is approached fast enough as
$\tau_0 - \tau(l) \sim \tau_0 e^{- \tau_0 l}$. These
correlations are nonmonotonic as a function of $T$ with an almost cusp
(rounded by $g$) at $T_c/2$, and increase below $T_c/2$.

Given the present result near $T_c/2$ possible topologies of
the flow in $d=2$ can be proposed as shown in Fig. \ref{fig:co}. In any case
the relevance of disorder at low temperature is a very good confirmation
of the presence of the moving glass phase at least at $T=0$ for which
temperature renormalization effects are absent.

However in $d=2$ thermal RG effect are obviously important and the question
of whether the MG phase exist at finite T arises. Two flow diagrams are
in principle possible. In one of them a low temperature phase exist (MG).
In that case an additional fixed point which controls the
transition is necessary. In the absence of such a fixed point
the moving glass will always be unstable due to temperature renormalization.
This last scenario is supported by the FRG results in $d=3-\epsilon$ and
by the next Section.

\subsection{Moving glass equation in $d=2+\epsilon$}

We now follow Goldschmidt et al. and try to continue
the above RG equations to $d=2+\epsilon$. The $\epsilon$
simply shifts the dimensions of the operators. We stress
that this is based on the assumption that the RG
functions are well behaved around $d=2$. It has been
used though in other cases such as the $O(n)$ model.

Taking into account the dimensions, one readily obtains
using the same reduced variables, to lowest order:

\begin{eqnarray}  \label{rgfull}
\frac{d \overline{g}}{d l} &=& - (\epsilon + \tau) \overline{g}
- b \overline{g}^2
\nonumber\\
\frac{d \tau}{d l} &=& - 2 \epsilon + \overline{g} \\
\frac{d \eta}{d l} &=& 0 \nonumber
\end{eqnarray}

with $b=B/(4 C_1^2)$ being a universal (regularization
independent) number (by analogy with CO).

These equations now have a new fixed point at
$\overline{g}=2 \epsilon$ and $\tau=-(1+2 b)\epsilon$
(note that $B/C_1$ is universal). To lowest order
the eigenvalues are:

\begin{eqnarray}
\lambda_{\pm}= - b \epsilon \pm i \sqrt{2 \epsilon}
\end{eqnarray}

Thus without needing to compute the coefficient $b$
we know that there is a fixed point, and we know the
leading behaviour of the eigenvalues
$\lambda_{\pm} \sim \pm i \sqrt{\epsilon}$. Such spiralling
fixed points have been obtained in other problems
(e.g Ref. \cite{boyanovsky_cardy_supra}).
However to know whether the
fixed point is attractive or repulsive one needs to
know the real part, which is controlled by the
$O(g^2,\tau^2,g \tau)$ terms in the RG equation.
For instance, the RG equation will contain at least:

\begin{eqnarray}  \label{rgfull}
\frac{d \overline{g}}{d l} &=& - (\epsilon + \tau) \overline{g}
- b \overline{g}^2
\nonumber\\
\frac{d \tau}{d l} &=& (- 2 \epsilon + \overline{g} + c \overline{g}^2)(1+\tau)
\end{eqnarray}

Inspection shows that $b$ actually controls the leading
behaviour of the real part. So this is the only non linear
term we need to compute. Results from FRG (see
\cite{appendixc}) and static CO
lead to expect that $b>0$, but to settle the question
and obtain the value
of $b$ an actual calculation along the lines of
\cite{carpentier_ledou_triangco} is needed.

It is tempting to associate this fixed point to a
finite temperature moving glass fixed point
(analogous to the new finite $T$ fixed points
found in recent manifold studies \cite{ledou_wiese_ranflow}).

Finally we not that at this fixed point one will have $z=2$
since we find that $d \eta/dl=0$.
(thus there will be large but finite barriers).

\section{Towards a complete description of elastic flows}

In this Section we investigate Models II and Model III.

\subsection{Study of the complete dynamical equation in elastic
limit (Model II)} \label{frgcomplete}

In this Section we come back to the Model II which 
contains both degrees of freedom transverse $u_y$ and
along motion $u_x$ and the full time dependence of the
pinning force. If one describes the elastic flow of a solid,
i.e in a regime, or a range of scales where there are no
topological defects, this model improves on Model III
(see Section Intro for a discussion of the
regimes where it will be useful).
As discussed in Section \ref{model} it is still
an approximation, but a rather good one, 
of the full (but intractable) model of the elastic
flow (Model I). 

Since Model II is still quite difficult, our aim in
this Section is more limited than in Section \ref{renormalizationtransverse}.
We show that the main features of
the transverse physics of the Moving Glass embedded in
Model III are also present when the degrees
of freedom $u_x$ along the motion are added.
In fact we show explicitly that in Model II the RG equations for along $y$
remain {\it identical} to the one of
Model III !
The two important issues we discuss are the one
of the existence or not of an extra temperature 
generated by motion, and of the generation of a
static ``random force''. We perform perturbation theory up to second order
in the disorder and examine the new terms generated
as well as the divergences. We develop an approach
which allows to treat {\it all harmonics} 
$\Delta^{\alpha \beta}_K$ of the disorder correlator.
First, we find that a static 
``random force'' is generated in the direction of motion. This may
seem surprising at first, because first order perturbation
theory gives a pinning force along $x$ which rapidly oscillates.
However, as our calculation shows, to second order
the various washboard frequency harmonics interfere
to produce a static random force. Second, we identify the
divergences in perturbation theory and follow the evolution
of the full disorder correlator under renormalization.
We show that up to some details the resulting picture is 
close - if not identical - to the one given by Model III. We work
at $T=0$ but the approach can be extended to $T>0$
along the lines of Section \ref{renormalizationtransverse}.

\subsubsection{general properties}

The equation of motion (\ref{startexp},\ref{defr})
keeping the time dependent terms reads:

\begin{equation}
{(R^{-1})}^{\alpha \beta}_{rt r't'} u^{\beta}_{r't'} = f_{\alpha} - \eta v_{\alpha}
+ F_{\alpha} (r,t,u_{rt})
\end{equation}

More specifically we will be interested in the model
for a triangular lattice ($n=2$ component model) with the force
applied along a symmetry direction. The equation 
of motion reads for lines in $d=3$:

\begin{eqnarray}
&& \eta_{xx} \partial_t u_x + \eta_0 v \partial_x u_x
+ (c_{66} \nabla^2 + c_{44} \partial_z^2) u_x 
+ (c_{11}-c_{66}) \partial_x (\partial_x u_x + \partial_y u_y)
= f - \eta_{xx} v + F^{pin}_x(r,t,u) \\
&& \eta_{yy} \partial_t u_y + \eta_0 v \partial_x u_y
+ (c_{66} \nabla^2 + c_{44} \partial_z^2) u_y 
+ (c_{11}-c_{66}) \partial_y (\partial_x u_x + \partial_y u_y)
= F^{pin}_y(r,t,u) \\
\end{eqnarray}

and setting $c_{44}=0$ for points in $d=2$. Note that $\eta_{xy}=0$ 
from the symmetry ($u_y \to - u_y$, $y \to -y$). We have allowed for
different $\eta_{xx}$ and $\eta_{yy}$ since, even if they start identical
(and equal to $\eta_0$) they will not remain so under renormalization. 
The statistical tilt symmetry ensures that the elastic coefficients
and $\eta_0 v$ remains uncorrected. Note that in later calculations
it will be convenient to rescale the $z$ direction,
setting $z=\sqrt{c_{44}/c_{66}} z'$.

The correlator of the pinning force can be written as:

\begin{equation}  \label{periodic}  
\overline{ F_{\alpha} (r,t,u) F_{\beta} (r',t',u') }
=  \delta^d(r-r') \Delta^{\alpha \beta}(u - u' + v (t - t') )
= \delta^d(r-r') \sum_K \Delta^{\alpha \beta}_K
e^{ - i K . (u - u' + v (t - t') ) }
\end{equation}

It contains all lattice harmonics $K$. Due to the modes with $K_x \ne 0$
it is an explicit {\it periodic function of time} with frequencies 
all integer multiples of the washboard frequency $2\pi \omega_0=v/a$.
In addition it contains non linear {\it static} components $K_x=0$, 
$K_y \ne 0$ (which lead to the Model III treated in Section 
\ref{renormalizationtransverse}).
Finally it contains a static $u$-independent component,
$\Delta_{K=0}$ which we call - by definition -
the static ``random force''. We believe this is the ``correct''
definition of the random force. Note however that one must distinguish
it from $\Delta(u=0)$. If one was to compute the
displacements correlator to order $\Delta^2$ one would
need $\Delta(u=0)$ to second order, which is a different calculation.
The important point is that the static ``random force''
is strictly zero in the bare model $\Delta_{K=0}(0)=0$,
but that it is generated in perturbation theory, as we show below.

The idea behind the method presented here is that strictly at $T=0$ all the time
dependence {\it remains strictly periodic} to all orders in perturbation theory
in the disorder (plus a static part).
Only frequencies multiple of $\omega_0$ can be generated. Indeed
to lowest order $F^{pin}_y(x,y,t,u=0)$ is periodic in time, and yields
a $u$ periodic. Iterating perturbation theory thus leads
only to periodic $u$ and $F^{pin}_y(x,y,t,u)$ with integer multiples of
$\omega_0$. This property allows us to
construct a closed RG scheme of the above model with a renormalized
disorder which remains of the form \label{periodic}

An immediate consequence is that {\it no temperature is generated} when
$T=0$ at the start. Temperature is
defined as the zero frequency limit of the incoherent noise
(or using the MSR formalism the vertex function
$\eta T = \Gamma_{\hat u \hat u}(q=0, \omega \to 0^{+}))$.
Thus here one has:
\begin{equation}  \label{periodic0}  
\eta \delta T = \int d\tau \sum_{K \neq 0}  (\Delta_{ren})^{\alpha \beta}_K
e^{ - i K . v \tau  } = 0 
\end{equation}
The static random force $K=0$ mode leads to a $\delta(\omega)$ part in
$u$ and thus is distinct from the temperature.
The fact that no temperature is generated strictly at $T=0$ is
a rather strong property of the elastic flow. In physical
terms in the $T=0$ elastic laminar periodic flow all particles
strictly replace each others after time $\tau_0=a/v$,
i.e $R_{i_x,i_y}(t+\tau)=R_{i_x+1,i_y}(t+\tau)$ where $i=i_x,i_y$ are integer
labels for the particles. It is clearly
possible that this laminar periodic flow becomes {\it unstable}
to chaotic motion. It is still very interesting to
investigate this laminar periodic flow. We will thus proceed
here assuming here that instabilities to chaos happen only
at finite large enough disorder, or at large enough scale.
We reserve the study of the stability of this flow
(chaos, non perturbative effects, etc..) to future study.
Finally note that this periodic flow is allowed 
by the assumed absence of topological
defects in the system. Dislocations, if present would
probably ruin periodicity and introduce a small additional
temperature (though this remains to be investigated).
Finally, it is interesting to note that the present considerations
are in stark contradiction with the Koshelev-Vinokur arguments
\cite{koshelev_dynamics} about the generation of a ``shaking temperature''.

We start by establishing the possible form for the disorder
correlator (at any order in perturbation theory) based on
the symmetries of the problem (Model II). This is highly necessary
here because the bare disorder 
is {\it potential} $\Delta^{\alpha \beta}_{K}(0) =
g_K K_{\alpha} K_{\beta}$, but it does not remain of this
form in perturbation theory ! We are interested
here in the case when the velocity is along a the
lattice direction. Our analysis here is very general, and we will
specify when we apply it to the case of a triangular
lattice. More details are contained in the Appendix
using the more rigorous MSR formalism.

The symmetries are as follows. First one can exchange
$t$ with $t'$ and $u$ with $u'$ in (\ref{periodic}) 
and relabel the disorder term. This gives
$\Delta^{\alpha \beta}_{-K}(v)=\Delta^{\beta \alpha}_K(v)$.
This is by construction and from the specific dependence
in $t$ and $t'$ of the disorder. Second, the action must be real
and thus $\Delta^{\alpha \beta}_{-K}(v)=\Delta^{\alpha \beta}_{K}(v)^{*}$.
Third, the symmetry $T_y$ ( $u_y \to - u_y$, $y \to - y$, $\hat{u}_y \to - \hat{u}_y$)
yields that
$\Delta^{yy}_{K_x,- K_y}(v) = \Delta^{yy}_{K_x,K_y}(v)$,
$\Delta^{xx}_{K_x,- K_y}(v) = \Delta^{xx}_{K_x,K_y}(v)$,
$\Delta^{xy}_{K_x,- K_y}(v) = - \Delta^{xy}_{K_x,K_y}(v)$,
$\Delta^{yx}_{K_x,- K_y}(v) = - \Delta^{yx}_{K_x,K_y}(v)$.
Similarly, because of $T_x$ one finds
$\Delta^{yy}_{- K_x,K_y}(-v) = \Delta^{yy}_{K_x,K_y}(v)$,
$\Delta^{xx}_{- K_x,K_y}(-v) = \Delta^{xx}_{K_x,K_y}(v)$,
$\Delta^{xy}_{- K_x,K_y}(-v) = - \Delta^{xy}_{K_x,K_y}(v)$,
$\Delta^{yx}_{- K_x,K_y}(-v) = - \Delta^{yx}_{K_x,K_y}(v)$.
Note that the global symmetry $T_x T_y$ implies that
$\Delta^{\alpha \beta}_{-K}(v)=\Delta^{\alpha \beta}_{K}(-v)$.
Thus one finds that one can split the disorder correlator into:
\begin{equation}
\Delta^{\alpha \beta}_K(v) = \Delta^{\alpha \beta}_{S,K}(v) + \Delta^{\alpha \beta}_{A,K}(v)
\end{equation}
where $\Delta^{\alpha \beta}_{S,K}(v)$ is real,
symmetric in $\alpha \beta$, even in $K$,
and even in $v$
and $\Delta^{\alpha \beta}_{A,K}(v)$ is imaginary,
antisymmetric in $\alpha \beta$, odd in $K$ and odd in $v$. This
naturally leads to the decomposition 
\begin{eqnarray}  \label{param}
&& \Delta^{\alpha \beta}_{S,K}(v) = \Delta^K_1 v^2 \delta_{\alpha \beta} +
\Delta^K_2 K_{\alpha} K_\beta + \Delta^K_3 v_{\alpha} v_\beta  \\
&& \Delta^{\alpha \beta}_{A,K}(v) = i \Delta^K_4 ( v_{\alpha} K_\beta
- v_\beta K_{\alpha})
\end{eqnarray}
where all $\Delta^K_i$ are even in $K$ and $v$ and real. The bare 
disorder has only $\Delta_2^K$ non zero and thus posses the extra
symmetry $\Delta^{\alpha \beta}_{-K}=\Delta^{\alpha \beta}_K$ or equivalently
$(u \to -u, v \to -v)$. Because of the convection term $v \partial_x u$
in the equation of motion, this additional symmetry will not hold to
higher orders in perturbation theory. It is natural to suppose that
to any fixed order in perturbation theory the $\Delta^K_i$ are
regular when $v \to 0$. Thus in the limit $v=0$ one recovers 
a strictly potential problem (all terms except $\Delta_2^K$
vanish).

Finally note that Model III is a particular case of
Model II which corresponds to the follwoing choice
of bare parameters: (i) isotropic response $c_{11}=c_{66}$ 
(ii) $\Delta^{x y}_K=0$. Then clearly the equations along
$x$ and $y$ decouple.

Another particular case that we note ``en passant'' is to start from 
$\Delta^{\alpha \beta}_{K}= g_K K_{\alpha} K_{\beta}$
and elastic elasticity. Then the equations are
only coupled through the time dependent part of
the non linear pinning force along $y$ which depends
on $u_x$ (one has $\Delta^{x y}_K \ne 0$). It would be interesting 
to check whether this is enough to change the behaviour.

At $T=0$ the lowest order corrections to the disorder
come from second order perturbation theory. The 
calculation of the effective action to second order
for the most general model is performed in the Appendix \ref{appendixc}.
We first give the full, awsome looking expression for it (\ref{full}),
and then will extract some of its main features. A full
consistent treatment is left for future work.
The result of Appendix \ref{appendixc} is that the correction to the disorder is:

\begin{eqnarray}  \label{full}
&& \delta \Delta_K^{\alpha \beta} = - K_{\gamma} K_{\delta}
\Delta_K^{\alpha \beta} \sum_{K'} \int_{q}
\Delta_{K'}^{\rho \lambda}
R^{\gamma \rho}(-i v K', -q)
R^{\delta \lambda}(i v K',q) \\
&& + \frac{1}{2} \sum_{K'} \int_{q}
(K-K')_{\gamma} (K-K')_{\delta}
\Delta_{K-K'}^{\alpha \beta}
[ \Delta_{K'}^{\rho \lambda}
R^{\gamma \rho}(-i v K', -q)
R^{\delta \lambda}(i v K',q) + \Delta_{-K'}^{\rho \lambda}
R^{\gamma \rho}(i v K', -q)
R^{\delta \lambda}(- i v K',q) ] \\
&& - \sum_{K'} \int_{q}
K'_{\delta} (K'-K)_{\gamma}
\Delta_{K'}^{\alpha \rho}
\Delta_{K'-K}^{\beta \lambda}
R^{\gamma \rho}(i v K', q)
R^{\delta \lambda}(i v (K' -K),q) \\
&&
+ K_{\delta} \Delta_{K}^{\alpha \rho}
\sum_{K'} \int_{q}
K'_{\gamma} \Delta_{K'}^{\beta \lambda}
R^{\gamma \rho}(i v (K+K'), q)
R^{\delta \lambda}(i v K',q)
 - K_{\gamma} \Delta_{-K}^{\beta \lambda}
\sum_{K'} \int_{q}
K'_{\delta}
\Delta_{K'}^{\alpha \rho}
R^{\gamma \rho}(i v K', q)
R^{\delta \lambda}(i v (K'-K),q)
\end{eqnarray}

\subsubsection{generation of the static random force}

Setting $K=0$ in the above formula one gets:

\begin{eqnarray}  \label{ranforce}
&& \delta \Delta_0^{\alpha \beta} =    \sum_{K} \int_{q}
K_{\gamma} K_{\delta}
[ \frac{1}{2} (\Delta_{-K}^{\alpha \beta} + \Delta_{K}^{\alpha \beta})
\Delta_{K}^{\rho \lambda}
R^{\gamma \rho}(-i v K, -q)
R^{\delta \lambda}(i v K,q) -
\Delta_{K}^{\alpha \rho}
\Delta_{K}^{\beta \lambda}
R^{\gamma \rho}(i v K, q)
R^{\delta \lambda}(i v K,q)]
\end{eqnarray}

This is the general expression for the static random force
correlator, i.e a $u$ independent 
gaussian random term $F(r)$ in the original equation of
motion.

The first remark is that the random force cannot have crossed correlations, i.e
$\delta \Delta_0^{x y}(v) = 0$. Thus are two independent random forces
one along $x$ of strength $\Delta_0^{xx}$ and one along $y$ of strength
$\Delta_0^{yy}$. This is consistent with formula (\ref{param}) which gives
$\Delta_K^{x y}(v) = i \Delta^K_4 K_y v$ and vanishes for $K=0$.
It can also be seen explicitly on the
above expression which is found to be symmetric in $\alpha \beta$
(using all the above symmetries). Since we know that
$\Delta_0^{x y}(v)$ must be antisymmetric in $x y$ (from the
above $K_y \to -K_y$ transformation) it must vanish.

Since the above expression is still complicated
as it does involve {\it all disorder harmonics},
one must carefully distinguish between:

(i) the static random force generated to lowest order in the bare disorder
$O(\Delta^2)$ (i.e at the initial stage of the RG). To this order
one can use the {\it bare disorder} in (\ref{ranforce}) and
the resulting perturbative expression of the random force 
(evaluated below) is found to be well behaved and without IR divergences.
Of course once even a small finite  random force is generated 
it is relevant by power counting and must be taken into account
(though it does not feedback in the RG for the non linear disorder).

(ii) the static random force generated to higher orders in
perturbation theory (or at the next stages of
the RG). There we find IR divergences. This means that
non trivial corrections originating from the
non linear disorder will have to be also taken into account
to estimate the random force generated. This is done
in the next subsection.

Thus we start by giving the expression of the 
random force generated from the bare disorder (i).

Setting $\Delta^{\alpha \beta}_{K}= g_K K_{\alpha} K_{\beta}$
in (\ref{ranforce}) one finds:

\begin{eqnarray}
\delta \Delta_0^{\alpha \beta} =
\sum_K \int_{q} K_{\alpha} K_{\beta} g^2_K
((K.R(-i v K, -q).K) (K.R(i v K,q).K) - ( K.R(i v K, q).K)^2 )
\end{eqnarray}

Note that it does vanish for $v=0$ as it should
since the problem becomes potential in that limit.

One can first specify this formula for the case $c_{11}=c_{66}$.
We also set $\eta_{xx}=\eta_{xy}$ (which is consistent since
we are just looking at the lowest order contribution in
perturbation theory). This yields:
\begin{eqnarray}  \label{isotrop}
\delta \Delta_0^{\alpha \beta} =
\sum_K \int_{q, BZ} K^4 K_{\alpha} K_{\beta} g^2_K
\frac{ 2 (\eta_0 v)^2 (K_x + q_x)^2 }{
(c^2 q^4 + (\eta_0 v)^2 (K_x + q_x)^2)^2}
\end{eqnarray}

One thus finds that a random force is indeed generated,
i,e there is a positive $\Delta_0^{xx}$ and  $\Delta_0^{yy}$.
One checks that indeed $\Delta_0^{xy}=0$.
The integral for $\Delta_0^{yy}$ is infrared divergent,
as discussed above (which is natural from the analysis
in Section ()) and will be examined in the next Section.

Let us estimate the magnitude of the static
random force generated along $x$. From the above
expression (\ref{isotrop}) one finds in the case
$d_z=0$ relevant for point lattices (in $d=1,2,3$):
\begin{eqnarray}
\Delta_0^{xx}  \approx 
\frac{C_d}{(\eta_0 v)^2} \sum_{K,|K_x| \ge K_m} K^4 g_K^2 +
\frac{C'_d}{(\eta_0 v)^{2-d/2} c^{d/2}} \sum_{K,|K_x| < K_m} K^4 |K_x|^{d/2} g_K^2
\end{eqnarray}
where $K_m = |K_0| \max (1,v_{cr}/v)$. There is a 
crossover velocity $\eta_0 v_{cr} \sim c \pi/a$.
One has defined $C_d=2 S_d/(d (2 \pi)^d)$ and 
$C'_d= S_d (4-d) \pi/(8 (2 \pi)^d \sin(\pi d/4))$
and $n=2$ for triangular lattices. This yields
for $r_f \ll a$:
\begin{eqnarray}
\Delta_0^{xx}  \approx 
\frac{g^2 a^{n-d}}{(\eta_0 v)^2 r_f^{4+n}}
\min( 1 , (\frac{v a}{v_{cr} r_f})^{d/2} )
\end{eqnarray}

In the case $c_{66} \ll c_{11}$ one can simply retain
only the transverse mode (thus setting $c=c_{66}$
in the above formulae. 
In the case $d_z=1$ (relevant for line lattices)
we will only give the large $v$ estimate
(valid for $v \gg v_{cr}=c_{66} \pi/a$).
It reads:

\begin{eqnarray}
\Delta_0^{xx}  \approx 
\frac{1}{\sqrt{c_{44}} (\eta_0 v)^{3/2}} \sum_{K} K^4 |K_x|^{1/2} g_K^2 
\sim \frac{g^2 a^n}{\sqrt{c_{44}} (\eta_0 v)^{3/2} r_f^{n+4+1/2}} 
\end{eqnarray}

Let us stress again that we have defined the random force
as $\Delta_{K=0}^{\alpha \beta}$ (which we believe is
the ``correct'' definition). If one was to compute the
displacements correlator $u u$ to order $\Delta^2$, in 
order to show that the displacements ``feel'' the random force
(e.g and grow unboundedly in the $x$ or $y$ direction) one would
need $\Delta(u=0)$ to second order (and to be consistent the
response function corrected to order $\Delta$)
which is a different calculation. While it will 
make little difference in an order of magnitude estimates for
the effect of the $x$ random force, it is drastically
different along $y$ (one quantity IR diverges and the other
does not - see Section \ref{renormalizationtransverse}, $\Delta(u=0)$ has no divergence) !!

\subsubsection{RG study of Model II}

We will now look for the divergences in perturbation
theory which appear in Model II. We will address only the case $v>0$.
Let us look again at (\ref{full}). It contains infrared divergences 
of the same type that was discussed in Section \ref{renormalizationtransverse}.
These divergences occur only for $q-$momentum integrals
which are {\it both} (i) zero frequency integrals ($K.v=0$ terms)
(ii) involve $q$ and $-q$. These are of the form:
\begin{eqnarray}  \label{diverg}
D_{\gamma \rho,\delta \lambda} = \int_q G^{\gamma \rho}(-q) G^{\delta \lambda}(q)
\end{eqnarray}
where $G^{\alpha \beta}(q)=R^{\alpha \beta}(\omega=0,q)$ is the static
response function. Here it has the form:
\begin{eqnarray}
G^{\alpha \beta}(q)= \sum_I \frac{P^I(q_x,q_y)}{c_I(q) + i v q_x}
\end{eqnarray}
where $I=T,L$ index the transverse and longitudinal
projectors and elastic eigenenergies as defined in ().
The key point (for $n=2$, e.g triangular lattices)
is that one has here $q_x \sim q_y^2 \sim q_z^2$.
Thus all projector elements are subdominant for small $q$
(i.e kill the IR divergences) except the two elements
$P^L_{yy} \sim P^T_{xx} \sim q_y^2/(q_x^2 + q_y^2) \sim 1$.
Thus the only IR divergent elements among (\ref{diverg})
are $D_{yyyy},D_{xxyy},D_{yyxx},D_{xxxx}$. Explicit calculation of
the divergent parts gives:

\begin{eqnarray}   \label{divergents}
&& D_{yyyy} \sim \int_{q_y,q_z} \frac{1}{2 v (c_{11} q_y^2 + c_{44} q_z^2)}
\qquad  D_{xxxx} \sim \int_{q_y,q_z} \frac{1}{2 v (c_{66} q_y^2 + c_{44} q_z^2)} \\
&& D_{xxyy} = D_{yyxx} \sim \int_{q_y,q_z}
\frac{1}{v ((c_{11} + c_{66}) q_y^2 + 2 c_{44} q_z^2)}
\end{eqnarray}

Let us now analyze the consequences of these divergences.
>From (\ref{full}) the relevant corrections to disorder 
are:

\begin{eqnarray}   \label{towardsrg}
\delta \Delta_K^{\alpha \beta} =
D_{\gamma \rho,\delta \lambda}
\sum_{P=(0,P_y)}
( - K_{\gamma} K_{\delta}
\Delta_K^{\alpha \beta}
\Delta_{P}^{\rho \lambda} +
(K-P)_{\gamma} (K-P)_{\delta}
\Delta_{K-P}^{\alpha \beta}
\frac{1}{2} [ \Delta_{P}^{\rho \lambda}
+ \Delta_{-P}^{\rho \lambda} ]
\end{eqnarray}

Note that there is no component $\Delta_{P=(0,P_y)}^{xx}$ in the bare
action but that it is generated in perturbation theory. 
This expression simplifies because of the symmetries
discussed above and the fact that only the terms (\ref{divergents})
appear in (\ref{towardsrg}) (the term $\Delta_{P_y}^{\rho \lambda}$
always occurs in sums symmetrized over $\rho \lambda$ and
one can use that $\Delta_{P_y,P_x=0}^{xy} = - \Delta_{P_y,P_x=0}^{yx}$
to cancel all crossed $D_{xxyy}$ terms). From what remains one finally
obtains the following RG equations:

\begin{eqnarray}   \label{rg2}
&& \frac{d \Delta_K^{\alpha \beta}}{dl} = \epsilon \Delta_K^{\alpha \beta}
+ \sum_{P=(0,P_y)} (
A_{11} \Delta_{P_y}^{yy} ( - K_y K_y \Delta_K^{\alpha \beta} 
+ (K-P)_{y} (K-P)_{y} \Delta_{K-P}^{\alpha \beta} )
+ A_{66} \Delta_{P}^{xx} K_x K_x ( \Delta_{K-P}^{\alpha \beta} 
- \Delta_K^{\alpha \beta} )
\end{eqnarray}

with $A_{11} = 1/(4 \pi v \sqrt{c_{11} c_{44}})$ and
$A_{66} = 1/(4 \pi v \sqrt{c_{66} c_{44}})$ if one uses
the same regularisation as in Section \ref{renormalizationtransverse}
It turns out that these complicated looking
equations have some simplifying features.

The RG equations (\ref{rg2}) are thus 
the generalization to Model II of the RG equations 
of Model III and contain both the physics of $u_x$
and $u_y$. They show that in the Moving Bragg glass, non trivial
non linear effects also occur in the direction of motion.
A more detailed study of these equation will be given
elsewhere \cite{ledou_giamarchi_futur}. Here we give their
salient features. The RG equations (\ref{rg2}) exhibit remarkable features.
First, the subset of these equations for $\Delta^{yy}_{(0,K_y)}$
closes onto itself !! Indeed one finds:

\begin{eqnarray}   \label{rg2}
\frac{d \Delta_{0,K_y}^{yy}}{dl} = \epsilon \Delta_{0,K_y}^{yy}
+ \sum_{P_y} (
A_{11} \Delta_{(0,P_y)}^{yy} ( - K_y K_y \Delta_{0,K_y}^{yy} 
+ (K_y -P_y) (K_y -P_y) \Delta_{0,K_y-P_y}^{\alpha \beta} )
\end{eqnarray}

This is exactly the RG equation of the Moving glass Model III !!
Thus we have shown that the Model III describes correctly
the transverse physics, as announced, within Model II.

One can also write the above
RG equations as coupled differential equations
for three periodic functions of two variables
$\Delta^{\alpha \beta}(u_x,u_y)$ with $\alpha,\beta = xx,yy,xy$.
We will temporarily use the shorthand notation $u=u_y$ and $v=u_x$.
Let us denote $\Delta^{yy}_{K_x=0}(u)$ by $\Delta_1(u)$.
Let us absorb the factor $\epsilon/A_{11}$ in
the $\Delta$. These coupled RG equations become:

\begin{eqnarray}   \label{rg2}
&& \frac{d \Delta_{\alpha \beta}(u,v)}{dl} = \Delta_{\alpha \beta}(u,v)
+ ( \Delta_1(0) - \Delta_1(u) ) \partial_u^2 \Delta_{\alpha \beta}(u,v)
+
\gamma
( \Delta_2(0) - \Delta_2(u) ) \partial_v^2 \Delta_{\alpha \beta}(u,v)
\end{eqnarray}
with $\gamma=A_{66}/A_{11}$ and $\Delta_1(u) = \int dv \Delta_{yy}(u,v)$
and $\Delta_2(u) = \int dv \Delta_{xx}(u,v)$

Or equivalently:

\begin{eqnarray}   \label{rg22}
&& \frac{d \Delta_{K_x}(u)}{dl} = \Delta_{K_x}(u)
+ (( \Delta_1(0) - \Delta_1(u) ) \partial_u^2 - ( \Delta_2(0) - \Delta_2(u) ) \gamma K_x^2) 
\Delta_{K_x}(u)
\end{eqnarray}

(dropping temporarily the $\alpha \beta$ index). 

The function $\Delta_1(u)$ obeys the closed RG
equation:

\begin{eqnarray}
\frac{d \Delta_1(u)}{dl} = \Delta_1(u) + \Delta_1''(u) (\Delta_1(0) - \Delta_1(u))
\end{eqnarray}

and thus $\Delta_1(u)$ converges towards the Moving glass fixed point
$\Delta_1^*(u) - \Delta_1^*(0) = \frac{1}{2}  u (u-1)$.
Let us now examine the behaviour of the other components of the
disorder with $K_x=0$ (and $\alpha \beta$ = $xx$, $xy$).
They obey the equation:
\begin{eqnarray}
\frac{d \Delta_{\alpha \beta}(u)}{dl} = \Delta_{\alpha \beta}(u) + 
\Delta_{\alpha \beta}''(u) (\Delta_1(0) - \Delta_1(u))
\end{eqnarray}
One easily sees that $\Delta_{\alpha \beta}(u)$ also
becomes {\it non analytic} beyond the dynamical Larkin length
since:
\begin{eqnarray}  \label{second}  
\frac{d \Delta_{\alpha,\beta}''(0)}{dl} =
\Delta_{\alpha,\beta}''(0)(1 - \Delta''_1(0) )
\end{eqnarray}

and thus the divergence of 
$\Delta''_1(0) \to - \infty$ at the Larkin length
$l_c=\ln R^y_c$ (see Section \ref{renormalizationtransverse} ) implies that 
$\Delta_{\alpha \beta}''(0)$ also diverges. Thus 
$\Delta_{\alpha\beta}(u)$ does become non analytic.
Note that the solution of this equation is
simply $\Delta_2(l)=(\frac{\Delta_2}{\Delta_1}) \Delta_1(l)$.

It is then easy to show that $\Delta^{\alpha,\beta}(u) = C \Delta_1^*(u)$
is a stable fixed point solution
(up to the usual growing constant).
Indeed, inserting the fixed point
value $\Delta_1(u)$ in (\ref{second}) one finds exactly the
stability operator of the original fixed point. This was
discussed in Appendix \ref{stability} and the results there shows the
stability within the space of (non analytic) periodic functions.
The constant $C$ can be determined. Indeed one can also
use (\ref{second}) to study $g=\Delta_{\alpha \beta}''(0^{+})$.
One gets $g \sim \int_{l>l_c} dl (1 - \Delta''_1(0^{+},l) )$. The
exponential convergence of $\Delta_1$ towards its fixed point
implies that $g$ is a finite constant. This determines $C$.
Thus at the fixed point one can replace $\Delta_2(u)=C\Delta_1(u)$
in \ref{rg22}.

Thus we have shown that the RG equations in a Moving Bragg 
Glass decouple completely along $y$, giving back the
one of the generic moving glass studied in
Section \ref{renormalizationtransverse}. A complete study
of the system \ref{rg2} will be given elsewhere.

\subsection{Full model for the elastic flow (model I)} \label{beyond}

We now come back to the problem of establishing the correct
long wavelength hydrodynamic description of a moving structure
with some internal order described by a displacement field
$u_{\alpha}(r,t)$. The first step is to write an equation of
motion which contains all terms which are (i) allowed by symmetry
and (ii) a priori relevant in the long wavelength limit by
power counting. We carry this step here,
check that all these terms are indeed generated in
perturbation theory from the original equation of
motion \ref{labase}, and estimate their magnitude.
The second step, which is to solve the universal large distance
physics of such an equation turns out here to be a formidable task,
which goes beyond this paper. 

As we have discussed in Section \ref{model} the problem of driven lattices
posses some additional ``almost exact'' symmetries which 
allow to simplify the hydrodynamic description and to extract
some of the physics. This has lead us to study Model II (Section \ref{frgcomplete})
which posses the statistical tilt symmetry forbidding many terms,
and the simpler Model III (Section \ref{renormalizationtransverse}) (the Moving glass
equation) which we believe contains most of the physics of
moving structures, i.e the physics of the transverse degrees of freedom.

As discussed in Section \ref{model} the only exact symmetries of
the problem, for motion along a symmetry direction
of the moving structure ($x$ axis) are the spatial inversions along the directions
transverse to the velocity. Power counting shows
that the general form for the equation of motion in $d \leq 3$ is
(Model I):

\begin{eqnarray}   \label{general0}
\eta_{\alpha \beta} \partial_t u_{\beta}
+ L^\gamma_{\alpha \beta} \partial_\gamma u_{\beta}
- C^{\gamma \delta}_{\alpha \beta} \partial_\gamma \partial_\delta u_{\beta}
- K^{\delta \epsilon}_{\alpha \beta \gamma}
\partial_\delta u_{\beta} \partial_\epsilon u_{\gamma}
= F^{dis}_{\alpha}(r,u,t) + \zeta_{\alpha}(r,t) 
+ f_\alpha - \eta_{\alpha \beta} v_{\beta} + \delta f_\alpha
\end{eqnarray}

where the velocity $v$ is fixed by the convention 
that $d/dt \int_r u^{\gamma}_{rt}=0$ and $f$ is the applied
force. The KPZ terms are
allowed because $u_\alpha$ is dimensionless at the upper critical
dimension $d_{uc=3}$ in a power counting at $T=0$. 
This can also be seen by writing a MSG formulation of (\ref{general0}).
The above equation of motion
is not fully complete unless one specifies the relevant
disorder and thermal noise correlators. 
The thermal noise has a gaussian correlator
$\overline{\eta_{\alpha} \eta_{\beta}} = 2 (\eta T)_{\alpha \beta}
\delta(t-t') \delta^d(r-r')$, in general anisotropic.
The correlator of the pinning force $F^{dis}_{\alpha}(r,u)$
(which has zero average) is gaussian and of the form:
\begin{equation}
\overline{F^{dis}_{\alpha}(r,u,t) F^{dis}_{\beta}(r',u',t')} = 
\Delta_{\alpha \beta} (u-u'+v(t-t'))
\delta^d(r-r')
\end{equation} 
where $\Delta_{\alpha \beta}(u)$ is a periodic function
in the case of periodic structure. Note however that
in general, $\Delta_{\alpha \beta}(u)$ is not
a potential disorder.

Compared to Model II, cutoff effects which break the
exact statistical tilt symmetry allow new terms 
to be generated, such as linear terms which correct
the original convection term $v \partial_x u$
and non linear KPZ type terms. The linear terms
are obviously relevant and the non linear KPZ terms,
in presence of the disorder, are relevant for $d \leq 3$.
Thus once they are generated, even if their bare value
is very small it may grow under RG and become
important at large scale (a full solution of the
RG equations for (\ref{general0}) would be needed in order
to conclude). However one may guess that since the
statistical tilt symmetry is ``almost exact'',
the scale at which these new term are able to
change the physics compared to Model II and Model III 
(if they do) may be very large here. Finally note that there are
also small corrections to the elastic matrix.

The above approach consists in writing a model independent
equation (\ref{general0}) based on symmetry arguments.
It may be useful in proving the universality of the
behaviours of various structures. 
However, in many cases it is much more instructive to
start from a given simple model without disorder,
such as \ref{labase}, and to estimate the bare values of the new terms 
to first order in a perturbation theory in disorder.
Indeed, in the absence of disorder
the above equation of motion reduces to:

\begin{eqnarray}   \label{general}
\eta^0_{\alpha \beta} ( \partial_t u_{\beta} + v \partial_x u_{\beta})
= (C^0)^{\gamma \delta}_{\alpha \beta} \partial_\gamma \partial_\delta u_{\beta}
+ f_\alpha - \eta^0_{\alpha \beta} v_{\beta} + \zeta_{\alpha}(r,t) 
\end{eqnarray}

We have thus computed in the Appendix \ref{linear} the 
corrections to first order in
perturbation theory with respect to disorder
to all terms of the equation (\ref{general0}). 

Though the above equation (\ref{general}) looks formidable
many terms are zero from the exact inversion symmetry.
We thus now explicity specify the terms allowed in the equation
of motion, for the case of an elastic structure
described by a $n=2$ component displacement field
($u_x$, $u_y$). In $d=2$ and $d=3$ the equation along $y$ should be odd
under the inversion ($u_y \to - u_y$, $y \to -y$)
and also under ($z \to -z$) while the equation along $x$
must be even under these transformations. This yields
in $d=3$:

\begin{eqnarray}   \label{general}
&& \eta_{yy} \partial_t u_y + v_1 \partial_x u_y + v_2 \partial_y u_x =
(c_1 \partial_x^2 + c_2 \partial_y^2 + c_3 \partial_z^2) u_y 
+ c_4 \partial_x \partial_y u_x  \nonumber \\
&& ( a_1 \partial_x u_x + a_2 \partial_y u_y ) \partial_x u_y +
(a_3 \partial_x u_x + a_4 \partial_y u_y ) \partial_y u_x +
a_5 \partial_z u_x \partial_z u_y +
F^{dis}_y(r,u,t) + \zeta_y(r,t) \nonumber \\
&& \eta_{xx} \partial_t u_x + v_3 \partial_x u_x + v_4 \partial_y u_y =
(c_5 \partial_x^2 + c_6 \partial_y^2 + c_7  \partial_z^2) u_x 
+ c_8 \partial_x \partial_y u_y +
 a_6 (\partial_x u_x)^2 + a_7 (\partial_y u_x)^2 +
a_8 (\partial_x u_y)^2 + a_9 (\partial_y u_y)^2 
\nonumber \\
&&
+ a_{10} \partial_x u_x \partial_y u_y
+ a_{11} \partial_x u_y \partial_y u_x
+ a_{12} (\partial_z u_x)^2 +  a_{13} (\partial_z u_y)^2 +
+ F^{pin}_x(r,u,t) + f_x - \eta_{xx} v + \delta f_x + \zeta_x(r,t)
\end{eqnarray}

and the same in $d=2$ with $c_3=a_5=c_7=a_{12}=a_{13}=0$.

While the full analysis of (\ref{general}) goes beyond the present
we now present some arguments showing that the linear terms,
by themselves are unlikely to alter drastically the physics
of the moving glass. The physical interpretation of the linear terms
is that now the local velocity explicitly depend on the local strain
rates of the structure. Though this may generate instabilities
(see below), unless an instability occurs, this is
unlikely to alter the transverse physics. Indeed one can see
that small additional linear terms
do {\it not} remove the divergence in perturbation 
theory which was the hallmark of the moving glass. Let us
write $v_1=v + w$, $v_2=a w$, $v_3 = v - w$, $v_4 = b w$
and consider $w$ as small compared to $v$. Also we choose
for simplicity isotropic elasticity 
$c_1=c_2=c_4=c_5=c_7=c_3=c$, $c_8=c_4=0$. Then
the eigenvalues of the $\omega=0$ (static) response matrix are:

\begin{equation}
D^{\pm}(q)= i ( qx v \pm w \sqrt{q_x^2 + a b q_y^2} ) + c q^2
\end{equation}

and eigenvectors 
$(\delta u_x,\delta u_y)=(b q_y , qx  \pm  \sqrt{q_x^2 + a b q_y^2}$.
Note that one must have $a b >0$ otherwise an instability develops.
The perturbation theory result shows that indeed $a b >0$
at least to lowest order in the disorder.
One finds for instance that the integral which is the key
of the FRG equation for the transverse modes ()
reads:

\begin{equation}
\int_q G_{yy}(q) G_{yy}(-q) = \int_q \frac{c^2 q^4 + (v-w)^2 q_x^2}{
(c^2 q^4 + (v q_x + w \sqrt{q_x^2 + a b q_y^2})^2)
(c^2 q^4 + (v q_x - w \sqrt{q_x^2 + a b q_y^2})^2)}
\end{equation}

As is easily seen this integral is infrared divergent for $d \leq 3$,
logarithmically in $d=3$ and power like in $d=2$
(since $w$ is a small correction to $v$).
The divergences occur in two hyperplanes 
$q_x=\pm \frac{abw}{\sqrt{v^2-w^2}} q_y$ which are tilted
symmetrically with respect of the direction of motion.

Let us now reexamine the Moving Glass equation, i.e Model III,
and ask whether cutoff effects (absence of exact statistical
tilt symmetry) will generate new relevant terms.
By definition this equation involves only $u_y$ and thus
the only a priori relevant terms allowed by symmetry are:

\begin{eqnarray}   \label{general}
&& \eta_{yy} \partial_t u_y + v_1 \partial_x u_y =
(c_1 \partial_x^2 + c_2 \partial_y^2 + c_4 \partial_z^2) u_y
+ a_2 \partial_y u_y  \partial_x u_y + F^{pin}_y + \zeta_y
\end{eqnarray}

Now, we have shown in Section \ref{renormalizationtransverse} that at the Moving glass
fixed point $\partial_x \sim \partial_y^2$ and thus
the KPZ term $a_2$ is irrelevant by power counting.
Note that a cubic KPZ terms $(\partial_y u_y)^2 \partial_x u_y$ is 
allowed by symmetry but again irrelevant near $d=3$.
Thus the Moving Glass equation Model III is stable to cutoff
effects and perfectly consistent. This lead us to claim in
\cite{giamarchi_moving_prl} that while previous descriptions
of moving systems, such as manifolds driven
in periodic \cite{rost_dynamics} or disordered potentials
\cite{hwa_pld,krug_random_mobility},
focused on the generation of dissipative KPZ, such terms are
much less important in the Moving glass equation,
a problem which, because of periodicity, belongs to a new universality
class.

Finally one can also reexamine Model II,
the physics of which is presumably very similar to Model III
at least as far as the transverse degrees of freedom
are concerned. This will certainly hold below a 
(large) length scale max($L_{lin}$, $L_{KPZ}$). Above one
must worry about the new terms. 
Power counting at $d=3$ (where $u_y$ and $u_x$ are dimensionless)
in the equation for the transverse degrees of freedom $u_y$
(using that $\partial_x \sim \partial_y^2$ in the absence
of the new terms) shows that the only KPZ term marginally
relevant at the Model II fixed point in $d=3$ (and thus
the dangerous one) is the term $a_4 \partial_y u_y \partial_y u_x$.
Note also that the linear term $v_2 \partial_y u_x$ also becomes
relevant there and will change the counting. In the end it
is probable that all terms in equation (\ref{general}) will have to be
treated simultaneously to get the physics beyond 
max($L_{lin}$, $L_{KPZ}$).

Finally we note that the arguments given in the previous Section about the
fact that no temperature is generated at $T=0$ are unspoiled by the
new terms generated here in the equation of motion compared to Model II.
That these new terms may lead to other instabilities of the periodic
time ordered flow resulting in chaotic motion is clearly an interesting
possibility deserving further investigations.

\section{Conclusion}

In this paper we have studied the problem of moving structures 
(such as vortex lattices) in a disordered medium 
following the new physical approach developped in Ref.
\onlinecite{giamarchi_moving_prl}. The main new emphasis
in that approach is that because of degrees of freedom
transverse to motion, periodic structures have a radically 
different physics than more conventional driven manifolds.
The main consequence of our study is that the moving
structures remain different from a perfect structures
(e.g. a perfect crystal) at all velocities (for $d \le 3$
for uncorrelated disorder). In particular 
they still exhibit glassy behaviour. The moving configurations
can be generally described in terms of {\it static channels}
which are the easiest paths in which particles follow
each other in their motion.
We have introduced here several degrees of approximation
of this problem, embodied in several models. The 
simplest one, Model III, introduced in Ref. \onlinecite{giamarchi_moving_prl}
focuses only on the transverse degrees of freedom.
A more complex one Model II also contains degrees of freedom
along the direction of motion. We have 
studied these models using several renormalization group
techniques as well as physical arguments. All our
calculations and results confirm that 
focusing on the transverse degrees of freedom (Model III) gives the
main physics for this problem. Indeed we have shown that
the more complete Model II leads to the same physics 
as Model III. 

At zero temperature we have explicitely demonstrated that the 
physics of the moving glass is governed by a new non trivial
attractive disordered fixed point. Using the RG, we have 
explicitly demonstrated the existence of the transverse critical force
predicted in Ref. \onlinecite{giamarchi_moving_prl}, which is
related to the nonanalytic behavior of the renormalized disorder 
correlator at the fixed point. Its actual value, 
computed from the RG coincides with the estimate 
given in Ref. \onlinecite{giamarchi_moving_prl}
based on the existence of a dynamical Larkin length $R_c^y$.
We have also found that at $T=0$ no temperature is
generated because perfect time periodicity is maintained.

We have also shown that a static random force is generated both 
along and perpendicular to the directions of motion.
As a consequence relative displacements in both $x$ and $y$ directions grow
logarithmically in $d=3$, but algebraically in $d=2$. 
Thus in $d=3$ at weak disorder or at large velocity,
the moving glass retains quasi-long range order and 
divergent Bragg peaks. Since the decay of translational order
is very slow in $d=3$ we predict that a glassy moving structure 
with quasi long range order and perfect topological order
in all directions exists: the Moving Bragg Glass.
The determination of its physical properties
is the main result of this paper. This phase is the natural
continuation to non zero velocities of the static Bragg Glass.

We have investigated the effect of a non zero
initial temperature. We found that the moving Bragg Glass
survives at finite temperature as a phase distinct from
a perfect crystal and with properties continuously related 
to the zero temperature moving Bragg Glass. At low temperature
the moving Bragg Glass still exhibits a highly non-linear 
transverse velocity-transferse force characteristics with an
``effective transverse critical current'' (in the same sense
as for the longitudinal critical current). Note however
that at $T>0$ the FRG calculation indicates that the
asymptotic behaviour is linear but with a strongly suppressed
transverse mobility at low temperature.

In addition the existence of elastic channels provides a new and precise
way to look at the problem of generation of dislocations in moving
structures. The natural transition is now a decoupling of the channels
with dislocations decoupling the adjacent layers. It is indeed
easier to decouple the channels via shear deformations than to destroy
the channel structure altogether. This leads to expect another moving 
glass phase which keeps a periodicity along $y$, which has been termed 
Moving Transverse Glass. Since it retains a periodicity along the 
direction perpendicular to motion it shares the properties of moving glasses,
and in particular it exhibit a non zero transverse critical force
at $T=0$.

We have given predictions for the phase diagram of moving systems. 
It shows that the existence of the Bragg glass phase in the 
statics has profound 
implications on the dynamical phase diagram as well. Indeed it is natural 
to connect continuously the static Bragg glass (at $v=0$) 
to the moving Bragg glass (at $v >0$). Thus there should be 
a wide range of velocities (down from the
creep region to the fast moving region) where
effects associated with transverse periodicity
(such as the transverse critical force)
should be observed. We have analyzed the crossovers between 
the Bragg glass properties and the Moving Glass properties
in the region where the velocity is not large.

Further experimental consequences should be investigated in details
for vortex systems in motion. A direct measurement of the
transverse $I-V$ characteristics at low temperature
would be of great interest. But consequences for the 
phase diagrams should be explored too. It was predicted in 
\cite{giamarchi_vortex_long} that the static Bragg glass
should undergo a transition into an amorphous glassy state
upon increase of disorder. As discussed in 
\cite{giamarchi_diagphas_prb} the field induced transition
observed in many experiments is the likely candidate
for such a transition. A similar prediction can
be made in the dynamics, namely that the moving Bragg
glass will experience a field induced transition into
an amorphous moving phase. A detailed investigation of these
transitions may help understand the nature of the
high field pinned phase. Indeed the nature of the transition
away from the Bragg glass may change once the system moves
if the moving amorphous phase is different from
the static amorphous phase. If the ``vortex glass'' phase
\cite{fisher_vortexglass_long,fisher_vortexglass_short}
exists at all in the statics in $d=3$, one may expect that
it would not survive as smoothly as the Bragg glass
once the system is set in motion. Another consequence is that
there should be a first order melting transition at weak disorder
or at large $v$ upon raising temperature from a Moving Bragg
glass to a flux liquid. Finally, it would be interesting to
investigate whether the anomalous response
to transverse forces could have an impact
on the anomalous Hall angle. As we have discussed,
other experimental systems, such as Wigner crystals,
seem to be a promising arena to investigate the
physics presented here. Finally it would be
interesting to reinvestigate more complex
CDW systems such as double $Q$ or triple $Q$.

Another direct experimental consequence, in the case of correlated
disorder is that one should observe a ``Moving Bose glass''.
Static columnar disorder in vortex systems is strong but at large velocity
one should expect that the effective disorder becomes weaker.
Thus the Bose glass driven at low temperature should have
interesting properties such as discussed here. The resulting moving Bose glass
should exhibit a transverse critical force and retain
a transverse Meissner effect in the direction perpendicular to motion.

The novel properties of periodic driven systems discussed in this paper
also suggest many other directions of investigation. 
As for the statics
one outstanding problem is to treat properly dislocations
in the moving glass system. A controlled calculation may seem
out of reach, but on the other hand 
the existence of elastic channels suggests a new and precise
way to look at the problem of generation of dislocations in moving
structures and may provide a starting point \cite{scheidl_dislocations_moving}.
Solving this issue starting
from large velocity is already a formidable task,
but could help us to understand what happens close to the threshold.
Indeed here again only simple cases, inspired from the manifold or CDW with
scalar displacements and no transverse periodicity, have been
considered previously.
As in the statics it is possible that the physics is modified in a quite
surprising way, and certainly all the issues about critical behavior
close to threshold, dynamics reordering, elastic to
plastic motion transitions, will have to be reconsidered. These
issues are of major theoretical concern but also of large
practical importance. Finally we note that though Model I remains
to be tackled in order to reach a complete description 
of the lattice elastic flow, we have shown that the extra
linear terms do not seem to change drastically the
main features of perturbation theory. The KPZ terms remain
to be treated, but an interesting possibility would be that
again because of periodicity their effect would be weaker
than expected.

Another interesting issue is to understand to which extent
a moving, or more generally a non potential system can be glassy.
This concept may seem doomed from the start
since one could conclude that the constant dissipation in the
system would tend to kill glassy properties. However
there too the situation may be more subtle and leave
room for unexpected behavior. We have proposed the moving glass
as a first physical realization of a ``dissipative glass'',
i.e. a glass with a constant dissipation rate in the steady state.
Other realizations of non potential glassy systems have been
studied since, in spin systems \cite{cugliandolo_nonpotential_prl}
or for elastic manifolds in random flows such as polymers
\cite{ledou_polymer_longrange,ledou_wiese_ranflow}.

It is important to characterize these
glassy effects in driven systems. Too close analogy
with glassiness in the statics could be misleading.
As we have discussed it is possible that the existence
of a transverse critical force leads to history dependence effects.
These should be checked. The role of the temperature in
moving systems and its relation to entropy production remains puzzling.
It is natural to expect, as in other related non potential systems
\cite{cugliandolo_nonpotential_prl,ledou_polymer_longrange,ledou_wiese_ranflow}
that the absence of fluctuation dissipation theorem
leads to a generation of a temperature. This is of course what
happens in the RG approach presented here. This heating effect
however is very different from the ``shaking temperature'' 
(since it disappears at $T=0$ in the Moving Bragg glass)
and rather is likely to be related to the entropy production.
Hopefully the methods introduced
here should allow to understand this relation better.
We have found within the FRG that at 
finite temperature the 
physics is controlled by a new non trivial finite 
temperature moving glass fixed point. This result
is strengthened by the fact that another non zero 
$T>0$ fixed point has also been obtained in
the problem of randomly driven polymers, a problem
which does have a dissipative glassy phase
\cite{ledou_wiese_ranflow}. The problem of understanding
dissipative glassy systems is also related to
the study of general {\it non hermitian} random operators.
Indeed non potential dynamical problems
(including e.g the moving glass) are described by 
a Fokker Planck operator whose spectrum is not
necessarily real (by contrast with potential problems
which are purely relaxational).
These Fokker Planck problems with complex spectrum,
( which could be called ``dynamical non hermitian quantum mechanics''!!)
are related to problems which have received a renewed
interest recently (such as vortex lines with tilted
columnar defects \cite{hwa_splay_prl,chen_kpz_dynamics},
spin relaxation in random magnetic fields \cite{mitra_pld_rmn}
diffusion of particles and polymers in random flows
\cite{chalker_diffusion,ledou_diffusion_particle,ledou_wiese_ranflow}.
Exploring this connexion, as well as the very interesting question of 
the classification of these glasses and the 
study of their physical properties is still largely open.

Finally other fascinating open questions remain.
The $T=0$ Moving Bragg glass fixed point, at least within 
Model II, is a time periodic state. The general question
of the stability of periodic attractors towards chaotic
motion is still very much open. It is related to
problems of time coupling and decoupling 
in non linear dynamics, such as synchronisation of
oscillators in josephson arrays, which has been
studied extensively recently
\cite{duwel_strogatz_resonance,strogatz_all_prl,watanabe_strogatz_constants,%
watanabe_strogatz_prl,hadley_beasley_josephson,bonilla_bimodal}
or synchronization by disorder
\cite{wiesenfeld_strogatz_disordered,mousseau_prl}.
The relation between instability to chaos
and possible non perturbative generation of
a temperature is also intriguing. Indeed one
important issue is whether dislocations when present will 
generate an additional temperature or chaos.
Finally, the general question of dynamical elastic instabilities
is also related to recently investigated questions about 
solid friction. It would be interesting to investigate in solid
friction quantities analogous to the transverse response 
and the transverse critical force once the solid is in motion.

We are pleased to acknowledge many useful and extensive discussions
with F. Williams, D. Geshkeinbeim, L. Gurevich
F. Pardo, M. Marchevsky, P. Kes, M. Ocio, A. Kapitulnik,
M. Charalambous, C. Simon. S. Bhattacharya E. Andrei

\appendix

\section{Derivation of the equation of Motion} \label{nonham}

We derive here the contiuum version of the equation of motion for case
where it does not derive from an Hamiltonian. It is then more simple to
perform the continuum limit
in the dynamical Martin-Siggia-Rose (MSR) action. One introduces for
each particle a conjugate field $(i \hat{u}_j)$ and
an interpolating field similar to the one for $u_i(t)$ such that
$\hat{u}(R_j^0 + v t,t) = \hat{u}_j(t)$.
Using these two fields the MSG action reads
\begin{eqnarray}
S &=& \sum_j (i \hat{u}_j) [
\eta \frac{d u_j(t)}{dt} + \frac{\delta H_{el}}{\delta u_j}
- \int_r \partial V(r) \delta(r - R_i^0+vt+u_j(t) ) +
F - \eta v + \zeta_j(t) ] \\
 &=& \int_r \sum_j \delta(r - R_j^0+vt) (i \hat{u}(r,t)) ( \partial_t +
v.\partial_r )u(r,t) -  \\
& & \int_r  (i \hat{u}(r,t))\partial V(r) \sum_j \delta(r - R_j^0+vt+u_j(t) ) +
F - \eta v + \zeta(R_j(t),t)
\end{eqnarray}
Since the interpolating fields are smooth at the scale of the
interparticle distance they have no Fourier components outside the
Brillouin zone and the the kinetic term can be written in an exact
manner
\begin{equation}
\int_{q,BZ} (i \hat{u}(-q,t)) [ ( \partial_t + v.\partial_q ) u(q,t)
+ c(q) u(q,t) ]
\end{equation}
Using the Fourier decomposition of the density
\begin{equation}
\sum_i \delta(r - R_i^0 -vt - u_i(t)) = \text{det}(\partial_\alpha
\phi_\beta (r))
\sum_K e^{ i K ( r - v t - u(\phi(r,t)+v t,t)) }
\end{equation}
where $\phi$ is the labelling field (\ref{label}). The pinning force is
thus given by
\begin{equation} \label{fpin}
F_{\text{pin}} = - \int_r  (i \hat{u}_\alpha(\phi(r,t)+v t,t))
\partial_\alpha V(r) det(\partial_\gamma \phi_\beta (r))
\sum_K e^{ i K ( r - v t - u(\phi(r,t)+v t,t)) }
\end{equation}
Up to now we have made only exact transformations.
Note that the disorder term depends of $u(\phi(r,t)+v t,t)$
rather than $u(r,t)$. This amount in using the origin of the
displacements instead of the actual position of the vortex to label the
displacement fields. In the elastic limit $u(r,t) \ll r$ since
$|u_i - u_{i+1}| \ll a$ where $a$ is the lattice spacing.
One is therefore entitled to replace
$u(\phi(r,t)+v t,t)$ by $u(r,t)$
$\hat{u}(\phi(r,t)+v t,t)$ by $\hat{u}(r,t)$.
An integration by part of (\ref{fpin}) gives back the action coming from
the pinning force (\ref{pinham}) and two additional terms of the form
\begin{equation}
S_a = \int_r V(r) \partial_\alpha \hat{u}_\alpha det(\partial_\gamma \phi_\beta (r))
\sum_K e^{ i K ( r - v t - u(\phi(r,t)+v t,t))}
+ \hat{u}_\alpha \rho_0 \partial_\beta \partial_\alpha u_\beta
+ \hat{u}_\alpha \sum_K e^{ i K ( r - v t - u(\phi(r,t)+v t,t))} i
K_\beta \partial_\alpha u_\beta
\end{equation}
Such term coming from the fact that the continuum limit was performed
after the functional derivation with respect to $u$, contains only
higher gradients, and are negligeable in the elastic limit. They can
in principle however generate relevant terms. All the allowed relevant
terms will be examined in detail in section \ref{beyond}.

\section{First order perturbation theory} \label{appendixb}

\subsection{general analysis}

In this Appendix we study the perturbation theory in the disorder and
compute the effective
action $\Gamma[u,\hat{u}]$ to lowest order (i.e first order) in the
interacting part $S_{int}$, using the standard formula:
\begin{equation} \label{cumulant}
\Gamma[u,\hat{u}] = S_0[u,\hat{u}] + \langle S_{int}[u+\delta u, \hat{u} + \delta  \hat{u}]
 \rangle_{\delta u, \delta \hat{u}}
\end{equation}
where the averages in (\ref{cumulant}) over $\delta u$, $\delta \hat{u}$ are taken with
respect to the free quadratic action
$S_0$ given in (\ref{freemsg}). We will remain as general as possible, in order to treat
several problems and cases simultaneously, and will specify only
at the end to particular cases.
We will thus choose the following disorder term (as it appears
in the MSR action (\ref{totalmsg})).
\begin{eqnarray}
S_{int} = - \frac{1}{2} \int_{r r' t t'} (i \hat{u}^{\alpha}_{rt})
(i \hat{u}^{\beta}_{r't'})
\Delta^{\alpha \beta}(u_{rt} - u_{r't'} + v (t - t'), r-r')
\end{eqnarray}
These allows to treat several problems. It allows to treat
short range correlated disorder keeping the cutoff dependence
which allows to generate the extra linear and KPZ terms (see
section). It also allows to treat correlated
disorder. The disorder correlator will be chosen as:
\begin{eqnarray}
\Delta^{\alpha \beta}(u_{rt} - u_{r't'} + v (t - t'), r-r')
= \sum_K \Delta^{\alpha \beta}_K(r-r') e^{- i K.(u_{rt} - u_{r't'} + v (t - t'))}
\end{eqnarray}
where the symbol $\sum_K$ denotes a discrete sum of lattice
harmonics for a {\it periodic} problem and a continuous sum
$\sum_K \equiv \int d^dk/(2 \pi)^d$ for
a {\it non periodic} manifold.
For the model (\ref{copinfor}) one has:
\begin{eqnarray}
\Delta^{\alpha \beta}_K(r-r') = K_\alpha K_\beta g(r-r') e^{i K (r-r')}
\end{eqnarray}
this is the {\it bare} starting correlator (it will itself be corrected
and will not remain under this form, see below).
In the case of Model II (\ref{copinfor2}), i.e the continumm limit
of the above model, one can replace:
\begin{eqnarray}
g(r-r') e^{i K (r-r')} \to g_K \delta(r-r')
\end{eqnarray}
This is because the scale at which the displacement
field varies is large compared to the correlation length of
the disorder (see discussion of Section ()).
Since we know that non potential terms may be generated
under RG (from FDT violation) we will from the start rather study the
continuous model:
\begin{eqnarray}
\Delta^{\alpha \beta}_K(r-r') \to \Delta^{\alpha \beta}_K \delta(r-r')
\end{eqnarray}
We will work at finite $T$ and also specify to $T=0$.
With these definitions one finds:
\begin{eqnarray} \nonumber \label{msgfirstorder}
\Gamma[u,\hat{u}] & = & S_0 +
\int_{r t} (i \hat{u}^{\alpha}_{rt}) \Sigma^{\alpha}_{rt}[u]
- \frac{1}{2} \int_{r r' t t'}
(i \hat{u}^{\alpha}_{rt}) (i \hat{u}^{\beta}_{r't'})
D^{\alpha \beta}_{rt,r't'}[u]
\end{eqnarray}
with:
\begin{eqnarray}
&& \Sigma^{\alpha}_{rt}[u] = - \int_{r't'}
\sum_K R^{\gamma \beta}_{rt,r't'}
(-i K_\gamma) \Delta^{\alpha \beta}_K(r-r')
e^{- i K.(u_{rt} - u_{r't'} + v(t-t'))}  e^{- \frac{1}{2} K.B_{rt,r't'}.K} \\
&& D^{\alpha \beta}_{rt,r't'}[u] =
\sum_K \Delta^{\alpha \beta}_K(r-r')
e^{- i K.(u_{rt} - u_{r't'} + v(t-t'))}  e^{- \frac{1}{2} K.B_{rr',tt'}.K}
\end{eqnarray}
We have used that $\Delta^{\beta \alpha}_{-K} (r'-r)=\Delta^{\alpha \beta}_K(r-r')$
which comes from simple relabeling in ().
Using time and space translational invariance of the
bare action one can also write:
\begin{eqnarray}
&& \Sigma^{\alpha}_{rt}[u] = \int_{r't'} \Sigma^{\alpha}(u_{rt} - u_{rt'}, t-t', r-r') \\
&& \Sigma^{\alpha}(u_{rt} - u_{r't'}, t-t', r-r') = -
\sum_K R^{\gamma \delta}_{r-r',t-t'}
(-i K_\gamma) \Delta^{\alpha \delta}_K(r-r')
e^{- i K.(u_{rt} - u_{r't'} + v(t-t'))}  e^{- \frac{1}{2} K.B_{r-r',t-t'}.K} \\
&&  D^{\alpha \beta}_{rt,r't'}[u]
= D^{\alpha \beta}(u_{rt} - u_{r't'}, t-t',r-r') = \sum_K \Delta^{\alpha \beta}_K(r-r')
e^{- i K.(u_{rt} - u_{r't'} + v(t-t'))}  e^{- \frac{1}{2} K.B_{r-r',t-t'}.K}
\end{eqnarray}
At $T=0$ it reads simply:
\begin{eqnarray}
\Gamma[u,\hat{u}] = S[u,\hat{u}]
 -  \int_{r r' t t'} (i \hat{u}^{\alpha}_{rt}) R^{\gamma \beta}_{rt,r't'}
\Delta^{\alpha \beta ; \gamma}(u_{rt} - u_{r't'} + v (t - t'), r-r')
\end{eqnarray}
Thus to this order the effect of temperature
amounts to replace everywhere formally:
\begin{eqnarray}
\Delta^{\alpha \delta}_K(r-r') \to \Delta^{\alpha \delta}_K(r-r') e^{- \frac{1}{2} K.B_{r-r',t-t'}.K}
\end{eqnarray}
Temperature has thus two important effects (i) it generates a time dependence
(ii) it smoothes out the disorder. One can already see that there will be two
important different cases. Either (high enough dimension) there is a time persistent part
to the correlator $\lim_{t \to \infty} B_{r,t} = B_{\infty} < + \infty$, in which case
the disorder is smoothed out. Or $\lim_{t \to \infty} B_{r,t} = \infty$ and
the disorder gets smaller at larger scales (low dimension).
Another interpretation of the above result is that the corrected
equation of motion includes (i) a new, non random, time retarded force
$ \Sigma_{\alpha}[u]$ (see () ) (ii) a corrected pinning force which
has an extra time dependence.
\begin{equation}
{(R^{-1})}^{\alpha \beta}_{rt r't'} u^{\beta}_{r't'} = \Sigma_{\alpha}[u] 
+ \tilde{F}_{\alpha} (r,t,u_{rt}) + f_{\alpha} - \eta_{\alpha \beta} v_{\beta}
\end{equation}
where the new (time dependent) pinning force correlator is:
\begin{equation}
\overline{ \tilde{F}_{\alpha} (r,t,u) \tilde{F}_{\beta} (r',t',u') }
= \sum_K \Delta^{\alpha \beta}_K(r-r') e^{- i K.(u_{rt} - u_{r't'} + v (t - t'))}
e^{- \frac{1}{2} K.B_{r-r',t-t'}.K}
\end{equation}
We now separate the relevant contributions in this new, complicated
non linear, equation
of motion. We also define:
\begin{eqnarray}
&& \Sigma^{\alpha \beta}(u_{rt} - u_{rt'}, t-t', r-r')
= \frac{ \delta \Sigma^{\alpha}_{rt} [u]}{\delta u^{\beta}_{r't'}}
=  R^{\gamma \delta}_{rt,rt'} (
\langle \Delta^{\alpha \delta ; \gamma \beta}(u_{rt} - u_{rt'}) \rangle
- \delta_{tt'} \int_{t''}
R^{\gamma \delta}_{rt,rt''}
\langle \Delta^{\alpha \delta ; \gamma \beta}(u_{rt} - u_{rt''}) \rangle )
\end{eqnarray}
On the functional expression (\label{msgfirstorder}) we can identify the corrections
to various terms. The first thing to do is to obtain the corrected response
and correlation functions. For that one simply has to
expand (\label{msgfirstorder}) in powers of $u$ and $\hat{u}$
up to quadratic order. This yields $\Gamma^{1}$ and
$\Gamma^{2}$ respectively the linear and quadratic part.
The linear term proportional to $\hat{u}$
gives the correction to the force $\delta f^{\alpha} = - \Sigma^{\alpha}[u=0]$.

\subsubsection{linear part of the effective action: correction to the
force (or velocity)}
The linear term in the effective action in (\ref{totalmsg}) becomes:
\begin{eqnarray}
&& \int_{rt} (i \hat{u}^\alpha_{rt})
- ( f_{\alpha} - \eta_{\alpha \beta} v_{\beta} + \delta f_{\alpha}(v))
\end{eqnarray}
where the correction to the force is given by:
\begin{eqnarray}
\delta f_{\alpha}(v) =  \int_{r,t} R^{\gamma \beta}_{r,t}
\sum_K
(-i K_\gamma) \Delta^{\alpha \beta}_K(r) e^{-i K.v t} e^{- \frac{1}{2} K.B_{r,t}.K}
\end{eqnarray}
At $T=0$ this can also be written as:
\begin{eqnarray}
\delta f_{\alpha}(v) &=& - \int d\tau dr
R^{\gamma \delta}(r,\tau) \Delta^{\alpha \delta ; \gamma}(v \tau, r)
\end{eqnarray}

\subsubsection{quadratic part of the effective action: correction to
the response and correlation fonction}

The quadratic part of the effective action reads:

\begin{eqnarray} \nonumber \label{msgfirstorder2}
\Gamma^2[u,\hat{u}] & = & S^2_0 + \int_{r r' t t'}
 (i \hat{u}^{\alpha}_{rt})
\Sigma^{\alpha \beta}(0,t-t',r-r') u^\beta_{r't'}
- \frac{1}{2} \int_{r r' t t'}
(i \hat{u}^{\alpha}_{rt}) (i \hat{u}^{\beta}_{r't'})
D^{\alpha \beta}(0,t-t',r-r')
\end{eqnarray}

\bigskip

(i) {\it response function}

\medskip

The correction to the response fonction is thus:
\begin{equation}
{(\delta R^{-1})}^{\alpha \beta}(q,\omega)= -
\int_{r t} (1 - e^{i(q.r + \omega t)})
R^{\gamma \delta}(r,t) K_\gamma K_\delta \Delta^{\alpha \delta}_K(r)
e^{- i K.v t} e^{- \frac{1}{2} K.B_{r,t}.K}
\end{equation}
at $T=0$ it reads:
\begin{equation}
{(\delta R^{-1})}^{\alpha \beta}(q,\omega)=
\int_{r t} (1 - e^{i(q.r + \omega t)})
R^{\gamma \delta}(r,t) \Delta^{\alpha \delta ; \gamma \beta}(v t, r)
\end{equation}
It is useful to perform the small $q,\omega$ expansion
to obtain the corrections to the friction coefficient, the
linear terms and the elastic matrix. From the general equation
of motion () one has:
\begin{eqnarray}
{(\delta R^{-1})}^{\alpha \beta}(q,\omega)=
(i \omega) \delta \eta_{\alpha \beta} +
(i q_\rho) \delta L^\rho_{\alpha \beta} +
q_\rho q_\sigma \delta C^{\rho \sigma}_{\alpha \beta} + h.o.t
\end{eqnarray}
One finds:
\begin{eqnarray}   \label{resultapp}
&& \delta \eta_{\alpha \beta} =
\int_{rt} t R^{\gamma \delta}_{rt} K_\gamma K_\beta
\Delta^{\alpha \delta}_K(r) e^{- i K.v t} e^{- \frac{1}{2} K.B_{r,t}.K} \\
&& \delta L^\rho_{\alpha \beta} =
\int_{rt} r^\rho R^{\gamma \delta}_{rt} K_\gamma K_\beta
\Delta^{\alpha \delta}_K(r) e^{- i K.v t} e^{- \frac{1}{2} K.B_{r,t}.K} \\
&& \delta C^{\rho \sigma}_{\alpha \beta} = \frac{1}{2}
\int_{rt} r^\rho r^\sigma R^{\gamma \delta}_{rt} K_\gamma K_\beta
\Delta^{\alpha \delta}_K(r) e^{- i K.v t} e^{- \frac{1}{2} K.B_{r,t}.K} \\
\end{eqnarray}
This is a very general expression which we will
particularize to special cases below. At $T=0$ one has:
\begin{equation}
\delta \eta^{\alpha \beta}(v) = - \int d\tau dr \tau
R^{\gamma \delta}(r,\tau) \Delta^{\alpha \delta ; \gamma \beta}(v \tau, r)
\end{equation}

Note that $d \delta f_{\alpha}(v)/d v_{\beta}  =  - \delta \eta_{\alpha \beta}(v)$.
In the limit $v \to 0$ one finds $\delta f_{\alpha}(v) \sim 
- \delta \eta_{\alpha \beta} v_{\beta}$
using an integration by part, provided the function $\Delta$ is analytic.
Non analytic $\Delta$ yields a critical force.

\bigskip

(ii) {\it correlation functions}

\medskip

The complete $\hat{u} \hat{u}$ term in the quadratic part of the effective action
is $D^{\alpha \beta}(0,t-t',r-r')$. It allows to compute the corrected
correlation functions using the corrected response function:

\begin{eqnarray}
< u^\alpha_{-q,-\omega} u^\beta_{q,\omega} >
= [ R^{(1)}(\omega,q).(2 \eta T + D(0,\omega,q)).R^{(1)}(-\omega,-q)) ]_{\alpha \beta} \\
R^{(1)}(\omega,q) = \frac{1}{R^{-1}(\omega,q)  + \delta R^{-1}(\omega,q)}
\end{eqnarray}

where $.$ denotes the matrix multiplication of indices. Examining the large time
and space behaviours one finds that there are two important corrections:

(i) a correction to the temperature (from the equal time
piece):

\begin{eqnarray}
\delta (\eta T)_{\alpha \beta} = \frac{1}{2} \int_{r t} ( D^{\alpha \beta}(0,t,r)
- D^{\alpha \beta}(0,t=+\infty,r) )
\end{eqnarray}

(ii) a correction to a static random force. It is identified as
the time persistent part of the disorder:

\begin{eqnarray}
D^{\alpha \beta}(0) = \lim_{t \to \infty} \int_{r t} D^{\alpha \beta}(0,r)
\end{eqnarray}
It yields to a static part $\delta(\omega)$ in the displacement correlation.
Note however that there is additional important corrections
to the {\it non linear} part of the disorder, which we now
identify.

\subsubsection{non linear terms in the effective action: correction to
the disorder and generation of KPZ terms}

It turns out that it is important to follow not just the
random force but the complete non linear static part of the
disorder term. It is identified as:
\begin{eqnarray}
\lim_{t - t' \to \infty}
D^{\alpha \beta}(u_{rt} - u_{r't'}, t-t',r-r') =  \lim_{t - t' \to \infty}
\sum_K \Delta^{\alpha \beta}_K(r-r')
e^{- i K.(u_{rt} - u_{r't'} + v(t-t'))}  e^{- \frac{1}{2} K.B_{r-r',t-t'}.K}
\end{eqnarray}

Fianlly, the non linear KPZ terms can be easily seen to be generated
already to this order. Expanding  $\Sigma^{\alpha}_{rt}[u]$
to second order in the field $u$ one finds:

\begin{eqnarray}
\delta K^{\rho \sigma}_{\alpha \beta \gamma} 
= \frac{1}{2}
\int_{rt} r^\rho r^\sigma R^{\epsilon \delta}_{rt} K_\beta K_\gamma (i K_\epsilon)
\Delta^{\alpha \delta}_K(r) e^{- i K.v t} e^{- \frac{1}{2} K.B_{r,t}.K} \\
\end{eqnarray}

\subsection{explicit evaluation of the corrections in specific models}

\subsubsection{evaluation in Model II}

We first study the continuous Model II valid in the elastic
limit. It is obtained by the substitution:
\begin{eqnarray}
\Delta^{\alpha \beta}_K(r-r') \to \Delta^{\alpha \beta}_K \delta(r-r')
\end{eqnarray}
One finds first that:
\begin{eqnarray}
\delta L^\rho_{\alpha \beta} = 0 ~~,~~
\delta C^{\rho \sigma}_{\alpha \beta} = 0 ~~,~~
\delta K^{\rho \sigma}_{\alpha \beta \gamma} = 0
\end{eqnarray}
i.e that no linear, KPZ terms are generated, and that
there is no correction to the static part of the response fonction.
These are consequences of the statistical tilt symmetry
(see Section()). The only corrections are:
\begin{eqnarray}   \label{corrcont}
&& \delta f_{\alpha}(v) =  \int_{t} R^{\gamma \beta}_{r=0,t}
\sum_K
(-i K_\gamma) \Delta^{\alpha \beta}_K e^{-i K.v t} e^{- \frac{1}{2} K.B_{0,t}.K} \\
&& \delta \eta_{\alpha \beta} =
\int_{rt} t R^{\gamma \delta}_{r=0,t} K_\gamma K_\beta
\Delta^{\alpha \delta}_K e^{- i K.v t} e^{- \frac{1}{2} K.B_{0,t}.K} \\
\end{eqnarray}
For the bare problem one can further substitute $\Delta^{\alpha \delta}_K
= K_\alpha K_\beta g_K$. At $T=0$ it further simplifies as:
\begin{eqnarray}
&& \delta f_{\alpha}(v) =  \sum_K  (-i K_\gamma) \Delta^{\alpha \beta}_K
\int_{q} R^{\gamma \beta}_{q,\omega=K.v} \\
&& \delta \eta_{\alpha \beta} = \sum_K K_\gamma K_\beta
\Delta^{\alpha \delta}_K
\int_{q} \frac{\partial}{\partial \omega} R^{\gamma \delta}_{q,\omega} |_{\omega=K.v} \\
\end{eqnarray}
Note the symmetries $\eta_{\alpha \beta}(-v)=\eta_{\alpha \beta}(v)$ and
$\delta f_{\alpha}(-v)=-\delta f_{\alpha}(v)$.

Let us further specify to the problem to a periodic lattice.
The bare perturbation theory (i.e starting from $\eta_{\alpha \beta}=\eta_0$)
gives:

\begin{eqnarray} \label{hauger0}
&& \delta f_\alpha(v) = - \sum_K  \sum_{I=L,T} \int_{BZ}  dq
K_\alpha (K.P^I(q).K) g_{K} \frac{ v.(K+q) }{ c_I(q)^2 + (\eta_0 v.(K+q))^2 } \\
&& \delta \eta_{\alpha \beta} = \sum_K \sum_{I=L,T} \int_{BZ}  dq
K_\alpha K_\beta (K.P^I(q).K) g_{K}
\frac{1}{ ( c_I(q) + i \eta_0 v.(K+q))^2 }
\end{eqnarray}

One can also look at the dressed perturbation theory (i.e adding
from the start the terms which will be generated).

Suppose the velocity along a principal lattice direction $x$.
Then the symmetry $y \to -y$ ensures in (\ref{corrcont})
that to all orders $\eta_{x y}=0$. However $\eta_{x x}$ and $\eta_{yy}$
will in general be different. Thus the tensor to be inverted
will be (for $d=2$):

\begin{eqnarray}
(\eta_{yy} \delta_{\alpha \beta} + (\eta_{xx} - \eta_{yy}) e^\alpha_x e^\beta_x)
(i \omega) + i v q_x \delta_{\alpha \beta} +
c_L q_\alpha q_\beta + c_T (\delta_{\alpha \beta} q^2 - q_\alpha q_\beta)
\end{eqnarray}

which gives:

\begin{eqnarray}
\frac{P^T(q)_{\alpha \beta}}{c_T(q) + i \eta_{yy} \omega + i v q_x} +
\frac{P^L(q)_{\alpha \beta}}{c_L(q) + i \eta_{yy} \omega + i v q_x} +
\frac{(\eta_{yy} - \eta_{xx})
i \omega e^\alpha_x e^\beta_x }{c_T(q) + i \eta_{xx} \omega + i v q_x} +
\end{eqnarray}

The correlation functions can also be computed
using that here $D^{\alpha \beta}(0, t-t',r-r') = D^{\alpha \beta}(t-t') \delta(r-r')$.
Staring with the bare action and to lowest order
in disorder () yields at $T=0$:

\begin{eqnarray}
< u^\alpha_{-q,-\omega} u^\beta_{q,\omega} >
= \sum_K  \sum_{I=L,T, I'=L,T} \int_{q,BZ} (2 \pi) \delta(\omega - K.v) g_K
K_\gamma K_\delta
\frac{P^I_{\alpha \gamma}(q)}{c_I(q) + i \eta_0 (\omega + v.q)}
\frac{P^{I'}_{\beta \delta}(q)}{c_I'(q) - i \eta_0 (\omega + v.q)}
\end{eqnarray}

it is a sum of oscillating functions, plus a static one.

At $T>0$ the expression is more complicated. It reads:

\begin{eqnarray}
&& \delta < u^\alpha_{-q,-\omega} u^\beta_{q,\omega} >
= \sum_K  \sum_{I=L,T, I'=L,T}
\int_{q,BZ}  (2 \pi) \delta(\omega - K.v)
\frac{P^I_{\alpha \gamma}(q)}{c_I(q) + i \eta_0 (\omega + v.q)}
\frac{P^{I'}_{\beta \delta}(q)}{c_I'(q) - i \eta_0 (\omega + v.q)}
D_{\gamma \delta}(\omega) \\
&& +
\frac{P^I_{\alpha \gamma}(q)}{c_I(q) + i \eta_0 (\omega + v.q)}
\frac{P^{I'}_{\beta \delta}(q)}{c_I'(q) - i \eta_0 (\omega + v.q)}
2 i \eta T (\omega + v.q)
\delta \eta_{\gamma \delta}
\\
&& D_{\gamma \delta}(\omega) = \int_t g_K K_\gamma K_\delta
e^{- \frac{1}{2} K.B_{0,t-t'}.K}
e^{i \omega t - i K.v(t-t')}
\end{eqnarray}

\subsection{Corrections in Model I : linear and KPZ terms}  \label{linear}

We now show explicitly on the first order
perturbation theory that indeed linear terms of the form
$C^{\alpha \beta}_{\sigma} (i \hat{u}^{\alpha}_{rt}) \partial_{\sigma} u^{\beta}_{rt}$
with:
\begin{equation}
C^{\alpha \beta}_{\sigma} = - \int d\tau dr
r_{\sigma} R^{\gamma \delta}(r,\tau) \Delta^{\alpha \delta ; \gamma \beta}(v \tau, r)
\end{equation}
We now compute the various terms:
\begin{equation}
C^{\alpha \beta}_{\sigma} = - \sum_K
(i K_{\alpha}) (i K_{\beta}) (i K_{\gamma}) (i K_{\delta})
\int d\tau dr
r_{\sigma} R^{\gamma \delta}(r,\tau) g(r) e^{i K. (r - v \tau )}
\end{equation}
\begin{equation}
\delta f_{\alpha}(v) = \sum_K
(i K_{\alpha}) (i K_{\delta}) (i K_{\gamma}) e^{i K. (r - v \tau )}
\int d\tau dr g(r)
R^{\gamma \delta}(r,\tau) e^{i K. (r - v \tau )}
\end{equation}
\begin{equation}
\delta \eta^{\alpha \beta}(v) = - \sum_K
(i K_{\alpha}) (i K_{\beta}) (i K_{\gamma}) (i K_{\delta})
\int d\tau dr \tau g(r)
R^{\gamma \delta}(r,\tau)  e^{i K. (r - v \tau )}
\end{equation}
If one assumes isotropy to start with one has:
\begin{equation}
C^{\alpha \beta}_{\sigma} =  \sum_K
(i K_{\alpha}) (i K_{\beta}) K^2
\int d\tau dr
r_{\sigma} R(r,\tau) g(r) e^{i K. (r - v \tau )}
\end{equation}
\begin{equation}
\delta f_{\alpha}(v) = - \sum_K
(i K_{\alpha}) K^2
\int d\tau dr g(r)
R(r,\tau) e^{i K. (r - v \tau )}
\end{equation}
\begin{equation}
\delta \eta^{\alpha \beta}(v) =  \sum_K K^2
(i K_{\alpha}) (i K_{\beta})
\int d\tau dr \tau g(r)
R(r,\tau)  e^{i K. (r - v \tau )}
\end{equation}

Let us now study the case where $v$ is along a principal lattice direction,
assuming it remains so by renormalization (to be checked).
One finds by symmetry that only $C^{yy}_x=v_1$, $C^{yx}_x=v_2$, $C^{xx}_x=v_3$,
$C^{xy}_y=v_4$ are non zero.
Note that by symmetry $y \to -y$ $K_y \to -K_y$ one has
$\delta \eta_{xy}=0$, $\delta \eta_{yx}=0$, $\delta f_y=0$

One finds:

\begin{eqnarray}
C^{yy}_x= v_1 = \eta v - \sum_K \int_{\tau r} x K^2 K_y^2  R(r,\tau) g(r) e^{i K. (r - v \tau )}
\end{eqnarray}

\begin{eqnarray}
C^{yx}_y= v_2 = - \sum_K \int_{\tau r} y K^2 K_y K_x  R(r,\tau) g(r) e^{i K. (r - v \tau )}
\end{eqnarray}

\begin{eqnarray}
C^{xx}_x= v_3 = \eta v - \sum_K \int_{\tau r} x K^2 K_x^2  R(r,\tau) g(r) e^{i K. (r - v \tau )}
\end{eqnarray}

\begin{eqnarray}
C^{xy}_y= v_4 = - \sum_K \int_{\tau r} y K^2 K_y K_x  R(r,\tau) g(r) e^{i K. (r - v \tau )}
\end{eqnarray}

Note that $v_2=v_4$ exactly and that to lowest order in $v$
one has $v2=v3$ and $\eta+\delta \eta_{xx}(v=0)=\eta+\delta \eta_{yy}(v=0)$.

\begin{equation}
\delta \eta_{xx} (v) =  - \sum_K K^2 K_x^2
\int d\tau dr \tau g(r)
R(r,\tau)  e^{i K. (r - v \tau )}
\end{equation}

\begin{equation}
\delta \eta_{yy} (v) =  - \sum_K K^2 K_y^2
\int d\tau dr \tau g(r)
R(r,\tau)  e^{i K. (r - v \tau )}
\end{equation}

\begin{equation}
\delta f_x(v) =  - \sum_K K^2 i K_x
\int d\tau dr g(r)
R(r,\tau)  e^{i K. (r - v \tau )}
\end{equation}

Note that $v \eta(v)$ has a maximum - this may be
related to the two branches of solid friction
(dynamical friction smaller than static one).

\section{Dynamical effective action to second order and
analysis of divergences} \label{appendixc}

In this Appendix we obtain the perturbative expression
of the effective dynamical action to second order in disorder.
At each step we will remain as general as possible so that
our expressions can be applied to study a large class of models
and situations. Then we will then study particular situations
and identify the terms which correct the bare disorder
by performing a short distance or time expansion. We will
focus mainly on divergences occuring near $d=4$ (for $v=0$)
and $d=3$ (for $v >0$). A similar study in $d=2$ was
performed in \cite{carpentier_ledou_triangco}.
Note that the operators are local in $r$ but non local in time,
which makes the expansion more involved.

Note that we will study here a priori both the periodic
manifold case or the non periodic one. The only difference
is that in the periodic case one has discrete $\sum_K$
to be replaced by $\int_K$ in the continuous case.

The effective action to
second order in the interaction term is \cite{goldschmidt_dynamics_flux}:

\begin{equation}
-2 \Gamma^{(2)}[W] =  \langle S_{int}[W+\delta W]^2  \rangle_{\delta W}
-   \langle S_{int}[W+\delta W] \rangle^2_{\delta W} -
\langle \frac{\delta S_{int}[W+\delta W]}{\delta W} \rangle_{\delta W}  G
\langle \frac{\delta S_{int}[W+\delta W]}{\delta W} \rangle_{\delta W}
\end{equation}

with $W=(u,\hat{u})$ and $\delta W=(\delta u,\delta \hat{u})$
and a gaussian average over $\delta W$ is performed using
the bare quadratic action $S_0 + S_2$ in (\ref{action2}).
The last term merely ensures
that all one particle reducible diagrams be absent.

One thus has to study all possible Wick contractions of
the two vertex operators:

\begin{equation}
i \hat{u}^{\alpha_1}_{r_1 t_1} i \hat{u}^{\beta_1}_{r_1,t'_1}
\Delta_{\alpha_1 \beta_1}(u_{r_1,t_1} - u_{r_1,t'_1} + v (t_1 - t'_1))
\Delta_{\alpha_2 \beta_2}(u_{r_2 t_2} - u_{r_2,t'_2} + v (t_2 - t'_2))
i \hat{u}^{\alpha_2}_{r_2 t_2} i \hat{u}^{\beta_2}_{r_2,t'_2}
\end{equation}

imposing that at least two contractions joining the two vertices 1 and 2.

We will temporarily use the shorthand notation $U_{rtt'}=u_{rt} - u_{rt'} + v(t-t')$.
A tedious calculation gives, for the $i \hat{u} i \hat{u}$ term
in the effective action (in short notations, omitting
temporarily the $r$ variable and all integrals):

\begin{eqnarray}
&& -2 \Gamma =
2 (i \hat{u}^{\alpha_1}_{t_1}) (i \hat{u}^{\alpha_2}_{t_2})
\langle \Delta_{\alpha_1 \beta_1} (U_{t_1,t'_1})
\Delta_{\alpha_2 \beta_2; \gamma \delta} (U_{t_2,t'_2}) \rangle_c
R^{\delta \beta_2}_{t_2 t'_2}
( R^{\gamma \beta_1}_{t_2 t'_1} - R^{\gamma \beta_1}_{t'_2 t'_1} )
\nonumber \\
&& +
\frac{1}{2} (i \hat{u}^{\alpha_2}_{t_2}) (i \hat{u}^{\beta_2}_{t'_2})
\langle \Delta_{\alpha_1 \beta_1} (U_{t_1,t'_1})
\Delta_{\alpha_2 \beta_2; \gamma \delta} (U_{t_2,t'_2}) \rangle
(  R^{\gamma \alpha_1}_{t_2 t_1} - R^{\gamma \alpha_1}_{t'_2 t_1} )
(  R^{\delta \beta_1}_{t_2 t'_1} - R^{\delta \beta_1}_{t'_2 t'_1} )
+
\nonumber \\
&&
(i \hat{u}^{\alpha_1}_{t_1}) (i \hat{u}^{\alpha_2}_{t_2})
\langle \Delta_{\alpha_1 \beta_1; \delta} (U_{t_1,t'_1})
\Delta_{\alpha_2 \beta_2; \gamma} (U_{t_2,t'_2}) \rangle
( R^{\gamma \beta_1}_{t_2 t'_1} - R^{\gamma \beta_1}_{t'_2 t'_1} )
( R^{\delta \beta_2}_{t_1 t'_2} - R^{\delta \beta_2}_{t'_1 t'_2} )
\nonumber \\
&& +
\langle\Delta_{\alpha_1 \beta_1; \delta} (U_{t_1,t'_1})
\Delta_{\alpha_2 \beta_2; \gamma} (U_{t_2,t'_2})\rangle_c
(i \hat{u}^{\alpha_1}_{t_1}) (i \hat{u}^{\beta_1}_{t'_1})
R^{\gamma \beta_2}_{t_2 t'_2}
( R^{\delta \alpha_2}_{t_1 t_2} - R^{\delta \alpha_2}_{t'_1 t_2} )
\end{eqnarray}

where the symbol $<F[u]>$ means $<F[u+\delta u]>_{\delta u}$ and
$<..>_c$ denotes a connected average between the vertices,
i.e $<F[u_1] G[u_2] >_c = <F[u_1] G[u_2] > - <F[u_1]> <G[u_2] >$.
Note that simplifications will occur in the particular case
$T=0$ since the connected terms then
vanish identically, and one can also drop
the averages $<F[u]>=F(u)$.

Using the assumption of time and space translational invariance it can be put
under the form:

\begin{eqnarray}
\Gamma =
- \frac{1}{2} \int_{r r' t_1 t_2} (i \hat{u}^{\alpha}_{r t_1}) (i \hat{u}^{\beta}_{r+r',t_2})
\delta \Delta^{\alpha \beta}_{r'}
\end{eqnarray}

as a sum of four terms:
$\delta \Delta = \sum_{i=1,4} \delta \Delta^{(i)}_{eff}$:

\begin{eqnarray} \label{terms}
\delta \Delta^{(1)}_{r'} &=&
2 R^{\delta \lambda}(\tau_2,0)
R^{\gamma \rho}(\tau_1,r')
\langle \Delta_{\beta \lambda; \gamma \delta} (U_{r+r', t_2,t_2-\tau_2})
\nonumber \\
&&
[ \Delta_{\alpha \rho} (U_{r,t_1,t_2 -\tau_1})
- \Delta_{\alpha \rho} (U_{r,t_1,t_2 -\tau_1 - \tau_2}) ] \rangle_c
\nonumber \\
\delta \Delta^{(2)}_{r''} &=&
\frac{1}{2} \delta(r'') R^{\gamma \rho}(\tau,- r')
R^{\delta \lambda}(\tau',- r')
\langle \Delta_{\alpha \beta ; \gamma \delta} (U_{r,t_1,t_2})
\nonumber \\
&&
[ \Delta_{\rho \lambda}(U_{r+r', t_1-\tau, t_1-\tau'}) +
\Delta_{\rho \lambda}(U_{r+r', t_2-\tau, t_2-\tau'}) -
\nonumber \\
&&
\Delta_{\rho \lambda}(U_{r+r', t_1-\tau, t_2-\tau'}) -
\Delta_{\rho \lambda}(U_{r+r', t_2-\tau, t_1-\tau'}) ] \rangle
\nonumber \\
\delta \Delta^{(3)}_{r'} &=&
R^{\gamma \rho}(\tau_2, r')
R^{\delta \lambda}(\tau_1,- r') [
\langle \Delta_{\alpha \rho; \delta}(U_{r,t_1,t_2 - \tau_2})
\Delta_{\beta \lambda; \gamma}(U_{r+r',t_2, t_1 - \tau_1}) \rangle -
\nonumber \\
&&
\langle \Delta_{\alpha \rho; \delta}(U_{r,t_1, t_1 -\tau_1 - \tau_2})
\Delta_{\beta \lambda; \gamma}(U_{r+r',t_2, t_1 - \tau_1}) \rangle -
 \nonumber \\
&& \Delta_{\alpha \rho; \delta}(U_{r,t_1, t_2 - \tau_2})
\Delta_{\beta \lambda; \gamma}(U_{r+r',t_2, t_2 - \tau_1 - \tau_2}) \rangle ]
\nonumber \\
\delta \Delta^{(4)}_{r''} &=& \delta(r'')
R^{\gamma \rho}(\tau_2, 0)
R^{\delta \lambda}(\tau_1,- r')
\langle \Delta_{\alpha \beta; \delta} (U_{r,t_1, t_2})
\nonumber \\
&& [
\Delta_{\lambda \rho; \gamma}(U_{r+r', t_1-\tau_1, t_1-\tau_1-\tau_2}) -
\Delta_{\lambda \rho; \gamma}(U_{r+r', t_2-\tau_1, t_2-\tau_1-\tau_2}) ] \rangle_c
\end{eqnarray}

Note that some terms have vanished from the Ito time discretization property
$ R^{\gamma \beta_1}_{t'_2 t'_1} R^{\delta \beta_2}_{t'_1 t'_2} =0$

We have written these terms in that form for simplicity,
but one must keep in mind that in addition they
must be symmetrized under $\alpha \to \beta$ and
$r \to -r$ when necessary.

Up to now this is very general. We will now consider several cases.

\subsection{static degrees of freedom at zero temperature}

In this subsection we first set $T=0$, and thus
$\delta \Delta^{(1)}=\delta \Delta^{(4)}=0$ and
other averages can be dropped. Here we will also only study
static disorder and we will thus drop the $v(t-t')$ terms
thus setting $U_{rtt'} = u_{rt} - u_{rt'}$ in all above formulae.
The bare response function $R^{\alpha \beta}(\tau,r)$
remains however arbitrary. In the moving lattice problem this amounts
to restrict ourselves to the modes $K_x=0$, which is
of interest for studying the {\it transverse components}
$u.v=0$ which see only a static disorder
(assuming they can be decoupled) and $v$ still appears in the
response function. Since we will keep $u$ as a vector with
arbitrary number of components, the equations that we will obtain can
be applied to other problems with static disorder (e.g the
usual manifold case $v=0$, periodic or not, non potential
problems etc..).

The only remaining terms are:

\begin{eqnarray}
\delta \Delta^{(2)}_{r''} &=&
\frac{1}{2} \delta(r'') \int_{r',\tau,\tau'}
R^{\gamma \rho}(\tau,- r')
R^{\delta \lambda}(\tau',- r')
\Delta_{\alpha \beta ; \gamma \delta} (u_{r t_1} - u_{r t_2})
\nonumber \\
&&
[ \Delta_{\rho \lambda}(u_{r+r', t_1-\tau} - u_{r+r', t_1-\tau'}) +
\Delta_{\rho \lambda}(u_{r+r', t_2-\tau} - u_{r+r', t_2-\tau'}) -
\nonumber \\
&&
\Delta_{\rho \lambda}(u_{r+r', t_1-\tau} - u_{r+r', t_2-\tau'}) -
\Delta_{\rho \lambda}(u_{r+r', t_2-\tau} - u_{r+r', t_1-\tau'}) ]
\nonumber \\
\delta \Delta^{(3)}_{r'} &=&
R^{\gamma \rho}(\tau_2, r')
R^{\delta \lambda}(\tau_1,- r') [
\Delta_{\alpha \rho; \delta}(u_{r,t_1} - u_{r, t_2 - \tau_2})
\Delta_{\beta \lambda; \gamma}(u_{r+r',t_2} - u_{r+r', t_1 - \tau_1})  -
\nonumber \\
&&
 \Delta_{\alpha \rho; \delta}(u_{r,t_1} - u_{r, t_1 -\tau_1 - \tau_2})
\Delta_{\beta \lambda; \gamma}(u_{r+r',t_2} - u_{r+r', t_1 - \tau_1}) -
 \nonumber \\
&& \Delta_{\alpha \rho; \delta}(u_{r,t_1} - u_{r, t_2 - \tau_2})
\Delta_{\beta \lambda; \gamma}(u_{r+r',t_2} - u_{r+r', t_2 - \tau_1 - \tau_2})  ]
\nonumber \\
\end{eqnarray}

It is then easy to perform a short time, short distance operator
expansion in the variables $r',\tau_1,\tau_2$. This yields,
up to higher order irrelevant gradient terms, the following
total correction to the random force correlator:

\begin{eqnarray}
\delta \Delta_{\alpha \beta}(u) &=& \Delta_{\alpha \beta ; \gamma \delta}(u)
( \Delta_{\alpha' \beta'}(0)  - \Delta_{\alpha' \beta'}(u) )
\int_r G_{\gamma \alpha'}(r) G_{\delta \beta'}(r)
\nonumber \\
&& - \Delta_{\alpha \alpha' ; \delta}(u)
\Delta_{\beta \beta' ; \gamma}(u)
\int_r G_{\gamma \alpha'}(r) G_{\delta \beta'}(-r)
\end{eqnarray}

where we have defined the static response
$G(r) = \int_0^{\infty} d\tau R(\tau,r)$. Note that this
formula is valid for a large class of models. It does {\it not}
suppose for instance that the random force correlator is
the second derivative of a random potential.

It is important to note that the condition that the random force
is the gradient of a potential, i.e,
$\Delta_{\alpha \beta}(u) = - \partial_\alpha \partial_\beta R(u)$
where $R(u)$ is the correlator of the random potential
(not to be confused with the response function !).
is true only when $G(r) = G(-r)$. Indeed, in that case, assuming
the symmetry that $G_{\alpha \beta}(r)=G_{\beta \alpha}(r)$
one finds:

\begin{eqnarray}
\delta R(u) =
\frac{1}{2} R_{; \gamma \delta}(u) R_{;\alpha' \beta'}(u) 
- R_{; \gamma \delta}(u) R_{;\alpha' \beta'}(0)
\int_r G_{\gamma \alpha'}(r) G_{\delta \beta'}(r)
\end{eqnarray}

If $G(r) \ne G(-r)$ a non potential part is generated to
the disorder. If $u$ has only $n=1$ component it will
remain a derivative. If we study models with $n > 1$ component fields and
a non FDT response function we will generate nonpotential
disorder.

Higher derivatives terms have been neglected. In the periodic
case, one of them is the so-called annihilation term
(see  \cite{carpentier_ledou_triangco}) which
produces a random force term:

\begin{eqnarray}
\Gamma =
- \frac{1}{2} \int_{r t_1 t_2} (i \hat{u}^{\alpha}_{r t_1})
\nabla^2_{r} (i \hat{u}^{\beta}_{r,t_2})
\end{eqnarray}

This term is important in $d=2$ (the so called Cardy Ostlund
term) but irrelevant for $d>2$.

\subsection{full dynamical problem at zero temperature}

In this subsection we still set $T=0$ leading to the
same simplifications as in the previous subsection, but
we keep the $v(t-t'$ terms. So we are studying the
full dynamical problem of a driven lattice (i.e with
transverse and longitudinal displacement fields).

The effective action is the sum of the following two terms:

\begin{eqnarray}
&& \Gamma_1 =
- \frac{1}{4} \int_{r r' t t' \tau \tau'}
(i \hat{u}^{\alpha}_{r t}) (i \hat{u}^{\beta}_{r t'})
R^{\gamma \rho}(\tau,r')
R^{\delta \lambda}(\tau',r')
\Delta_{\alpha \beta ; \gamma \delta} (u_{r t} - u_{r t'} + v(t-t'))
\nonumber \\
&&
[ \Delta_{\rho \lambda}(u_{r-r', t-\tau} - u_{r-r', t-\tau'} + v(\tau'-\tau)) +
\Delta_{\rho \lambda}(u_{r-r', t'-\tau} - u_{r-r', t'-\tau'} + v(\tau'-\tau)) -
\nonumber \\
&&
\Delta_{\rho \lambda}(u_{r-r', t-\tau} - u_{r-r', t'-\tau'}+v(t-t'+\tau'-\tau)) -
\Delta_{\rho \lambda}(u_{r-r, t'-\tau} - u_{r-r', t-\tau'} +v(t'-t+\tau'-\tau)) ]
\nonumber \\
&& \Gamma_2 = - \frac{1}{2}
(i \hat{u}^{\alpha}_{r t}) (i \hat{u}^{\beta}_{r+r', t'})
R^{\gamma \rho}(\tau, r')
R^{\delta \lambda}(\tau',- r')
[
\Delta_{\alpha \rho; \delta} (u_{r,t} - u_{r, t' - \tau} + v(t-t'+\tau))
\nonumber \\
&&
( \Delta_{\beta \lambda; \gamma} (u_{r+r',t'} - u_{r+r', t - \tau'} + v(t'-t+\tau'))
- \Delta_{\beta \lambda; \gamma} (u_{r+r',t'} - u_{r+r', t' - \tau - \tau'} +v(\tau+\tau')) )
 \nonumber \\
&& - \Delta_{\alpha \rho; \delta} (u_{r,t} - u_{r, t - \tau - \tau'}+v(\tau+\tau'))
\Delta_{\beta \lambda; \gamma} (u_{r+r',t'} - u_{r+r', t - \tau'}+v(t'-t+\tau')) ]
\end{eqnarray}

We can now perform a short distance and time expansion
and compute the correction to the
random force correlator. Expressed as
$\Delta_{\alpha \beta}(U)$, with
$U=u-u'+v(t-t')$ it reads:

\begin{eqnarray}
\delta \Delta_{\alpha \beta}(U) &=& \int_{q,\tau\tau'}
\Delta_{\alpha \beta ; \gamma \delta} (U)
R^{\gamma \rho}(\tau,q)
R^{\delta \lambda}(\tau',-q)
\nonumber \\
&&
[ \Delta_{\rho \lambda}(v(\tau'-\tau)) - \frac{1}{2} (
\Delta_{\rho \lambda}(U+v(\tau'-\tau)) +
\Delta_{\rho \lambda}(-U + v (\tau'-\tau)) ) ] +
\nonumber \\
&& R^{\gamma \rho}(\tau, q)
R^{\delta \lambda}(\tau',q)
[
\Delta_{\alpha \rho; \delta} (U + v \tau))
( \Delta_{\beta \lambda; \gamma} (-U + v \tau'))
- \Delta_{\beta \lambda; \gamma} (v(\tau+\tau')) )
 \nonumber \\
&& - \Delta_{\alpha \rho; \delta} (v(\tau+\tau'))
\Delta_{\beta \lambda; \gamma} (-U+ v \tau')) ]
\end{eqnarray}

It is also convenient to study the Fourier transform
of the correlator:

\begin{eqnarray}
\Delta_{\alpha \beta}(U) = \sum_K \Delta^{\alpha \beta}_K e^{i K U}
\end{eqnarray}

and to compute the correction to $\Delta^{\alpha \beta}_K$.
We will express it using the response function in
$R^{\gamma \rho}(s=i \omega, q)$
spatial Fourier transform and time Laplace
transform. It is the sum of two contributions.
The contribution of $\Gamma_1$ is:

\begin{eqnarray}
&& \delta \Delta_K^{\alpha \beta} = (- i K_{\gamma}) (- i K_{\delta})
\Delta_K^{\alpha \beta} \sum_{K'} \int_{q}
\Delta_{K'}^{\rho \lambda}
R^{\gamma \rho}(-i v K', -q)
R^{\delta \lambda}(i v K',q) \\
&& - \frac{1}{2} \sum_{K'} \int_{q}
(- i (K-K')_{\gamma}) (- i (K-K')_{\delta})
\Delta_{K-K'}^{\alpha \beta}
[ \Delta_{K'}^{\rho \lambda}
R^{\gamma \rho}(-i v K', -q)
R^{\delta \lambda}(i v K',q) \nonumber \\
&+& \Delta_{-K'}^{\rho \lambda}
R^{\gamma \rho}(i v K', -q)
R^{\delta \lambda}(- i v K',q) ]
\end{eqnarray}

The contribution of $\Gamma_2$ is:

\begin{eqnarray}
&& \delta \Delta_K^{\alpha \beta} =
\sum_{K'} \int_{q}
(- i K')_{\delta}) (- i (K'-K)_{\gamma})
\Delta_{K'}^{\alpha \rho}
\Delta_{K'-K}^{\beta \lambda}
R^{\gamma \rho}(i v K', q)
R^{\delta \lambda}(i v (K' -K),q)
\\
&&
\delta \Delta_K^{\alpha \beta} =
(i K)_{\delta} \Delta_{K}^{\alpha \rho}
\sum_{K'} \int_{q}
(- i K')_{\gamma} \Delta_{K'}^{\beta \lambda}
R^{\gamma \rho}(i v (K+K'), q)
R^{\delta \lambda}(i v K',q)
\nonumber \\
&& - (i K)_{\gamma} \Delta_{-K}^{\beta \lambda}
\sum_{K'} \int_{q}
(- i K')_{\delta}
\Delta_{K'}^{\alpha \rho}
R^{\gamma \rho}(i v K', q)
R^{\delta \lambda}(i v (K'-K),q)
\end{eqnarray}

\subsection{study at finite temperature}

In this subsection we study the case of finite $T>0$.
We start with the static disorder case (corresponding to
subsection () above. The study is rather tedious and we will
skip some details. Since that situation has already been analyzed
(but applied to the different case of a periodic manifold
in $d=2$) we refer to \cite{carpentier_ledou_triangco}
for further details. We will concentrate mostly on what
is needed for the analysis near $d=d_u$ ($d_u=4$ for $v=0$ and $d_u=3$
for $v>0$).

The result (see \cite{carpentier_ledou_triangco}) is
that the short distance expansion of the effective action
up to second order in disorder produces
a $i \hat{u} i \hat{u}$ term which can be written as:

\begin{eqnarray} \label{definition}
\int_{r,t_1,t_2} \delta \Delta^{\alpha \beta}_K e^{-\frac{1}{2} K.B_{0,t_1-t_2}.K }
(i \hat{u}^{\alpha}_{r t_1})
(i \hat{u}^{\beta}_{r,t_2})
e^{i K (u_{r t_1} - u_{r t_2})}
\end{eqnarray}

which is thus of the same form as the first order term and
which thus corrects it. Here again, other operators (such as
higher gradients) are produced,
but they are irrelevant near $d_u$.

We will be using extensively the assumed
symmetries $\Delta^{\alpha \beta}_{K} = \Delta^{\beta \alpha}_{K}
= \Delta^{\alpha \beta}_{-K}$. We are {\it not} using the potential
condition that $\Delta^{\alpha \beta}_{K} \sim K_\alpha K_\beta$ since this is
wrong in general (see discussion above).

We find that the corrections
$\Delta^{\alpha \beta}_K$ are a priori the following,
starting with the terms which will not give a contribution:

(i) the terms with connected averages (1) and (4) give:

\begin{eqnarray}
&& \delta \Delta^{\alpha \beta}_K =
- 2 \Delta^{\alpha \rho}_K \sum_{K'} K'_\gamma K'_\delta
\Delta^{\beta \lambda}_{K'}
R^{\delta \lambda}_{0,\tau_2}
R^{\gamma \rho}_{r,\tau_1}
e^{-\frac{1}{2}  K'.B_{0,\tau_2}.K' }
( e^{K.(C_{r,\tau_1} - C_{r,\tau_1-\tau_2}).K'}
- e^{K.(C_{r,\tau_1+\tau_2} - C_{r,\tau_1}).K'}  ) \\
&& + K_\delta
\Delta^{\alpha \beta}_K \sum_{K'} K'_\gamma
\Delta^{\lambda \rho}_{K'}
R^{\gamma \rho}_{0,\tau_2}
R^{\delta \lambda}_{-r,\tau_1}
e^{-\frac{1}{2}  K'.B_{0,\tau_2}.K' } (
e^{- K.(C_{r,-\tau_1} - C_{r,- \tau_1 - \tau_2}).K'} -
e^{K.(C_{r,-\tau_1} - C_{r,- \tau_1 - \tau_2}).K'} )
\nonumber
\end{eqnarray}

One can check that this term will not produce a divergence.
(check).

(ii) the last two terms of $\delta \Delta^{(3)}$ give:

\begin{eqnarray}
&& \delta \Delta^{\alpha \beta}_K =
\sum_{K'}
( K_\delta K'_\gamma (\Delta^{\alpha \rho}_K \Delta^{\beta \lambda}_{K'})|_{\alpha \beta}
- K_\gamma K'_\delta (\Delta^{\alpha \rho}_{K'} \Delta^{\beta \lambda}_{K})|_{\alpha \beta})
R^{\gamma \rho}_{r,\tau_2}
R^{\delta \lambda}_{-r,\tau_1}
e^{-\frac{1}{2}  K'.B_{0,\tau_1 + \tau_2}.K' }
e^{- K.(C_{r,-\tau_1} - C_{r,\tau_2}).K'}
\end{eqnarray}

where $(..)_{\alpha \beta}$ means symmetrization over the
indices $\alpha, \beta$.
This term was unlikely to produce a divergence for the same
reason as above, but in any case
it does not since it vanishes ! Indeed
one sees on this expression that this quantity vanishes
because of the symmetry $\tau_1 \to \tau_2$ which makes the summmand
over $K'$ odd under $K' \to -K'$.

(iii) finally, the terms which will produce divergences are
the term $\delta \Delta^{(2)}$ and the first term of $\delta \Delta^{(3)}$.
They give a total contribution:

\begin{eqnarray}
&& \delta \Delta^{\alpha \beta}_{P} = \sum_{K,K'=P-K} (
K_\gamma K_\delta \Delta^{\alpha \beta}_K \Delta^{\rho \lambda}_{K'}
R^{\gamma \rho}_{- r,\tau}
R^{\delta \lambda}_{-r,\tau'}
e^{K.(2 C_{0,0} - C_{r,- \tau} - C_{r,- \tau'} ).K'} \\
&& + K'_\gamma K_\delta \Delta^{\alpha \rho}_K \Delta^{\beta \lambda}_{K'}
R^{\gamma \rho}_{r,\tau}
R^{\delta \lambda}_{-r,\tau'}
e^{K.(2 C_{0,0} - C_{r,\tau} - C_{r,- \tau'} ).K'} )
\\
&&
- P_\gamma P_\delta
\Delta^{\alpha \beta}_P \sum_{K'}
\Delta^{\rho \lambda}_{K'}
R^{\gamma \rho}_{- r,\tau}
R^{\delta \lambda}_{-r,\tau'}
e^{-\frac{1}{2}  K'.B_{0,\tau' -\tau}.K' }
\cosh{P.(C_{r,- \tau} - C_{r,- \tau'}).K'}
\end{eqnarray}

we are using extensively the assumed
symmetries $\Delta^{\alpha \beta}_{K} = \Delta^{\beta \alpha}_{K}
= \Delta^{\alpha \beta}_{-K}$. We are {\it not} using the potential
condition that $\Delta^{\alpha \beta}_{K} \sim K_\alpha K_\beta$ since this is
wrong in general (see discussion above).

This term will produce a divergence at $d_u$. It is simply the finite
temperature generalization of () above. The idea is that since
$B(r,\tau)$ is finite (and cutoff dependent) at large $r$, $\tau$
and since at $T=0$ the infrared divergence came from the
large $r$, $\tau$ values, the new IR divergence is the same as the
old one, with a coefficient obtained by simply
by taking the large $r$, $\tau$ limit in the exponential factors.

Near $d_u$ the large time or space limit of $B(r,\tau)=2(C_{0,0} - C_{r,\tau})$
is proportional to the temperature:

\begin{eqnarray}
\lim_{max(r,t) \to \infty} B^{\beta \alpha}(r,\tau) =
2 C^{\beta \alpha}_{0,0} = B_\infty \delta_{\alpha \beta} =
2 T \int_q ( \frac{P^T_{\beta \alpha}(q)}{c_T(q)} + \frac{P^L_{\beta \alpha}(q)}{c_T(q)} )
\end{eqnarray}

If one assumes a circular cutoff $\Lambda=2 \pi/a$ (case $v=0$), one finds:

\begin{eqnarray}
B_\infty = 2 T \frac{S_d}{(2 \pi)^d}
2 T (\frac{1}{d c_L} + \frac{d-1}{d c_T}) \int_0^\lambda q^{d-1} dq \frac{1}{q^2}
\end{eqnarray}

For isotropic elasticity we simply obtain

\begin{eqnarray}
B_\infty = \frac{2 T}{c} \frac{S_d}{(d-2) (2 \pi)^2} a^{2-d}
\end{eqnarray}

The final divergent contribution will be:

\begin{eqnarray}
&& \delta \Delta^{\alpha \beta}_{P} = \sum_{K,K'=P-K} (
K_\gamma K_\delta \Delta^{\alpha \beta}_K \Delta^{\rho \lambda}_{K'}
R^{\gamma \rho}_{- r,\tau}
R^{\delta \lambda}_{-r,\tau'}
+ K'_\gamma K_\delta \Delta^{\alpha \rho}_K \Delta^{\beta \lambda}_{K'}
R^{\gamma \rho}_{r,\tau}
R^{\delta \lambda}_{-r,\tau'} )
e^{K.B_\infty.K'}
\\
&&
- P_\gamma P_\delta
\Delta^{\alpha \beta}_P \sum_{K'}
\Delta^{\rho \lambda}_{K'}
R^{\gamma \rho}_{- r,\tau}
R^{\delta \lambda}_{-r,\tau'}
e^{-\frac{1}{2}  K'.B_{\infty}.K' }
\end{eqnarray}

\subsection{second order corrections to the mobility}

The second order renormalization of the mobility
comes from the $i \hat{u}$ terms. We will only give them in
the one component case. They are given by:

\begin{eqnarray}
-2 \delta S_{eff} &=& i \hat{u_{t_1}} (
- 2 \Delta'(u_{t_1} - u_{t'_1}) \Delta''(u_{t_2} - u_{t'_2})
R_{t'_2 t_2} (R_{t_2 t'_1} - R_{t'_2 t'_1})
(R_{t_1 t'_2} - R_{t'_1 t'_2})
\nonumber \\
&&
- 2 \Delta''(u_{t_1} - u_{t'_1}) \Delta'(u_{t_2} - u_{t'_2})
R_{t_1 t'_1} R_{t'_2 t_2} (R_{t_1 t'_2} - R_{t'_1 t'_2})
 +
\nonumber \\
&&
\Delta''(u_{t_1} - u_{t'_1}) \Delta'(u_{t_2} - u_{t'_2})
(R_{t_1 t_2} - R_{t'_1 t_2})
(R_{t_1 t'_2} - R_{t'_1 t'_2})
(R_{t_2 t'_1} - R_{t'_2 t'_1}) +
\nonumber \\
&&
\Delta'''(u_{t_1} - u_{t'_1}) \Delta(u_{t_2} - u_{t'_2})
 R_{t_1 t'_1}
(R_{t_1 t_2} - R_{t'_1 t_2})
(R_{t_1 t'_2} - R_{t'_1 t'_2}) )
\end{eqnarray}

Note that one additional contraction vanishes since it
involves $R_{t_2 t'_2} R_{t'_2 t_2}=0$.

A preliminary investigation shows that no IR divergences are 
found either to second order (which would mean $theta=1$
to $O(\epsilon^2)$), but this has to be confirmed by
a more detailed analysis.

\section{Analysis in $d=2$}

\label{appendix2d}

We start from the MSR dynamical action corresponding to
equation (\ref{eqd2}) is:

\begin{eqnarray} \label{totaldco}
S[u,\hat{u}] & = &  S_0[u,\hat{u}] + S_2[u,\hat{u}] + S_{int}[u,\hat{u}] \nonumber \\
S_0[u,\hat{u}] & = & \int_{r r' t t'} ~ i \hat{u}_{rt}  (R^{-1})_{rt r't'}  u_{r't'}
- \int_{r t} \eta T  i \hat{u}_{rt} i \hat{u}_{rt} \nonumber \\
S_2[u,\hat{u}] & = &  \frac{1}{2} \int_{q,t,t'}
(i \hat{u}_{q,t}) (i \hat{u}_{-q,t'}) (\Delta q^2 + \Delta_0)
\nonumber \\
S_{int}[u,\hat{u}] & = &
- \int dr dt dt' \frac{1}{2} (i \hat{u}_{rt}) (i \hat{u}_{rt'})
\Delta(u_{rt} - u_{rt'})
\end{eqnarray}
where
\begin{equation}
R^{-1} = \eta \partial_t - c_x \partial_x^2 - c_y \partial_y^2
+ v \partial_x
\end{equation}
is the two dimensional propagator defined in (\ref{defr}).

In the absence of disorder the free action $S_0$ yields the following
correlators $C$ in (\ref{lesc})
($\Delta=0$ in (\ref{totaldco})):
\begin{eqnarray} \label{lescorrel}
R(r,t) &=& \int_q e^{i q r} \eta^{-1} e^{- (c_x q_x^2 + c_y q_y^2 + i v q_x) t/\eta}
\theta(t)
= \int_{q,\omega} e^{i q r + i \omega t} \frac{1}{c_x q_x^2 + c_y q_y^2 + i v q_x + i \eta \omega}
\nonumber \\
C(r,t) &=& \int_q \frac{T}{c_x q_x^2 + c_y q_y^2} e^{- ( c_x q_x^2 + c_y q_y^2 ) |t|/
\eta }
e^{i q r - i v q_x t/\eta}
= \int_{q,\omega} e^{i q r + i \omega t } \frac{2 T \eta}{|c_x q_x^2 + c_y q_y^2
+ i v q_x + i \eta \omega|^2}
\nonumber \\
B(r,t) &=& \int_q \frac{2 T}{c_x q_x^2 + c_y q_y^2} (1 -
e^{- ( c_x q_x^2 + c_y q_y^2 ) |t|/\eta } \cos(q r + v q_x t/\eta) )
\nonumber \\
&=& \int_{q,\omega} (1 - \cos(q r + \omega t ))
\frac{4 T \eta}{| c_x q_x^2 + c_y q_y^2 + i v q_x + i \eta \omega|^2}
\end{eqnarray}
Note that for $v >0$ $C(r,t) \neq C(-r,t)$.

In the presence of disorder one studies perturbation theory
expanding in the interaction term $S_{int}$ using the quadratic
part $S_0 + S_2$ as the bare action. The disorder has a quadratic
part $S_2$ which is purely static and is immaterial in the
perturbation theory. Indeed the response function of
$S_0 + S_2$ is identical to the one of $S_0$ and
the correlation function is changed as:
$C_{q,t} \to C_{q,t} +
{C_{stat}}_{q,t}$ with:
\begin{equation}
{C_{stat}}_{q,t} = \frac{\Delta q^2 + \Delta_0}{c^2 q^4 +  v^2 q_x^2}
\end{equation}
which is purely static and does not appear in any diagram of
perturbation theory. Thus to establish the dynamical RG equations
one can consider that $S_0$ is used as the bare action.

As in section~\ref{appendixb}, we compute the
effective action $\Gamma[u,\hat{u}]$ in perturbation of $S_{int}$.
To lowest order one gets
\begin{equation}
\Gamma[u,\hat{u}] = S_0[u,\hat{u}] + S_2[u,\hat{u}] +
\langle S_{int}[u+\delta u, \hat{u} + \delta  \hat{u}]
 \rangle_{\delta u, \delta \hat{u}}
\end{equation}
which gives
\begin{equation} \label{effectco}
\Gamma[u,\hat{u}] = S_0 + S_2 -
\int_{r t t'} R_{rt,rt'} (i \hat{u}_{rt})
\langle \Delta'(u_{rt} - u_{rt'}) \rangle
- \frac{1}{2} \int_{r t t'} (i \hat{u}_{rt}) (i \hat{u}_{rt'})
\langle \Delta(u_{rt} - u_{rt'}) \rangle
\end{equation}
where as usual in $\langle F[u] \rangle$ one has split the fields
and use $S_0$ to make the
averages $\langle F[u+ \delta u] \rangle_{\delta u}$.
The average in (\ref{effectco}) is easy to perform and gives
\begin{equation} \label{deltasdco}
\delta \Gamma = \int_{r t t'} (i \hat{u}_{rt})~g \sin(u_{rt} - u_{rt'})
R_{r t r t'} e^{ - \frac{1}{2} B_{r t r t'} }
- \frac{1}{2} (i \hat{u}_{r t}) (i \hat{u}_{r t'})~
g \cos(u_{rt} - u_{rt'}) e^{ - \frac{1}{2} B_{r t r t'} }
\end{equation}
where:
\begin{equation} \label{lesc}
B_{r t r' t'} = C_{r t r t} + C_{r' t' r' t'}
- C_{r t r' t'} - C_{r' t' r t}
\end{equation}
The expression (\ref{deltasdco}) immediately
yields the corrections to first order in
$g$ for the friction coefficient, the temperature,
and the disorder given in the text.
The correction to $\eta$ comes from a
a gradient expansion in time which yields a correction
to the term $i\hat{u}_{rt} \eta \partial_t u_{rt}$.
The correction to $\eta T$ from a correction to
the $i\hat{u}_{rt} i\hat{u}_{rt}$
and the correction to disorder from the long time limit
of the exponential.

To compute the RG equations from (\ref{corrections}), we have to decide
on a regularization scheme. Here we choose
to take an infrared regulator by defining a large time
$t_{max}$ but no infrared regulator in momentum $q$.
The ultraviolet cutoff is enforced via a gaussian
cutoff in momentum, i.e we define:

\begin{equation} \label{bregul}
B(r,t,a) = \int_q  \frac{d^2 q}{(2 \pi)^2}
\frac{2 T}{c q^2}
(1 - e^{- c q^2 \mu |t|} e^{i q r + i v q_x \mu t} ) e^{-a^2 q^2}
\end{equation}
where the mobility $\mu = 1/\eta$ has been introduced.
(\ref{bregul}) can be readily evaluated as:
\begin{equation} \label{bcomplic}
B(r,t,a) = \frac{2 T}{c} \int_{a^2}^{+ \infty} ds
\int_q \frac{d^2 q}{(2 \pi)^2}
(e^{- s q^2}  - e^{- (s + c \mu t ) q^2 } \cos(v q_x \mu t) )
= \frac{T}{2 \pi c} \int_{0}^{\frac{c \mu t}{c \mu t+a^2}} \frac{du}{u} (\frac{1}{1-u}
- e^{- \frac{y^2 + (x+ v \mu t)^2}{4 c \mu t}   u} )
\end{equation}
where in the intermediate stage we have integrated over
$q$ and performed an intermediate change of
variable $u=c \mu t/(c \mu t + s)$.
Using
\begin{eqnarray}
\int_{0}^{z} \frac{du}{u} (1 -
e^{- r u} ) = C + \ln(r z) - Ei[- r z]) \qquad\qquad
Ei[x] = \int_{-\infty}^{x} e^t \frac{dt}{t}  & & \\
Ei[- x]  \sim_{x \ll 0} C
+ \log (-x) - x + {{{x^2}}\over 4} + O(x^3) & & \nonumber
\end{eqnarray}
one obtains for (\ref{bcomplic})
\begin{equation} \label{bsimple}
B(r,t,a) =  \frac{T}{2 \pi c}
( Log[\frac{c \mu |t| +a^2}{a^2}] + C + Log[\frac{y^2 + (x+ v \mu t)^2}{4(c \mu |t| +a^2)}]
- Ei[\frac{- (y^2 + (x+ v \mu t)^2) }{4(c \mu |t| +a^2)}] )
\end{equation}

\section{stability of FRG equation}
\label{stability}

One can show that the fixed point solution is locally {\it attractive}
within the space of periodic functions on [0,1]. One can in
fact diagonalize it completely. This allows to study
crossovers in precise way.

The function $\overline{\Delta}(u) = \Delta(0) - \Delta(u)$
is positive on the interval [0,1] and satisfies the RG equation:

\begin{eqnarray}
\frac{d}{dl} \overline{\Delta}(u) =
(1 + \overline{\Delta}''(u)) \overline{\Delta}(u)
\end{eqnarray}

with the conditions $\overline{\Delta}(0)= \overline{\Delta}(1) =0$.

We have checked numerically that analytic initial functions (i.e
with zero odd derivatives at $0$) converge
towards the non analytic fixed point $\overline{\Delta}^*(u)=u (1-u)/2$.
The stability analysis is performed by writing
$\overline{\Delta}(u)=\overline{\Delta}^*(u)+ \delta(u)$. One then
has to solve the eigenvalue problem:

\begin{eqnarray}
\frac{1}{2} u (1-u) \delta''(u) = - \lambda \delta(u)
\end{eqnarray}

The eigenfunctions are such that:

\begin{eqnarray}
\frac{1}{2} u (1-u) \delta_n''(u) = - \frac{1}{2} n (n-1) \delta_n(u)
\end{eqnarray}

One can also define the variable $u = (1+v)/2$. Then
the eigenfunctions are the Jacobi polynomials $\delta_n(u) = P^{-1.-1}_n(v)$
(see Abramowitz and Stegun p 779) and form an orthonormal complete
set. They can be written as:

\begin{eqnarray}
\delta_n(u) = \frac{(-1)^n}{n!} u (1-u) \frac{d^n}{d u^n} [ u(1-u) ]^{n-1}
= \frac{(-1)^n}{2^n n!} (1-v^2) \frac{d^n}{d v^n} (1-v^2)^{n-1}
\end{eqnarray}

Because of the $u \to -u$ symmetry, which due to periodicity
becomes $u \to (1-u)$ symmetry, we can restrict ourselves to
$n$ an even and non zero positive integer.
The lowest eigenmodes are thus $\delta_2(v)= (v^2-1)/4$ (eigenvalue $-1$),
$\delta_4(v)= 3(1-6v^2 + 5 v^4)/16$ (eigenvalue $-6$),
$\delta_6(v)= 5(v^2-1)(1-14v^2+ 21 v^4)/32$ (eigenvalue $-15$), etc..
Note that they satisfy $\delta_n(v=-1) = \delta_n(v=1) =0$ as requested.
Note that all these eigenfunctions are non analytic
(though by combining several one may get analytic ones). Rendering the initial function
non analytic is presumably the role of the non linearity. This equation
is interesting since it is the simplest case on which one can work out
the full stability spectrum and it may enlight us about the generation of
non analyticity in these type of RG equations.

\section{parametrization of moving structures}

\label{parametrization}

Moving structures can be generally parametrized by
their internal space $D$, the number of components
$n$ of the displacements fields (characterizing its deformations - or the number of 
components of the order parameter) and the embedding space $d$.
We denote for convenience by the same
symbol the space itself and its dimension.

In the statics one can distinguish several cases.
The problem of manifolds in random potentials
has been studied for e.g 
(i) fully oriented manifolds where $D$ and $n$ are
orthogonal ($d=D+n$) such as directed polymers
or interfaces (ii) isotropic manifolds $n=d$
such as self avoiding chains in random potentials
(iii) problem of lattices where usually 
$d=D$ and $n \le d$. Lattices with $D<d$ are
possible in principle, such as flat but fluctuating
tethered mebranes $D=n<d$ or isotropic tethered membranes
$D<d=n$ or any intermediate case (the so-called tubules).

In the driven dynamics, let us call $x$
the direction in the embedding space
along which the system is driven. 

One can distinguish the following cases:

(A) the structure is elastic (i.e not liquid)
in the direction where it is driven. Then
there is a displacement $u_x$ along $x$ and 
$x$ also belongs to the $n$-space. There are
two subcases:

(A1) $x$ also belongs to the internal space $D$.
This is the problem of manifolds driven {\it along
one of their internal dimension}, to which 
the moving glass studied here belongs.

A general parametrization in that case would be
\begin{equation}
D=(x, y_1, z)
\qquad n=(u_x,u_y=(u_{y_1},u_{y_2}))
\qquad d=(x,y_1,y_2,z)
\end{equation}

it allows for manifolds with $D<d$ which do not entirely
fill space (i.e with ``height'' degrees of freedom $u_{y2}$).
Then a parametrization of the dimensions (and the subspaces)
is:

\begin{equation}
D=1+d_1+d_z
\qquad n=1+d_1+d_2
\qquad d=1+d_1+d_2+d_z
\end{equation}

where $d_1$ and $d_2$ are the number of components of
$u_{y_1}$ and $u_{y_2}$ respectively, and $d_z$ the
number of components of $z$. In this paper we
have mainly studied the case $d=D$ ($d_2=0$)
but with $d_1>0$. Note that a single $Q$ CDW has $d_1=d_2=0$ ($u_y=0$)
and $d=D$.

(A2) $x$ does not belong to the internal space $D$.
This is the problem of manifolds driven
{\it perpendicular to their internal dimension}.
The general parametrization in that case is:

\begin{equation}
D=(y_1,, z)
\qquad n=(u_x,u_y=(u_{y_1},u_{y_2}))
\qquad d=(x=u_x,y_1,y_2,z)
\end{equation}

and thus $D=d_1+d_z$, $n=1+d_1+d_2$ and $d=1+d_1+d_2+d_z$.
It also contains the case of a driven order parameter $u$
which does not couple at all to internal space
(such as a vector spin order parameter). Indeed in that 
particular case one can define the ``embedding
space'' as the sum $d=D+n$ (and thus $u_{y_1}=0$ ).

(B) The structure is a {\it liquid} in the
direction where it is driven. Then
$x$ belongs to $D$-space but not 
to $n$ space. Then one sets $u_x=0$
in the case (A1) above, i.e $n=d_1 + d_2$. 
The parametrization is thus:

\begin{equation}
D=(x, y_1, z)
\qquad n=(u_{y_1},u_{y_2})
\qquad d=(x,y_1,y_2,z)
\end{equation}

with $D=1+d_1+d_z$, $n=d_1+d_2$, $d=1+d_1+d_2+d_z$. The Transverse
Moving Glass studied here is one example 
with $d_2=0$ and ($d_1=1$, $d_z=0$) in 
$d=2$ and ($d_1=1$, $d_z=1$) in $d=3$
(a moving line lattice giving a smectic sheets
structure of channels)
and ($d_1=2$, $d_z=0$) in $d=3$
(a moving point lattice giving a line crystal
structure of channels). Note that as for any
liquid scalar density fluctuations should in
principle be also incorporated in a
complete description.

\section{Hartree method}

For completeness we give here the Hartree equations
exact in the large $N$ limit, for Model III generalized 
to $N$ components. The equations at $v=0$ were
derived and analyzed in \cite{ledou_polymer_longrange}
(see also \cite{cugliandolo_ledoussal_manifold,cugliandolo_ledoussal_zerod}.
These equations will be analyzed further in
a future publication.

The Hartree equations are:

\begin{eqnarray}
\partial_t R_{k t t'}
&=& 
-( k^2 + i v k_x ) R_{k t t'}
+ 4 \int_0^t ds V_2''(B_{t s}) R_{t s}
( R_{k t t'} - R_{k s t'} ) 
\nonumber 
\\
\partial_t C_{k t t'}
&=& 
- (k^2 + i k_x v) C_{k t t'} 
+ 2 \int_0^{t'} ds \;  V_1'(B_{t s})  R_{-k t' s}
\nonumber \\
&&
+ 4 \int_0^t ds V_2''(B_{t s}) R_{t s}
( C_{k t t'} - C_{k s,t'} )
+ 2 T \; R_{-k t' t}
\end{eqnarray}
where $R_{t t'} = \int_k R_{k t t'}$, 
$B_{t t'} = \int_k B_{k t t'}$ and
$B_{k t t'}=C_{k t t} + C_{k t' t'} -
C_{k t t'} - C_{- k t t'}$, noting that
$C_{k t t'}=C_{-k t' t}$.

where $V_2$ contains only the potential part
of disorder while $V_1$ contain all disorder
(see \cite{ledou_polymer_longrange} for 
definitions). One can look for a time-translational invariant solution
of these equations: $R_{k t t'}=r_k(t-t')$,
$C_{k t t'}=c_k(t-t')$ \cite{ledou_polymer_longrange}
(in the statics this is the equivalent
of a replica symmetric solution). It can be written
in Fourier transform version:

\begin{equation}
r_k(\omega) = \frac{1}{ i \omega + k^2 + i v k_x + \Sigma(0) - \Sigma(\omega)} \qquad
c_k(\omega) = \frac{ 2 T + D(\omega)}{ 
| i \omega + k^2 + i v k_x + \Sigma(0) - \Sigma(\omega) |^2}
\end{equation}

Note that $c_k(\omega) = c_{-k}(-\omega)$. We have defined:

\begin{equation}
\Sigma(\omega) = - 4 \int_{-\infty}^{+\infty} dt e^{i \omega t} 
V_2''(b(t)) r(t)   \qquad 
D(\omega) = 4 \int_{-\infty}^{+\infty} dt e^{i \omega t} V_1'(b(t)) 
\end{equation}

where $r(t)= \int_k r(t)$ and $b(t) = \int_k b_k(t) =
\int_k (2 c_k(t=0)  - c_k(t) - c_{-k}(t))$. Note that
$b_k(t) = \int_{\omega} (1-e^{i \omega t}) (c_k(\omega)
+ c_{-k}(\omega)$ and $b(t) = \int_{\omega} (1- \cos(\omega t)) c(\omega)$
where $c(\omega)=\int_k c_k(\omega)$ is an even function of $\omega$.

A very superficial analysis of the above equation would indicate
that this non periodic problem has asymptotically linear response 
and is not glassy for $v>0$ for $N = \infty$
(while it has non linear asymptotic response for $v=0$ both at $N$ finite
\cite{ledou_polymer_longrange} and $N$ infinite \cite{ledou_wiese_ranflow})
Indeed the response function seems to be massive since
integrating over $k_x$ one has:

\begin{equation}
r(\omega) = \frac{1}{2} \int_{k_y}  
\frac{1}{ (k_y^2 + i \omega + \Sigma(0) - \Sigma(\omega) +
\frac{v^2}{4})^{1/2} }
\end{equation}

Thus the response to an applied force being $F/V = ( 1 - \Sigma'(0) )$
would be linear at least at the most naive level (for a more detailed behaviour 
one must add a small transverse force and follow the methods
of \cite{horner_particle_driven}). This is related to the absence of divergence
for $\eta$ noticed in the FRG Section \ref{renormalizationtransverse}.
Further investigations would be necessary however before reaching a 
conclusion. One should makes sure that no transition occur in the
above equations (such can happen in the case $v=0$). Also,
it is possible that the glassy physics found in Section \ref{renormalizationtransverse}
which comes from a renormalization of the disorder may not be fully
captured here by the most naive large $N$ limit.

\section{Direct method of Renormalization}

We present here a simple method to obtain the 
FRG equations for Model III at $T=0$, based on
a mode elimination by hand. This approach is
less rigorous than the full calculation of the
effective action performed in (\ref{appendixc}).
However it is quite direct and illustrate 
a simple way how non potential disorder
is generated in this problem. 

We start from:

\begin{equation}
{(G^{-1})}^{\alpha \beta}_{r r'} u^{\beta}_{r'} = F_{\alpha} (r,u_{r})
\end{equation}

and separate $u \to u + \delta u$. The equation for the
fast modes $\delta u$ is:

\begin{equation}
{(G^{-1})}^{\alpha \beta}_{r r'} \delta u^{\beta}_{r'} = F_{\alpha} (r,u_{r} + \delta u_{r})
\approx F_{\alpha} (r,u) + F_{\alpha;\beta } (r,u) \delta u^{\beta}_{r} 
\end{equation}
to lowest order. Solving to the same order one has:

\begin{equation}
\delta u^{\alpha}_{r} = G^{\alpha \beta}_{r r_1} F_{\beta}(r_1,u)
+ G^{\alpha \beta}_{r r_1} F_{\beta ; \gamma}(r_1,u) 
G^{\gamma \delta}_{r_1 r_2} F_{\delta}(r_2,u)
\end{equation}

The equation for the slow modes is now:

\begin{eqnarray}
{(G^{-1})}^{\alpha \beta}_{r r'} u^{\beta}_{r'} & = & F_{\alpha} (r,u_{r} + \delta u_{r})
\approx F_{\alpha} (r,u) + F_{\alpha;\beta } (r,u) \delta u^{\beta}_{r} 
+ \frac{1}{2} F_{\alpha;\beta \gamma} (r,u) \delta u^{\beta}_{r} \delta u^{\gamma}_{r}
\nonumber \\
&&
\approx F_{\alpha} + F^{(2)}_{\alpha} + F^{(3)}_{\alpha} + F^{(4)}_{\alpha}
\end{eqnarray}

One finds:

\begin{eqnarray}
F^{(2)}_{\alpha} &=&
F_{\alpha;\beta}(r,u) G^{\beta \gamma}_{r r_1} F_{\gamma}(r_1,u)
\nonumber \\
F^{(3)}_{\alpha} &=&
F_{\alpha;\beta}(r,u) G^{\beta \gamma}_{r r_1} F_{\gamma;\delta}(r_1,u)
G^{\delta \lambda}_{r_1 r_2} F_{\lambda}(r_2,u)
\nonumber \\
F^{(4)}_{\alpha} &=&
\frac{1}{2}
F_{\alpha;\beta \gamma}(r,u) G^{\beta \delta}_{r r_1} F_{\delta}(r_1,u)
G^{\gamma \lambda}_{r r_2} F_{\lambda}(r_2,u)  \label{forces}
\end{eqnarray}

It is easy to see that if $G_{r,r'} \ne G_{r',r}$ the above
forces are {\it non potential} even if the original
one is the gradient of a potential. One can now compute the
new disorder correlator $\Delta_{\alpha \alpha'}(u-u',r-r')$ which
has several contributions:

\begin{eqnarray}
\Delta^{(2)}_{\alpha \alpha'} &=& \overline{ F^{(2)}_{\alpha}(r,u) F^{(2)}_{\alpha'}(r',u') }
= - \delta_{r r'} \Delta_{\alpha \alpha' ; \beta \beta' }(u-u')
\Delta_{\gamma \gamma'}(u-u') G^{\beta \gamma}_{r r_1} 
G^{\beta' \gamma'}_{r r_1} - \nonumber \\
&&
\Delta_{\alpha' \gamma ; \beta' }(u-u')
\Delta_{\alpha \gamma' ; \beta }(u-u')
G^{\beta \gamma}_{r r'} 
G^{\beta' \gamma'}_{r' r}  \nonumber \\
\Delta^{(3)}_{\alpha \alpha'} &=& \overline{ F_{\alpha'}(r',u') F^{(3)}_{\alpha}(r,u) }=
- \Delta_{\alpha' \lambda}(u-u') 
\Delta_{\alpha \gamma ; \beta \delta}(0)  G^{\beta \gamma}_{r r}
G^{\delta \lambda}_{r r'} \nonumber \\
\Delta^{(4)}_{\alpha \alpha'} &=& \overline{ F_{\alpha'}(r',u') F^{(4)}_{\alpha}(r,u) }=
\frac{1}{2} ( \delta_{r r'}
\Delta_{\alpha \alpha' ; \beta \gamma}(u-u')
\Delta_{\delta \lambda}(0)
G^{\beta \delta}_{r r_1}
G^{\gamma \lambda}_{r r_1} +
\nonumber \\
&&
\Delta_{\alpha' \delta}(u-u')
\Delta_{\alpha \lambda ; \beta \gamma}(0)
G^{\beta \delta}_{r r'}
G^{\gamma \lambda}_{r r}
+ 
\Delta_{\alpha' \lambda}(u-u')
\Delta_{\alpha \delta ; \beta \gamma}(0) 
G^{\beta \delta}_{r r}
G^{\gamma \lambda}_{r r'}) 
\label{correlators}
\end{eqnarray}

Putting everything together, in the long
wavelength limit:

\begin{eqnarray}
\delta \Delta_{\alpha \alpha'}(u-u') \delta_{r r'} \sim \Delta^{(2)}_{\alpha \alpha'} +
\Delta^{(3)}_{\alpha \alpha'} + \Delta^{(3)}_{\alpha' \alpha} +
\Delta^{(4)}_{\alpha \alpha'} + \Delta^{(4)}_{\alpha' \alpha}  
\end{eqnarray}

one notes that $\Delta^{(3)}$ cancels the last two terms of 
$\Delta^{(4)}$ and thus finds:

\begin{eqnarray}
&& \delta \Delta_{\alpha \beta}(u-u') = 
\Delta_{\alpha \beta ; \gamma \delta}(u)
( \Delta_{\alpha_2 \beta_2}(0)  - \Delta_{\alpha_2 \beta_2}(u) ) 
\int_r \delta G_{\gamma \alpha_2}(r) \delta G_{\delta \beta_2}(r)
\nonumber \\
&& - \Delta_{\alpha \alpha' ; \delta}(u)
\Delta_{\beta \beta' ; \gamma}(u) 
\int_r \delta G_{\gamma \alpha'}(r) \delta G_{\delta \beta'}(-r)
\end{eqnarray}

One thus recovers the result of (\ref{appendixc}).
The time-dependent problem at $T=0$ can also be analyzed in the same 
way. All the above equations up to (\ref{forces}) are identical
with $r$ replaced by $r t$ and $G_{r r'}$ replaced by $R_{r t r' t'}$.
In (\ref{correlators}) one gets integrals over times 
and again the zero-frequency $G_{r r'}$.


\begin{thebibliography}{100}
\bibitem{footnote_balents}
The authors of \cite{balents_mglass_comment_reply} did criticize us for setting
  $u_x=0$ and use the resulting Model III to describe the physics of the moving
  glass. As we pointed out to them, they have missed the important point of
  physics that irrespective of the behaviour along $x$ the physics is
  controlled by the moving glass equation.
\bibitem{footnote_hauger}
It is interesting to note that some breakdown of the perturbation theory was
  noticed \cite{schmidt_hauger} in the perturbative calculation of the
  displacements at fourth order. The condition for such terms to be negligible
  was noticed to be violated for a pinning force perpendicular to the flow
  velocity. However no direct physical consequences were investigated at that
  time. Indeed divergences in the perturbation theory are well hidden, and for
  instance do not appear in the velocity expansion. At lowest order, as can be
  seen from (\ref{hauger}) they are killed by terms like $v q$ in the numerator
  that vanish for directions orthogonal to the displacements.
\bibitem{footnote_solutions}
There is however a degenerate family of discontinuous solutions which may have
  some physical interpretation.
\end{thebibliography}

\end{document}